\documentclass[a4paper,11pt]{article}
\pdfoutput=1 
\usepackage{jheppub}	
\usepackage[utf8]{inputenc}
\usepackage[english]{babel}
\usepackage{makeidx}
\usepackage{amsfonts}
\usepackage{enumerate}
\usepackage{mathrsfs}
\usepackage{tensor}
\usepackage{xcolor,tikz,pgfplots}
\usepackage{graphicx}

\usepackage{cancel}

\usepackage[autostyle]{csquotes}

\usepackage{youngtab}
\usepackage{caption}
\usepackage{subcaption}
\newcommand\subcap[1]{\phantomcaption%
       \caption*{\textbf{\figurename~\thefigure\thesubfigure:} #1}} 

\usepackage{braket}

\usetikzlibrary{matrix,calc,positioning,decorations.markings,decorations.pathmorphing,decorations.pathreplacing}
\usetikzlibrary{arrows,cd}
\usetikzlibrary{shapes.misc}
\usetikzlibrary {shapes.geometric}
\usetikzlibrary {shapes.multipart}
\usepackage{bbding}
\tikzset{
    >=stealth',
    punkt/.style={
           rectangle,
           rounded corners,
           draw=black, very thick,
           text width=7.2em,
           minimum height=2em,
           text centered},
    punkt2/.style={
           rectangle,
           rounded corners,
           draw=black!20!red, very thick,
           text width=7em,
           minimum height=2em,
           text centered},
    punktL/.style={
           rectangle,
           rounded corners,
           draw=black!20!red, very thick,
           text width=8.8em,
           minimum height=2em,
           text centered},
    pil/.style={
           ->,
           thick,
           shorten <=2pt,
           shorten >=2pt,},
    pil2/.style={
           <->,
           thick,
           shorten <=2pt,
           shorten >=2pt,},
    pil3/.style={
           -,
           thick,
           shorten <=2pt,}}

\usepackage{float}
\usepackage{booktabs}

\usepackage{bm}
\usepackage{anyfontsize}

\newcommand{\NN}{\mathcal{N}}

\numberwithin{equation}{section}

\title{{\fontsize{15.1pt}{17pt}\selectfont Multi-planarizable quivers, orientifolds, and conformal dualities}}

\author[a]{\fontsize{12.5pt}{13pt}\selectfont Antonio Amariti,}
\author[b]{Massimo Bianchi,}
\author[c,d]{Marco Fazzi,} 
\author[e,f]{Salvo Mancani,}
\author[f]{Fabio Riccioni,}
\author[a,g]{Simone Rota}

\affiliation[a]{INFN, Sezione di Milano, Via Celoria 16, I-20133 Milano, Italy}
\affiliation[b]{Dipartimento  di  Fisica,  Universit\`a  di  Roma  ``Tor  Vergata'' \& INFN, Sezione di Roma ``Tor Vergata'', Via della Ricerca Scientifica 1, I-00133 Roma, Italy}
\affiliation[c]{Department of Physics and Astronomy, Uppsala University, SE-75120 Uppsala, Sweden}
\affiliation[d]{Dipartimento di Fisica, Universit\`a di Milano-Bicocca \& INFN, Sezione di Milano-Bicocca, Piazza della Scienza 3, I-20126 Milano, Italy}
\affiliation[e]{Dipartimento di Fisica, Università di Roma ``La Sapienza'', Piazzale Aldo Moro 2, I-00185 Roma, Italy}
\affiliation[f]{INFN, Sezione di  Roma, Piazzale Aldo Moro 2, I-00185 Roma, Italy}
\affiliation[g]{Dipartimento di Fisica, Universit\`a degli Studi di Milano, Via Celoria 16, I-20133 Milano, Italy}

\emailAdd{antonio.amariti@mi.infn.it,  marco.fazzi@physics.uu.se,  simone.rota@mi.infn.it, massimo.bianchi@roma2.infn.it, salvo.mancani@uniroma1.it, Fabio.Riccioni@roma1.infn.it}

\abstract{
We study orientifold projections of families of four-dimensional $\mathcal{N}=1$ toric quiver gauge theories. 
We restrict to quivers that have the unusual property of being associated with multiple periodic planar diagrams which give rise, in general, to inequivalent models. A suitable orientifold projection relates a subfamily of the latter by conformal duality. That is, there exist exactly marginal deformations that connect the projected models. The deformations take the form of a sign flip in some of the superpotential interactions, similarly to the $\beta$-deformation of $\mathcal{N}=4$ SYM. 
Our construction generalizes previous results on the orientifold projections of the PdP$_{3b}$ and PdP$_{3c}$ singularities.
}

 \clearpage

\begin{document}

\maketitle

\section{Introduction}

Unexpected relations between different conformal field theories can be often explained in terms of symmetries and dualities.
This is because they can connect non-perturbative regimes where the spectrum reorganizes in non-trivial ways.
A key role is played by the conformal manifold \cite{
Kol:2002zt,Benvenuti:2005wi,Kol:2010ub,Green:2010da}, because the motion on the space of marginal deformations allows to connect apparently unrelated models, for example cases with different amount of visible supersymmetry and global symmetry, or lagrangian theories and non-lagrangian ones.
For this reason in the past few years finding conformally dual theories has become an active field of research \cite{Razamat:2019vfd,Zafrir:2019hps,Zafrir:2020epd,Razamat:2020gcc,Razamat:2020pra,Etxebarria:2021lmq,Razamat:2022gpm,Amariti:2022tbd}. Many examples of conformally dual theories have been found by explicit analysis of the conformal manifold in the case of 4d $\mathcal{N}=1$ superconformal field theories (SCFTs). In most of these examples, orthogonal and symplectic gauge groups (which we will both refer to as real henceforth) and tensor matter fields are required, and it is natural to think that constructing similar models from the string theory perspective requires the use of orientifold planes.

Motivated by such results, a somewhat different strategy has been recently adopted in \cite{Antinucci:2020yki,Antinucci:2021edv,Amariti:2021lhk,Amariti:2022dyi},  applying orientifold projections to  toric quiver gauge theories.  These projections have been performed by using the approach of \cite{Franco:2007ii}, where the action of the orientifold on the gauge theory is read from the identifications imposed on the dimer by consistent choices of fixed points or fixed lines. 
In some cases the conformal dualities have a direct interpretation in terms of Seiberg or (inherited) S-duality \cite{Amariti:2021lhk}, but in general such an interpretation is not possible and the final models have to be thought of as alternative descriptions of the same conformal manifold.

The examples worked out in \cite{Antinucci:2021edv,Amariti:2021lhk,Amariti:2022dyi}, corresponding to orientifolds of $L^{aba}$ and $L^{aba}/\mathbb{ Z}_2$, are of this latter type. The parent models (i.e. before the orientifold projection) describe stacks of D3-branes probing different Calabi--Yau threefolds (CY$_3$'s) and, only after the projection and for suitable choices of ranks, the models have the same conformal and 't Hooft anomalies and superconformal indices. The claim is that the models are also conformally dual and the marginal deformations that relate them are associated to quadratic superpotential deformations.\footnote{Note that these mass terms are relevant in the parent theories \cite{Bianchi:2014qma,Bianchi:2020fuk}, but become marginal after the orientifold projection.} Then, by integrating out such \emph{conformal masses}, the final models become identical.

Another similar behavior was anticipated in \cite{Antinucci:2020yki}. In this case the two models before the projection are associated to D3-branes probing two different singularities, namely (cones over) the pseudo del Pezzo surfaces PdP$_{3b}$ and PdP$_{3c}$. After a fixed-line and fixed-point projection respectively, again with a suitable choice of ranks, it turns out that  the two models 
have the same quiver structure, field content,  anomalies and index. These models differ only by the superpotential interactions, i.e. they differ only by the point on the conformal manifold that they describe, or equivalently there is an exactly marginal deformation connecting them. This notion of conformal duality between these two models generalizes the one found for the $L^{aba}$ and $L^{aba}/\mathbb{ Z}_2$  orientifolds discussed in \cite{Antinucci:2021edv,Amariti:2021lhk,Amariti:2022dyi}, even though in the former there is not any (explicit) \emph{conformal mass} connecting the two theories. 

Motivated by this example, in this paper we look for a generalization of this behavior for infinite families of toric quiver gauge theories. We note that the crucial property at the core of the relation between PdP$_{3b}$ and PdP$_{3c}$  is that the two models can be represented with the same quiver even before the orientifold projection. The quiver admits  two different planarizations, corresponding to the two inequivalent toric descriptions. We then look for families of quivers admitting multiple planarizations, referring to quivers of such type as \emph{multi-planarizable}. We find an infinite family of models of such type: toric diagrams described by three parallel lines connecting the integer point lattices, two on the perimeter and a central one (see Fig. \ref{FIG3}).
We observe  that there always exist coincident quivers (in some Seiberg/toric dual phase) associated to different toric diagrams in this family. The recipe to obtain such quivers corresponds to moving one or more lattice points from one external line to the other.

We then study orientifold projections of these models, and by inspection we identify a subfamily that generalizes the construction discussed above for PdP$_{3b}$ and PdP$_{3c}$. This subfamily corresponds to the case where there is an even number of points on the perimeter. 

We show that after the orientifold (by fixed lines or fixed points, depending on the toric diagram) the models 
differ only by exactly marginal deformations. 
Furthermore we observe that acting with Seiberg duality and global symmetries we can always find a phase where the difference between these models corresponds to a sign or a set of signs in some superpotential interaction. 

This paper is organized as follows. In Sec. \ref{sec:toy} we introduce a toy model which contains all relevant features to explain the conformal dualities we are interested in. In section \ref{sec:gauging} we review the basic conformally dual pair obtained in \cite{Antinucci:2020yki}, namely PdP$_{3b}$ and PdP$_{3c}$. This allows us to reinterpret the conformal duality for these models after the orientifold projection along the lines of the discussion in section \ref{sec:toy}. Furthermore this section is helpful to set the language and the notations that we use  to 
embed, in section \ref{sec:embedding}, the field theories in string theory. Such an embedding is  realized  as an orientifold projection of a toric CY$_3$. In this section we first fix our general setup, by introducing the notion of multi-planarizable quivers and discuss how to obtain them by suitable flips of the dimer. Then we summarize the rules of the orientifold projections that  give rise to the unoriented conformally dual models.
In section \ref{sec:study} we move to a case-study belonging to the new infinite family of conformal dualities introduced in the paper. 
We conclude in section \ref{sec:conc} with some observations and future perspective. An appendix (section \ref{sec:apx}) supplements our analysis, and contains many more examples belonging to the same family. In appendix \ref{sec:apxcounting} we count the number of conformally dual models in each family.

\section{Toy models for conformal dualities}
\label{sec:toy}

In this section we construct pairs of $\NN = 1$ toy models that differ only by a marginal deformation on the conformal manifold. In particular, given a unique quiver we turn on different interaction terms such that the models are not the same. We begin with SQCD in the conformal window, and introduce singlets and interactions so that the global symmetry contains (at least) an $SU(2)$ factor. Turning on further interactions breaks this $SU(2)$ and yields different models. We find a behavior that resembles $\beta$-deformations in $\NN=4$ SYM \cite{Leigh:1995ep}, as the models differ only by a sign flip in a subset of terms.

We exhibit examples with unitary and real gauge factors, which will be embedded in string models in later sections.

\subsection{SQCD with singlets}

The simplest toy model in which the mechanism of conformal duality described above can be found is SQCD with a single unitary gauge group $SU(N)$ and $F$ flavors. We can break the flavor group $SU(F)$ in different patterns, by introducing singlets and turning on interactions. Let us start from the quiver drawn in Fig. \ref{fig:SQCD} with $F$ quarks $q$, $F$ anti-quarks $\widetilde{q}$ and no interaction, and choose $F$ such that the theory stays inside the conformal window, i.e. $F=2N$. Now break the non-abelian global symmetry from $SU(F) \times SU(F)$ into the pattern in Fig. \ref{fig:SQCDPattern0}, $SU(F_1) \times SU(F_2) \times SU(G_1) \times SU(G_2)$ with $F_1 + F_2 = G_1 + G_2 = F$. The quarks $q_1$, $q_2$ transform under $SU(F_1) \times SU(F_2)$, whereas the anti-quarks $\widetilde{q_1}$ and $\widetilde{q}_2$ under $SU(G_1) \times SU(G_2)$. Note that with no gauge singlets and $W=0$ the non-abelian global symmetry of this quiver is actually enhanced back to $SU(F) \times SU(F)$. In order to effectively break the flavor, we introduce gauge singlets $M_{11}$, $M_{12}$, $M_{21}$ and $M_{22}$, which interact with the quarks via a cubic superpotential. Denoting the color indices with $a$ and flavor indices with $f_{\ell}$, $g_j$ (${\ell},j=1,2$), we can write
\begin{align}\label{eq:SQCDW}
    W = h_{{\ell}j} \, {\left(M_{{\ell}j}\right)_{f_j}}^{g_{\ell}} {\left( \widetilde{q}_{\ell} \right)_{g_{\ell}}}^{a} {\left( q_j \right)_{a}}^{f_j} \; ,
\end{align} 
so that we can control the actual global symmetry by turning on subsets of $h_{{\ell}j}$ and imposing conditions on the ranks of the flavor groups. Henceforth we will omit color and flavor indices for clarity. For instance, the theory with only $h_{11} \neq 0$ and $h_{12} = h_{21} = h_{22} = 0$ has superpotential given by
\begin{align}
    W = h_{11} M_{11} \widetilde{q}_1 q_1 \; , 
\end{align}
and if we require further that $F_1 = G_1$ and $F_2 = G_2$, we obtain the quiver of Fig. \ref{fig:SQCDPattern1}. This model is studied in \cite{Barnes:2004jj}, where it is shown that it is dual to the case with $h_{11}=0$ and $h_{12},\, h_{21},\, h_{22} \neq 0$ with a bifundamental singlet, an adjoint singlet on the second node and the same requirement for the ranks of the flavor groups. 

An interesting case is the one given by $h_{11},\, h_{22} \neq 0$ and $h_{12}=h_{21}=0$, thus the interaction has the form
\begin{align}
    W = h_{11} M_{11} \widetilde{q}_{1} q_{1} + h_{22} M_{22} \widetilde{q}_{2} q_{2} \; ,
\end{align}
where two out of four couplings are non-zero. Requiring that $G_1 = G_2$, $F_1 = F_2$ and identifying singlets $M_{11} = M_{22} = M$ and couplings $h_{11} = h_{22} = h$, the resulting quiver is drawn in Fig. \ref{fig:SQCDPattern2} and the superpotential simplifies to
\begin{align}
    W = h M \widetilde{q}_{1} q_{1} + h M \widetilde{q}_{2} q_{2} \; .
\end{align}
Note that we can rotate the quarks $q_1$, $q_2$ and the anti-quarks $\widetilde{q}_1$, $\widetilde{q}_2$, signaling an $SU(2) \times SU(2)$ global symmetry that emerges only under the required condition. Rotating both quarks and anti-quarks gives the same superpotential, while rotating only one species is equivalent to choosing $h_{12}, \,h_{21}$ to be non-zero and the diagonal terms to be zero.

The final case has $h_{{\ell}j}\neq 0$ $\forall \, {\ell},\, j$. We can require that either $F_1 = F_2$, $G_1 = G_2$ so that there are four bifundamental singlets, or $F_1 = G_1$, $F_2 = G_2$ resulting in two conjugated bifundamental singlets and one adjoint singlets at each node \cite{Barnes:2004jj}.  

All patterns of global symmetry breaking discussed above are summarized in Tab. \ref{tab:SQCDPatterns}.

\begin{center}
	\begin{tabular*}{1\textwidth}{@{\extracolsep{\fill}}c|cc}
		\toprule
		Couplings & Superpotential & Global Symmetry  \\
		\midrule
		$h_{ij}=0$ & $W=0$ & $SU(F) \times SU(F)$, Fig. \ref{fig:SQCD} \\[7pt]
		$h_{11}\neq0$ & $W=h_{11}M_{11}\widetilde{q}_1 q_1$ & $SU(F_1)^2 \times SU(F_2)^2$, Fig. \ref{fig:SQCDPattern1} \\[7pt]
		$h_{11} = h_{22} = h \neq 0$ & $W=h M (\widetilde{q}_1 q_1 +\widetilde{q}_2 q_2)$ & $SU(F_1) \times SU(G_1=F_1) \times SU(2)^2$, Fig. \ref{fig:SQCDPattern2} 
		\\[7pt]
		$h_{{\ell}j} \neq 0$ & $W = h_{{\ell}j} M_{{\ell}j} \widetilde{q}_{\ell} q_j$ & $SU(F_1) \times SU(F_2) \times $ \\ & & $SU(G_1) \times SU(G_2)$, Fig. \ref{fig:SQCDPattern0} \\
		\bottomrule
	\end{tabular*}
	\captionof{table}{The various patterns for breaking the global symmetry $SU(F)\times SU(F)$ of SQCD introducing four singlets $M_{{\ell}j}$ and four couplings $h_{{\ell}j}$, ${\ell},j=1,2$, referring to Fig. \ref{fig:SQCDPattern0} with $F_1 + F_2 = F = G_1 + G_2$.}
	\label{tab:SQCDPatterns}
\end{center} 

Consider now the case with two couplings turned on, either $h_{11} = h_{22} = h \neq 0$ or $h_{12} = h_{21} = h \neq 0$. In particular, from the quiver in Fig. \ref{fig:SQCDPattern2} we focus on two models, obtained by turning on the first pair or the second one. Denoting them as model $A$ and $B$ respectively, their superpotentials read
\begin{align}\label{eq:WPattern2}
    W_A = h \, M (\widetilde{q}_1 q_1 + \widetilde{q}_2 q_2 ) \; , \nonumber \\[5pt]
    W_B = h \, M (\widetilde{q}_1 q_2 + \widetilde{q}_2 q_1 ) \; . 
\end{align}
As mentioned above, we can move from one model to the other by rotating the pairs of quarks, thanks to the $SU(2)$ global symmetry. It is useful for later purposes to change basis for the quarks as
\begin{align}\label{eq:SU2QuarkRotation}
    q_1 \to (Q_+ + Q_-) \; , \nonumber \\[5pt]
    q_2 \to (Q_+ - Q_-) \; , \nonumber 
\end{align}
so that the superpotentials in the two models become
\begin{align}
    W_A \to h \, M \left[ \widetilde{q}_1 (Q_+ + Q_-) + \widetilde{q}_2 (Q_+ - Q_-) \right] \; , \nonumber \\[5pt]
    W_B \to h \, M \left[ \widetilde{q}_1 (Q_+ - Q_-) + \widetilde{q}_2 (Q_+ + Q_-) \right] \; , 
\end{align}
and we can redefine the field $Q_- \to - Q_-$ in $W_A$ so that the two interactions match.
We add further flavors to this construction, in order to turn on additional interactions. In Fig. \ref{fig:SQCDPattern2}, add e.g. $SU(F_3)^2 \times SU(G_3)^2$, bifundamental matter fields $p$, $\widetilde{p}$, bifundamental singlets $l$, $r$, $f_1$, $f_2$, $\widetilde{f}_1$, $\widetilde{f}_2$ landing us onto the model in Fig. \ref{fig:SQCDPattern1Large}. Note that the ranks of the flavor groups must be chosen such that $2F_1 + F_3 = 2 G_1 + G_3$ for the theory to be free of gauge anomalies. Another interaction can be turned in both model $A$ and $B$, namely \begin{align}
    W_{\lambda} = \lambda \, p l \widetilde{q}_1 + \lambda \, r \widetilde{p} q_1 + \lambda \, \widetilde{q}_2 p f_1 f_2 + \lambda \, \widetilde{p} q_2 \widetilde{f}_2 \widetilde{f}_1\; ,
\end{align}
where $\lambda$ is the coupling.\footnote{In principle, there may be two different couplings, $\lambda_{(3)}$ for cubic terms and $\lambda_{(4)}$ for quartic, and the argument still holds. We pick $\lambda_{(3)} = \lambda_{(4)} = \lambda$ for the sake of brevity.} This interaction breaks the $SU(2)$ global symmetry, as we cannot rotate the quarks anymore. In the basis chosen earlier, we have
\begin{align}\label{eq:WPattern2Large}
    W_A + W_{\lambda} \to & \; h M \left[ \widetilde{q}_1 (Q_+ + Q_-) + \widetilde{q}_2 (Q_+ - Q_-) \right] \nonumber \\[5pt]
    &+ \lambda \, p l \widetilde{q}_1 + \lambda \, r \widetilde{p} (Q_+ + Q_-) + \lambda \widetilde{q}_2 p f_1 f_2 + \widetilde{p} (Q_+ - Q_-) \widetilde{f}_2 \widetilde{f}_1 \; , \nonumber \\[5pt]
    W_B + W_{\lambda} \to & \; h M \left[ \widetilde{q}_1 (Q_+ - Q_-) + \widetilde{q}_2 (Q_+ + Q_-) \right] \nonumber \\[5pt]
    &+ \lambda \, p l \widetilde{q}_1 + \lambda \, r \widetilde{p} (Q_+ + Q_-) + \lambda \widetilde{q}_2 p f_1 f_2 + \widetilde{p} (Q_+ - Q_-) \widetilde{f}_2 \widetilde{f}_1\; ,
\end{align}
and, if we proceed as before, the additional transformation $Q_- \to - Q_-$ does not send one model into the other. The new interaction is crucial as there is no field redefinition that transforms the two superpotentials into the same form. However, the two models preserve the same global symmetry. In particular, the constraints for the $R$-charges are compatible and if a conformal point exists for one model, the same happens for the other, where one of the couplings has acquired a minus sign. As long as the ranks of the gauge group and of the flavor groups are chosen such that the $\beta$-functions vanish, we expect that they live on the same conformal manifold. 

In order to see that, let us choose the ranks of the flavour groups as
\begin{align}\label{eq:rankassignement}
    G_1 &= N + 2 \; , \quad F_3 = N + 4 \; , \nonumber \\[5pt]
    F_1 &= N - 2 \; , \quad G_3 = N - 4 \; , 
\end{align}
and let us turn on the interaction terms one by one, following the flow at every step. When $W=0$ all of the fields are free, i.e. $R=2/3$, the central charge is $a = 11/24$ and the cubic interaction $W_A$ is marginal, while the quartic operators in $W_{\lambda}$ are irrelevant. Requesting that the $\beta$-functions of all the couplings vanish leads to a new conformal fixed point, whose central charge $a$ has decreased to $a \simeq 0.305$. The $R$-charges of gauge-invariant operators stay above the unitarity bound, so no accidental symmetries are generated along the flow. The story is repeated exactly with $W_B$, and both $W_A + W_{\lambda}$ and $W_B + W_{\lambda}$ have the same central charge. At this conformal point, the non-anomalous global symmetry of the two models match as well, with the same charges for the matter fields, see Tab. \ref{tab:ABglobalcharges} and therefore the 't Hooft anomalies match.
One can see that $W_B$ $(W_A)$ is classically marginal at the fixed point of model $A$ $(B)$. Furthemore, from Tab. \ref{tab:ABglobalcharges}, we can see that $W_B$ ($W_A$) is not charged under any of the global symmetries, hence it is exactly marginal \cite{Green:2010da} for the model $A$ ($B$). This identifies a direction on the conformal manifold along which we can move from one model to the other by turning on exactly marginal operators.

\begin{center}
	\begin{tabular*}{1\textwidth}{@{\extracolsep{\fill}}c|ccccccccccccc}
		\toprule
		 & $q_1$ & $q_2$ & $\widetilde{q}_1$ & $\widetilde{q}_2$ & $M$ & $p$ & $\widetilde{p}$ & $l$ & $r$ & $f_1$ & $f_2$ & $\widetilde{f}_1$ & $\widetilde{f}_2$ \\
		\midrule
		$U(1)_{_B}$ & 1 & 1 & $-1$ & $-1$ & 0 & 1 & $-1$ & 0 & 0 & 0 & 0 & 0 & 0 \\[4pt]
            $U(1)_1$ & 1 & 1 & 0 & 0 & $-1$ & $-\frac{1}{2}$ & $-\frac{3}{2}$ & $\frac{1}{2}$ & $\frac{1}{2}$ & 0 & $\frac{1}{2}$ & 0 & $\frac{1}{2}$ \\[4pt]
            $U(1)_2$ & 2 & 2 & $-1$ & $-1$ & $-1$ & $\frac{1}{2}$ & $-\frac{5}{2}$ & $\frac{1}{2}$ & $\frac{1}{2}$ & 1 & $-\frac{1}{2}$ & 1 & $-\frac{1}{2}$ \\[4pt]
            $U(1)_3$ & 0 & 0 & 0 & 0 & 0 & 0 & 0 & 0 & 0 & 1 & $-1$ & 1 & $-1$ \\[4pt]
		\bottomrule
	\end{tabular*}
	\captionof{table}{The global symmetry charges for the matter fields, both for $W_A + W_{\lambda}$ and for $W_B + W_{\lambda}$.}
	\label{tab:ABglobalcharges}
\end{center}

Furthermore, we shall see that the quiver in Fig. \ref{fig:SQCDPattern1Large} can be embedded in a dimer construction, where one can read off the low-energy gauge theory associated to a particular toric singularity. To be more precise, the flavor factors would be gauged and further interactions arise, since each field must appear twice in the superpotential with opposite sign,\footnote{This can be accounted for by inserting a factor $(-1)^{{\ell}+1}$ in front of the superpotential in Eq. \eqref{eq:SQCDW} and the subsequent ones.} as a consequence of the strong constraints given by the toric condition. We notice that in each case considered below, any pairs of  superpotentials of conformally dual models, obtained along these lines, do not admit any field redefinitions transforming one into another.

\begin{figure}
    \begin{subfigure}{0.45\textwidth}
    \centering{
     \begin{tikzpicture}[auto]
        \node [circle, draw=blue!50, fill=blue!20, inner sep=0pt, minimum size=5mm] (G) at (0,0) {$N$};
        \node[rectangle, draw=red!50, fill=red!20, inner sep=0pt, minimum size=4mm] (F1) at (2,0) {$F$};
        \node[rectangle, draw=red!50, fill=red!20, inner sep=0pt, minimum size=4mm] (F2) at (-2,0) {$F$};
        \draw (G) to node {$p$} (F1) [->, thick];
        \draw (G) to node {$\widetilde{q}$} (F2) [<-, thick];
     \end{tikzpicture}
     \vspace{50pt}
     \subcap{The quiver for SQCD with gauge group $SU(N)$, non-abelian global symmetry $SU(F) \times SU(F)$ and $W=0$.}\label{fig:SQCD}
    }
    \end{subfigure}
    \hfill
    \begin{subfigure}{0.45\textwidth}
    \centering{
     \begin{tikzpicture}[auto]
        \node [circle, draw=blue!50, fill=blue!20, inner sep=0pt, minimum size=5mm] (N) at (0,0) {$N$};
        \node[rectangle, draw=red!50, fill=red!20, inner sep=0pt, minimum size=4mm] (F1) at (-1.5,0) {$F_1$};
        \node[rectangle, draw=red!50, fill=red!20, inner sep=0pt, minimum size=4mm] (F2) at (1.5,0) {$F_2$};
        \node[rectangle, draw=red!50, fill=red!20, inner sep=0pt, minimum size=4mm] (G1) at (0,1.5) {$G_1$};
        \node[rectangle, draw=red!50, fill=red!20, inner sep=0pt, minimum size=4mm] (G2) at (0,-1.5) {$G_2$};
        \draw (N) to node {$q_1$} (F1) [->, thick];
        \draw (N) to node {$q_2$} (F2) [->, thick];
        \draw (G1) to node [swap] {$\widetilde{q}_1$} (N) [->, thick];
        \draw (G2) to node [swap] {$\widetilde{q}_2$} (N) [->, thick];
        \draw (G1.east) [bend left=40] to node {$M_{12}$} (F2) [<-, thick];
        \draw (G2.east) [bend right=40] to node [swap] {$M_{22}$} (F2) [<-, thick];
        \draw (G1.west) [bend right=40] to node [swap] {$M_{11}$} (F1) [<-, thick];
        \draw (G2.west) [bend left=40] to node {$M_{21}$} (F1) [<-, thick];
     \end{tikzpicture}
     \vspace{10pt}
     \subcap{The quiver for SQCD with gauge group $SU(N)$, non-abelian global symmetry $SU(F_1) \times SU(F_2) \times SU(G_1) \times SU(G_2)$, with $F_1 + F_2 = F = G_1 + G_2$, singlets $M_{ij}$ and $W=h_{ij}M_{ij}\widetilde{q}_iq_j$.}\label{fig:SQCDPattern0}
    }
    \end{subfigure}
    \\[20pt]
    \begin{subfigure}{0.45\textwidth}
    \centering{
         \begin{tikzpicture}[auto]
        \node [circle, draw=blue!50, fill=blue!20, inner sep=0pt, minimum size=5mm] (N) at (0,0) {$N$};
        \node[rectangle, draw=red!50, fill=red!20, inner sep=0pt, minimum size=4mm] (F1) at (-2,0) {$F_1$};
        \node[rectangle, draw=red!50, fill=red!20, inner sep=0pt, minimum size=4mm] (F2) at (2,0) {$F_2$};
        \draw ($(F1.north east)!0.3!(F1.south east)$) to node {$q_1$} ($(N.north west)!0.3!(N.south west)$) [->, thick];
        \draw ($(F1.north east)!0.7!(F1.south east)$) to node [swap] {$\widetilde{q}_1$} ($(N.north west)!0.75!(N.south west)$) [<-, thick];
        \draw ($(N.north east)!0.3!(N.south east)$) to node {$q_2$} ($(F2.north west)!0.3!(F2.south west)$) [->, thick];
        \draw ($(N.north east)!0.75!(N.south east)$) to node [swap] {$\widetilde{q}_2$} ($(F2.north west)!0.7!(F2.south west)$) [<-, thick];
        \draw (F1) to [out=140, in=50, looseness=8] (F1) [-, thick];
        \node [above=0.7cm of F1] (M) {$M_{11}$};
     \end{tikzpicture}
     \vspace{15pt}
     \subcap{The quiver for SQCD with gauge group $SU(N)$, global symmetry $SU(F_1)^2\times SU(F_2)^2$, adjoint singlet $M_{11}$ and interaction $W=h_{11}M_{11}\widetilde{q}_1 q_1$.}\label{fig:SQCDPattern1}
    }
    \end{subfigure}
    \hfill
    \begin{subfigure}{0.45\textwidth}
    \centering{
       \begin{tikzpicture}[auto]
        \node [circle, draw=blue!50, fill=blue!20, inner sep=0pt, minimum size=5mm] (G) at (0,0) {$N$};
        \node[rectangle, draw=red!50, fill=red!20, inner sep=0pt, minimum size=4mm] (F1) at (2,0) {$F_1$};
        \node[rectangle, draw=red!50, fill=red!20, inner sep=0pt, minimum size=4mm] (F2) at (-2,0) {$G_1$};
        \draw ($(G.north east)!0.3!(G.south east)$) to node {$q_1$} ($(F1.north west)!0.3!(F1.south west)$) [->, thick];
        \draw ($(G.north east)!0.75!(G.south east)$) to node [swap] {$q_2$} ($(F1.north west)!0.7!(F1.south west)$) [->, thick];
        \draw ($(G.north west)!0.3!(G.south west)$) to node [swap] {$\widetilde{q}_1$} ($(F2.north east)!0.3!(F2.south east)$) [<-, thick];
        \draw ($(G.north west)!0.75!(G.south west)$) to node {$\widetilde{q}_2$} ($(F2.north east)!0.7!(F2.south east)$) [<-, thick];
        \draw (F1.north)  [bend right=35] to node [swap] {$M_{11}$} (F2.north) [->, thick];
     \end{tikzpicture}
     \vspace{5pt}
     \subcap{The quiver for SQCD with gauge group $SU(N)$, non-abelian global symmetry $SU(F_1) \times SU(G_1=F_1)\times SU(2)^2$, bifundamental singlet $M_{11}= M_{22} = M$ and interaction $W=h M (\widetilde{q}_1q_1 + \widetilde{q}_2q_2)$.}\label{fig:SQCDPattern2}
    }
    \end{subfigure}
\end{figure}

\begin{figure}
    \centering{
     \begin{tikzpicture}[auto]
        \node [circle, draw=blue!50, fill=blue!20, inner sep=0pt, minimum size=5mm] (G) at (0,0) {$N$};
        \node[rectangle, draw=red!50, fill=red!20, inner sep=0pt, minimum size=4mm] (F1) at (2,0) {$F_1$};
        \node[rectangle, draw=red!50, fill=red!20, inner sep=0pt, minimum size=4mm] (G1) at (-2,0) {$G_1$};
        \node[rectangle, draw=red!50, fill=red!20, inner sep=0pt, minimum size=4mm] (G3) at (1,-1.5) {$G_3$};
        \node[rectangle, draw=red!50, fill=red!20, inner sep=0pt, minimum size=4mm] (F3) at (-1,-1.5) {$F_3$};
        \draw ($(G.north east)!0.3!(G.south east)$) to node {$q_1$} ($(F1.north west)!0.3!(F1.south west)$) [->, thick];
        \draw ($(G.north east)!0.75!(G.south east)$) to node [swap] {$q_2$} ($(F1.north west)!0.7!(F1.south west)$) [->, thick];
        \draw ($(G.north west)!0.3!(G.south west)$) to node [swap] {$\widetilde{q}_1$} ($(G1.north east)!0.3!(G1.south east)$) [<-, thick];
        \draw ($(G.north west)!0.75!(G.south west)$) to node {$\widetilde{q}_2$} ($(G1.north east)!0.7!(G1.south east)$) [<-, thick];
        \draw (F1.north)  [bend right=35] to node [swap] {$M$} (G1.north) [->, thick];
        \draw (F3) to node {$l$} (G1) [->, thick];
        \draw (G) to node [pos=0.8] {$p$} (F3) [->, thick];
        \draw (G3) to node [pos=0.3] {$\widetilde{p}$} (G) [->, thick];
        \draw (F1) to node {$r$} (G3) [->, thick];
        \node[rectangle, draw=red!50, fill=red!20, inner sep=0pt, minimum size=4mm] (F3b) at (-3.5,0) {$F_3$};
        \node[rectangle, draw=red!50, fill=red!20, inner sep=0pt, minimum size=4mm] (G3b) at (+3.5,0) {$G_3$};
        \draw (F3.west) [bend left=30] to node [pos=0.8] {$f_1$} (F3b.south) [->, thick];
        \draw (G3b.south) [bend left=30] to node [pos=0.15] {$\widetilde{f}_1$} (G3.east) [->, thick];
        \draw (F3b) to node {$f_2$} (G1) [->, thick];
        \draw (F1) to node {$\widetilde{f}_2$} (G3b) [->, thick];
     \end{tikzpicture}
     \caption{The quiver for SQCD with singlets with gauge group $SU(N)$ and non-abelian global symmetry $SU(F_1)\times SU(G_1) \times SU(F_3)^2 \times SU(G_3)^2$.}\label{fig:SQCDPattern1Large}
    }
\end{figure}

\subsection{Real gauge groups}\label{sec:ToyModelReal}

The previous mechanism of conformal duality can occur also in the case of a real gauge group. In order to see that, consider a single gauge group, either $SO(N)$ or $USp(N)$,\footnote{In our convention $USp(2) \cong SU(2)$, hence in $USp(N)$ $N$ is even.} with $2F$ flavors and a singlet in a tensor representation of the flavor group $SU(F)$. Let us choose the tensor to be symmetric for orthogonal gauge groups and antisymmetric for symplectic ones. The quiver associated to this theory is drawn in Fig. \ref{fig:ToyModelQuiver1},\footnote{Even though the gauge group is real, we keep drawing arrows for the bifundamental matter fields to show the chirality under $SU(F)$.} where $q_1$ and $q_2$ are the fundamental flavors, $T$ is the tensor, and the real group is drawn as a diamond (see Fig. \ref{fig:legend} for a legend about our notation for quivers.). We can turn on two interaction terms
\begin{align}\label{eq:WReal}
    & W_{A} = h \, \frac{1}{2} T (q_1^2 + q_2^2) \; , \\[7pt]
    & W_{B} = h \, T q_1 q_2 \; ,
\end{align}
where $h$ is the coupling, and we refer to the theory with $W_A$ ($W_B$) turned on as model $A$ ($B$). The constraints for the $R$-charges given by these interactions are compatible with each other. Redefine the fields as
\begin{align}\label{eq:Transformation1}
    & q_1 \to \frac{1}{\sqrt{2}} \left( Q_+ + Q_- \right) \; , \nonumber \\[5pt]
    & q_2 \to \frac{1}{\sqrt{2}} \left( Q_+ - Q_- \right) \; , 
\end{align}
so that the interaction terms become
\begin{align}
    & W_{A} = h \, \frac{1}{2} T (q_1^2+q_2^2) \to \frac{1}{2} T \left( Q_+^2 + Q_-^2 \right) \; , \nonumber \\[7pt]
    & W_{B} = h \, T (q_1 q_2) \to \frac{1}{2} T \left( Q_+^2 - Q_-^2 \right) \; ,
\end{align}
and they differ only by a relative sign. As before, we can still perform a second transformation on $W_A$ that acts as $Q_- \to i Q_-$ and the two theories end up having the same interaction. Hence, they are equivalent and we can move from one to the other with simple field redefinitions, similarly to the previous case.

\begin{figure}
    \centering{
    \begin{tikzpicture}[auto]
        \node [circle, draw=blue!50, fill=blue!20, inner sep=0pt, minimum size=5mm] (G) at (0,0) {$N$};
        \node [diamond, draw=blue!50, fill=blue!20, inner sep=0pt, minimum size=6mm] (R) at (5,0) {$N$};
        \node[rectangle, draw=red!50, fill=red!20, inner sep=0pt, minimum size=4mm] (F) at (3,-1.5) {$F$};
        \node [right=0.2cm of G] (g) {$\textrm{gauge}\quad SU(N)$};
        \node [right=0.2cm of R] (r) {$\textrm{gauge}\quad SO/USp(N)$};
        \node [right=0.2cm of F] (f) {$\textrm{flavor}\quad SU(F)$};
    \end{tikzpicture}
    }
    \caption{The notation we use for quivers: a blue circle for a $SU(N)$ gauge factor, a blue diamond for a $SO(N)$ or $USp(N)$ gauge factor and a red square for a flavor factor.}
    \label{fig:legend}
\end{figure}

Let us add two flavour groups and four bifundamental fields, $w$, $v$, $f_1$ and $f_2$ as in Fig.~\ref{fig:ToyModelQuiver2}. We can turn on an additional interaction in both models, as
\begin{align}
    W_{\lambda} = \lambda \, q_1 w v - \lambda \, q_2 w v  + \lambda \, q_1 w f_1 f_2 + \lambda \, q_2 w f_1 f_2 \; ,
\end{align}
so that model $A$ has superpotential $W_A + W_{\lambda}$ and model $B$ has $W_B + W_{\lambda}$. We note that the constraints for the $R$-charges are compatible, similarly as before, and we ask whether the two theories are equivalent. If we perform again the two transformations, i.e. Eq. \eqref{eq:Transformation1} for both models and subsequently $Q_- \to i Q_-$ for model $A$, we obtain
\begin{align}\label{eq:WRealLarge}
    W_{A} + W_{\lambda} &\to h \, \frac{1}{2} T \left( Q_+^2 - Q_-^2 \right) + \lambda \left[ \frac{1}{\sqrt{2}} \left( Q_+ + i Q_- \right) - \frac{1}{\sqrt{2}} \left( Q_+ - i Q_- \right) \right] w v \nonumber \\[5pt]
    & \; + \frac{1}{\sqrt{2}} \left( Q_+ + i Q_- \right) w f_1 f_2 + \frac{1}{\sqrt{2}} \left( Q_+ - i Q_- \right) w f_1 f_2 \; , \nonumber \\[7pt]
    W_{B} + W_{\lambda} &\to h \, \frac{1}{2} T \left( Q_+^2 - Q_-^2 \right) + \lambda \left[ \frac{1}{\sqrt{2}} \left( Q_+ + Q_- \right) - \frac{1}{\sqrt{2}} \left( Q_+ - Q_- \right) \right] w v \nonumber \\[5pt]
    & \; + \frac{1}{\sqrt{2}} \left( Q_+ + i Q_- \right) w f_1 f_2 + \frac{1}{\sqrt{2}} \left( Q_+ - i Q_- \right) w f_1 f_2  \; ,
\end{align}
and the two models are no longer trivially connected by a simple field redefinition, due to the presence of additional interactions. As before, the two theories preserve the same global symmetry, including the abelian $R$-symmetry and if a conformal point exists, e.g. by choosing ranks such that the $\beta$-functions vanish, it exists for both models. Let us choose the ranks as
\begin{align}
    F = N - 2 \, , \; \overline{F} = N - 4 \; , 
\end{align}
and study, as before, the dynamics of operators turning interactions one by one. With no interactions, all of the fields are free and the central charge is $a = 11/48$. Around this conformal point, the cubic operators are marginal and quartic ones are irrelevant. As in the previous section, it exists a consistent conformal point in the IR where the $\beta$-functions for all the couplings vanish, with $a \simeq 0.153$, reached by both $W_A + W_{\lambda}$ and $W_B + W_{\lambda}$. No accidental symmetries are generated along the flow. Again, the two models share the same global symmetry and the charges match, see Tab \ref{tab:OrthogonalABglobalcharges}, and $W_A$ is exactly marginal for model $B$, and viceversa. As before, the two models live on the same conformal manifold, even though they they flow to different points on the conformal manifold. 

\begin{center}
	\begin{tabular*}{0.7\textwidth}{@{\extracolsep{\fill}}c|ccccccc}
		\toprule
		 & $q_1$ & $q_2$ & $w$ & $v$ & $T$ & $f_1$ & $f_2$ \\
		\midrule
            $U(1)_1$ & 1 & 1 & $-3$ & 2 & $-2$ & 0 & 2 \\[4pt]
		$U(1)_2$ & 0 & 0 & $-1$ & 1 & 0 & 1 & 0 \\[4pt]
		\bottomrule
	\end{tabular*}
	\captionof{table}{The global symmetry charges for the matter fields, both for $W_A + W_{\lambda}$ and for $W_B + W_{\lambda}$.}
	\label{tab:OrthogonalABglobalcharges}
\end{center}

\begin{figure}
    \begin{subfigure}{0.4\textwidth}
    \centering{
    \begin{tikzpicture}[auto]
         \node [diamond, draw=blue!50, fill=blue!20, inner sep=0pt, minimum size=6mm] (G) at (0,0) {$N$};
         \node[rectangle, draw=red!50, fill=red!20, inner sep=0pt, minimum size=4mm] (F) at (0,2) {$F$};
         \node [left=0.2cm of G] {\small{$SO(N)$}};
         \node [left=0.2cm of F] {\small{$SU(F)$}};
         \draw (F) to [out=140, in=50, looseness=8] (F) [-, thick];
         \draw ($(F.south west)!0.3!(F.south east)$) to node [swap] {$q_1$} ($(G.north west)!0.25!(G.north east)$) [->, thick];
         \draw ($(F.south west)!0.7!(F.south east)$) to node {$q_2$} ($(G.north west)!0.75!(G.north east)$) [->, thick];
         \node [above=0.7 cm of F] {$T$};
    \end{tikzpicture}
    \subcap{The quiver model with a single real gauge and global symmetry $SU(2F)$. }\label{fig:ToyModelQuiver1}}
    \end{subfigure}
    \hfill
    \begin{subfigure}{0.4\textwidth}
    \centering{
    \begin{tikzpicture}[auto]
         \node [diamond, draw=blue!50, fill=blue!20, inner sep=0pt, minimum size=6mm] (G) at (0,0) {$N$};
         \node[rectangle, draw=red!50, fill=red!20, inner sep=0pt, minimum size=4mm] (F) at (0,2) {$F$};
         \node[rectangle, draw=red!50, fill=red!20, inner sep=0pt, minimum size=4mm] (FF) at (2,0) {$\overline{F}$};
         \node [left=0.2cm of G] {\small{$SO(N)$}};
         \node [left=0.2cm of F] {\small{$SU(F)$}};
         \node [right=0.2cm of FF] {\small{$SU(\overline{F})$}};
         \draw ($(F.south west)!0.3!(F.south east)$) to node [swap] {$q_1$} ($(G.north west)!0.25!(G.north east)$) [->, thick];
         \draw ($(F.south west)!0.7!(F.south east)$) to node {$q_2$} ($(G.north west)!0.75!(G.north east)$) [->, thick];
         \draw (F) to [out=140, in=50, looseness=8] (F) [-, thick];
         \draw (G) to node [swap] {$w$} (FF) [->, thick]; 
         \draw (FF) to node [swap] {$v$} (F) [->, thick]; 
         \node [above=0.7 cm of F] {$T$};
         \node[rectangle, draw=red!50, fill=red!20, inner sep=0pt, minimum size=4mm] (FFb) at (2,2) {$\overline{F}$};
         \node [right=0.2cm of FFb] {\small{$SU(\overline{F})$}};
         \draw (FF) to node [swap] {$f_1$} (FFb) [->, thick];
         \draw (FFb) to node [swap] {$f_2$} (F) [->, thick];
    \end{tikzpicture}
    \subcap{The quiver model with a single real gauge and global symmetry $SU(F) \times SU(\overline{F})^2$.}\label{fig:ToyModelQuiver2}}
    \end{subfigure}
\end{figure}

\subsection{Duality frames}

When the models we introduced in the previous subsections are embedded in larger quivers, the resulting theory contains more gauge group factors, fields and interactions. At any rate, two models that differ only by superpotential terms of the form $W_A$ and $W_B$ are in fact the same theory, but if an interaction that breaks explicitly the rotation among the quarks in both models is present, then they live on the same conformal manifold, at separate points where one of the couplings (or a subset thereof) has changed sign. However, the presence of extra structure can hide the superpotential mechanism we showed earlier. While the computation of protected quantities such as anomalies and indices would undoubtedly signal that the two theories end up on the same conformal manifold, it can be unclear whether they are actually the same theory or different points on the same manifold \cite{Antinucci:2020yki}. One would need to compare and map gauge-invariant operators of the models. In this regard, finding the proper duality frame can make the comparison easier.  

Below we will show prototypical examples that we will encounter in the rest of the paper. The idea is that after some duality operation, the superpotential can be brought to the form of Eqs. \eqref{eq:WPattern2Large}-\eqref{eq:WRealLarge}. Even though there are more gauge factors, the underlying message is the same as long as all factors lie in the conformal window. For instance, it may happen that the $SU(2)$ is not explicit, namely $\widetilde{q}_2$ and $q_2$ are not present in the theory, but a composite operator will take their place. There is a Seiberg-dual frame in which this is a fundamental field of the theory and one only needs to find this dual description. A common situation occurs as follows. Suppose we have the quiver in Fig. \ref{fig:DFrameUnitaryEl}, part of two larger models $A$ and $B$ with superpotential
\begin{align}\label{eq:WSeibergEl}
    &W_A = h \, M \widetilde{q}_1 q_1 + h \, M \widetilde{d} \widetilde{u} u d + \lambda \, p l \widetilde{d} \widetilde{u} + \lambda \, \widetilde{p} u d r + \ldots \; , \nonumber \\[5pt] 
    &W_B = h \, M \widetilde{q}_1 u d + h \, M \widetilde{d} \widetilde{u} q_1 + \lambda \, p l \widetilde{d} \widetilde{u} + \lambda \, \widetilde{p} u d r + \ldots \; ,
\end{align}
where the dots indicate the part of the superpotential that does not contain $\widetilde{d}$, $d$, $\widetilde{u}$ or $u$ and it is shared by models $A$ and $B$. Note that all the nodes in the quiver are gauge factors. If we dualize both nodes $G_4$ and $G_5$ in models $A$ and $B$, the dual quiver results in the one of Fig. \ref{fig:DFrameUnitaryMag}, where the dual quarks are $D$, $U$, $\widetilde{U}$, $\widetilde{D}$ and the mesons $u d = q_2$, $\widetilde{d} \widetilde{u} = \widetilde{q}_2$. As there will be more fields transforming under the nodes $G_4$ and $G_5$, more mesons will be present and other interaction terms will arise, but they do not spoil the argument. The superpotentials of the Seiberg-dual phases read
\begin{align}\label{eq:WSeibergMag}
    &W_A = h \, M \widetilde{q}_1 q_1 + h \, M \widetilde{q}_2 q_1 + \lambda \, p l \widetilde{q}_2 + \lambda \, \widetilde{p} q_2 r + q_2 \widetilde{D} \widetilde{U} + \widetilde{q}_2 U D + \ldots \; , \nonumber \\[5pt] 
    &W_B = h \, M \widetilde{q}_1 q_2 + h \, M \widetilde{q}_2 q_1 + \lambda \, p l \widetilde{q}_2 + \lambda \, \widetilde{p} q_2 r + q_2 \widetilde{D} \widetilde{U} + \widetilde{q}_2 U D + \ldots \; ,
\end{align}
and the quarks $q_i$ and $\widetilde{q}_j$ can be rotated as in Eq. \eqref{eq:SU2QuarkRotation}, hence, with the proper choice of ranks such that the two theories stay inside the conformal window, they live on the same conformal manifold. Looking back at the original phase (i.e. before dualization), despite the fact that the interactions differ in the two models (a cubic and a quintic in model $A$, and two quartic terms in model $B$), the two models flow to the same conformal manifold.

\begin{figure}
    \begin{subfigure}{0.45\textwidth}
    \centering{
     \begin{tikzpicture}[auto]
        \node [circle, draw=blue!50, fill=blue!20, inner sep=0pt, minimum size=5mm] (G) at (0,0) {$N$};
        \node[circle, draw=blue!50, fill=blue!20, inner sep=0pt, minimum size=5mm] (F1) at (2,0) {$F_1$};
        \node[circle, draw=blue!50, fill=blue!20, inner sep=0pt, minimum size=5mm] (G1) at (-2,0) {$G_1$};
        \node[circle, draw=blue!50, fill=blue!20, inner sep=0pt, minimum size=5mm] (G3) at (1,-1.5) {$G_3$};
        \node[circle, draw=blue!50, fill=blue!20, inner sep=0pt, minimum size=5mm] (F3) at (-1,-1.5) {$F_3$};
        \node[circle, draw=blue!50, fill=blue!20, inner sep=0pt, minimum size=5mm] (G4) at (-1,1.5) {$G_4$};
        \node[circle, draw=blue!50, fill=blue!20, inner sep=0pt, minimum size=5mm] (G5) at (1,1.5) {$G_5$};
        \draw (G) to node [pos=0.7] {$q_1$} (F1) [->, thick];
        \draw (G) to node [swap] {$\widetilde{q}_1$} (G1) [<-, thick];
        \draw (F1.south)  [bend left=40] to node [swap] {$M$} (G1.south) [->, thick];
        \draw (F3) to node {$l$} (G1) [->, thick];
        \draw (G) to node [pos=0.8] {$p$} (F3) [->, thick];
        \draw (G3) to node [pos=0.3] {$\widetilde{p}$} (G) [->, thick];
        \draw (F1) to node {$r$} (G3) [->, thick];
        \draw (G1) to node {$\widetilde{d}$} (G4) [->, thick];
        \draw (G4) to node {$\widetilde{u}$} (G) [->, thick];
        \draw (G5) to node {$d$} (F1) [->, thick];
        \draw (G) to node [pos=0.6, swap] {$u$} (G5) [->, thick];
        \node[left=0.8 cm of G1] (d1) {}; 
        \draw (d1) to node {$\ldots$} (G1) [->, thick, gray]; 
        \node[left=0.8 cm of F3] (d2) {}; 
        \draw (d2) to node [swap] {$\ldots$} (F3) [->, thick, gray];
        \node[left=0.8 cm of G4] (d3) {}; 
        \draw (d3) to node {$\ldots$} (G4) [->, thick, gray];
        \node[right=0.8 cm of G5] (d4) {}; 
        \draw (d4) to node [swap] {$\ldots$} (G5) [->, thick, gray]; 
        \node[right=0.8 cm of F1] (d5) {}; 
        \draw (d5) to node [swap] {$\ldots$} (F1) [->, thick, gray];
        \node[right=0.8 cm of G3] (d6) {}; 
        \draw (d6) to node {$\ldots$} (G3) [->, thick, gray];
     \end{tikzpicture}
     \subcap{Quiver for a part of a larger model. The connection to the remaining part is indicated by the dots.}\label{fig:DFrameUnitaryEl}
    }
    \end{subfigure}
    \hfill
        \begin{subfigure}{0.45\textwidth}
    \centering{
     \begin{tikzpicture}[auto]
        \node [circle, draw=blue!50, fill=blue!20, inner sep=0pt, minimum size=5mm] (G) at (0,0) {$N$};
        \node[circle, draw=blue!50, fill=blue!20, inner sep=0pt, minimum size=5mm] (F1) at (2,0) {$F_1$};
        \node[circle, draw=blue!50, fill=blue!20, inner sep=0pt, minimum size=5mm] (G1) at (-2,0) {$G_1$};
        \node[circle, draw=blue!50, fill=blue!20, inner sep=0pt, minimum size=5mm] (G3) at (1,-1.5) {$G_3$};
        \node[circle, draw=blue!50, fill=blue!20, inner sep=0pt, minimum size=5mm] (F3) at (-1,-1.5) {$F_3$};
        \node[circle, draw=blue!50, fill=blue!20, inner sep=0pt, minimum size=5mm] (G4) at (-1,1.5) {$G_4$};
        \node[circle, draw=blue!50, fill=blue!20, inner sep=0pt, minimum size=5mm] (G5) at (1,1.5) {$G_5$};
        \draw ($(G.north east)!0.3!(G.south east)$) to node [pos=0.7] {$q_1$} ($(F1.north west)!0.3!(F1.south west)$) [->, thick];
        \draw ($(G.north east)!0.7!(G.south east)$) to node [pos=0.7, swap] {$q_2$} ($(F1.north west)!0.7!(F1.south west)$) [->, thick];
        \draw ($(G.north west)!0.3!(G.south west)$) to node [swap] {$\widetilde{q}_1$} ($(G1.north east)!0.3!(G1.south east)$) [<-, thick];
        \draw ($(G.north west)!0.7!(G.south west)$) to node {$\widetilde{q}_2$} ($(G1.north east)!0.7!(G1.south east)$) [<-, thick];
        \draw (F1.south)  [bend left=40] to node [swap] {$M$} (G1.south) [->, thick];
        \draw (F3) to node {$l$} (G1) [->, thick];
        \draw (G) to node [pos=0.8] {$p$} (F3) [->, thick];
        \draw (G3) to node [pos=0.3] {$\widetilde{p}$} (G) [->, thick];
        \draw (F1) to node {$r$} (G3) [->, thick];
        \draw (G1) to node {$D$} (G4) [<-, thick];
        \draw (G4) to node {$U$} (G) [<-, thick];
        \draw (G5) to node {$\widetilde{D}$} (F1) [<-, thick];
        \draw (G) to node [pos=0.6, swap] {$\widetilde{U}$} (G5) [<-, thick];
        \node[left=0.8 cm of G1] (d1) {}; 
        \draw (d1) to node {$\ldots$} (G1) [->, thick, gray]; 
        \node[left=0.8 cm of F3] (d2) {}; 
        \draw (d2) to node [swap] {$\ldots$} (F3) [->, thick, gray, gray, gray];
        \node[left=0.8 cm of G4] (d3) {}; 
        \draw (d3) to node {$\ldots$} (G4) [->, thick, gray];
        \node[right=0.8 cm of G5] (d4) {}; 
        \draw (d4) to node [swap] {$\ldots$} (G5) [->, thick, gray, gray]; 
        \node[right=0.8 cm of F1] (d5) {}; 
        \draw (d5) to node [swap] {$\ldots$} (F1) [->, thick, gray, gray];
        \node[right=0.8 cm of G3] (d6) {}; 
        \draw (d6) to node {$\ldots$} (G3) [->, thick, gray];
     \end{tikzpicture}
     \vspace{13pt}
     \subcap{The quiver after dualization of nodes $SU(G_4)$ and $SU(G_5)$.}\label{fig:DFrameUnitaryMag}
    }
    \end{subfigure}
\end{figure}

It may also happen that we need to dualize only one model, if the interaction in the other one is already in the form of Eq. \eqref{eq:WSeibergMag}.\footnote{We need to be careful that the new phase does not exhibit any global symmetry enhancement.} As a consequence, we have two interesting types of conformal duality: either a pair of models with two cubic terms in $A$ and two quartic terms in $B$, or a pair of models with one cubic term and one quintic term in $A$ and two cubic terms in $B$.

\subsubsection*{Deconfinement of tensors}

\begin{figure}
\begin{subfigure}{0.45\textwidth}
\centering{
    \begin{tikzpicture}[auto]
         \node [diamond, draw=blue!50, fill=blue!20, inner sep=0pt, minimum size=6mm] (G) at (0,0) {$N$};
         \node [rectangle, draw=red!50, fill=red!20, inner sep=0pt, minimum size=4mm] (F) at (0,2) {$F$};
         \node [rectangle, draw=red!50, fill=red!20, inner sep=0pt, minimum size=4mm] (G1) at (2,0) {$G_1$};
         \node [circle, draw=blue!50, fill=blue!20, inner sep=0pt, minimum size=5mm] (G2) at (2,2) {$G_2$};
         \node [left=0.8 cm of G] (d1) {};
         \node [right=0.8 cm of G2] (d2) {};
         \node [below=0.2cm of G] {\small{$SO(N)$}};
         \node [left=0.8cm of F] {\small{$SU(F)$}};
         \node [below=0.3cm of G1] {\small{$SU(G_1)$}};
         \node [below=0.2cm of G2] {\small{$\quad SU(G_2)$}};
         \draw (F) to node [swap] {$q_1$} (G) [->, thick];
         \draw (G2) to node [pos=0.7] {$\widetilde{d}$} (G) [->, thick];
         \draw (F) to [out=140, in=50, looseness=8] (F) [-, thick];
         \draw (G2) to [out=140, in=50, looseness=8] (G2) [-, thick];
         \draw (G) to node [swap, pos=0.7] {$w$} (G1) [->, thick]; 
         \draw (F) to node {$u$} (G2) [->, thick]; 
         \draw (FF) to node [swap, pos=0.8] {$v$} (F) [->, thick]; 
         \node [above=0.7 cm of F] {$T_1$};
         \node [above=0.7 cm of G2] {$T_2$};
         \draw (d1) to node [swap] {$\ldots$} (G) [->, thick, gray];
         \draw (d2) to node [swap] {$\ldots$} (G2) [->, thick, gray];
    \end{tikzpicture}\vspace{10pt}
    \subcap{The quiver with tensors, part of a larger model.}\label{fig:DualityFrameRealGroups}}    
\end{subfigure}
\hfill
\begin{subfigure}{0.45\textwidth}
\centering{
    \begin{tikzpicture}[auto]
         \node [diamond, draw=blue!50, fill=blue!20, inner sep=0pt, minimum size=6mm] (G) at (0,0) {$N$};
         \node [rectangle, draw=red!50, fill=red!20, inner sep=0pt, minimum size=4mm] (F) at (0,2) {$F$};
         \node [rectangle, draw=red!50, fill=red!20, inner sep=0pt, minimum size=4mm] (G1) at (2,0) {$G_1$};
         \node [circle, draw=blue!50, fill=blue!20, inner sep=0pt, minimum size=5mm] (G2) at (2,2) {$G_2$};
         \node [diamond, draw=blue!50, fill=blue!20, inner sep=0pt, minimum size=5mm] (Gc) at (2,3.5) {$G_c$};
         \node [left=0.8 cm of G] (d1) {};
         \node [right=0.8 cm of G2] (d2) {};
         \node [below=0.2cm of G] {\small{$SO(N)$}};
         \node [left=0.2cm of F] {\small{$SU(F)$}};
         \node [below=0.3cm of G1] {\small{$SU(G_1)$}};
         \node [below=0.2cm of G2] {\small{$\quad SU(G_2)$}};
         \node [right=0.2cm of Gc] {\small{$USp(G_c)$}};
         \draw ($(F.south west)!0.3!(F.south east)$) to node [swap] {$q_1$} ($(G.north west)!0.25!(G.north east)$) [->, thick];
         \draw ($(F.south west)!0.7!(F.south east)$) to node {$q_2$} ($(G.north west)!0.75!(G.north east)$) [->, thick];       
         \draw (G) to node [pos=0.8] {$d$} (G2) [->, thick];
         \draw (F) to [out=140, in=50, looseness=8] (F) [-, thick];
         \draw (G) to node [swap] {$w$} (G1) [->, thick]; 
         \draw (G2) to node [swap] {$\widetilde{u}$} (F) [->, thick]; 
         \draw (FF) to node [swap, pos=0.8] {$v$} (F) [->, thick]; 
         \draw (G2) to node [swap] {$t_2$} (Gc) [->, thick];
         \node [above=0.7 cm of F] {$T_1$};
         \draw (d1) to node [swap] {$\ldots$} (G) [->, thick, gray];
         \draw (d2) to node [swap] {$\ldots$} (G2) [->, thick, gray];
    \end{tikzpicture}
    \subcap{The quiver with tensors, part of a larger model, where tensor $T_2$ has been deconfined into $\left(t_2\right)^2$ and the gauge node $SU(G_2)$ has been dualized.}\label{fig:DualityFrameRealGroupsDual}}    
\end{subfigure}
\end{figure}

The last case of interest concerns the presence of tensor fields. Suppose we have the quiver of Fig. \ref{fig:DualityFrameRealGroups}, part of two larger models $A$ and $B$ with superpotentials
\begin{align}\label{eq:WTensorsElectric}
    &W_A = h \, \frac{1}{2} T_1 \left( q_1^2 + u^2 \widetilde{d}^2 \right) + \lambda \, w v q_1 + \ldots \; , \nonumber \\[5pt]
    &W_B = h \, T_1 q_1 u \widetilde{d} + \lambda \, w v q_1 + \ldots \; .
\end{align}
We may proceed as before, but we need to dualize the node with the tensor field $T_2$. In order to do that, we need first to deconfine the tensor field \cite{Berkooz:1995km, Pouliot:1995me,Luty:1996cg}, which is seen as the result of the confinement of a real group. At this step, we need to be careful not to add extra gauge-invariant operators, which may arise due to the presence of the confining gauge factor. In particular, when $T_2$ is an antisymmetric tensor of $SU(\overline{G})$ and $\overline{G}$ is even, the confining gauge factor is symplectic\footnote{In the case of a symmetric tensor, the auxiliary gauge factor that confines is orthogonal. The moduli space of such a gauge theory must be studied carefully, since the origin may not be smoothed out \cite{Intriligator:1995id}. Whenever we need to deconfine a symmetric tensor, we restrict ourselves to the case in which the auxiliary gauge factor $SO(N+4)$ confines, generating a symmetric tensor of $SU(N)$.} and we may need to add the Pfaffian operator $W_{T_2} = \text{Pf}(T_2)$ to the superpotential, so that it is forced to vanish. However, the complete structure of the interaction may already impose this constraint, so a case-by-case analysis is actually needed.\footnote{In this paper we will study cases where those operators are irrelevant.}

After having deconfined, we are allowed to dualize the node $SU(G_2)$, resulting in the meson $u \widetilde{d} = q_2$, possibly among other composite operators that become fundamental in the dual phase. The dual theory has the quiver shown in Fig. \ref{fig:DualityFrameRealGroupsDual} and the superpotentials of the two models read
\begin{align}\label{eq:WTensorsMagnetic}
    &W_A = h \, \frac{1}{2} T_1 \left( q_1^2 + q_2^2 \right) + \lambda \, w v q_1 + q_2 \widetilde{D} U + \ldots \; , \nonumber \\[5pt]
    &W_B = h \, T_1 q_1 q_2 + \lambda \, w v q_1 + q_2 \widetilde{D} U + \ldots \; .
\end{align}
We realize we are back to the models described in Sec. \ref{sec:ToyModelReal}, hence the same argument holds. In this case, model $A$ with a cubic interaction and a quintic interaction, involving a tensor field, flows to the same conformal manifold of model $B$ with a quartic term involving a tensor field. 

\section{Gauging the flavour: PdP$_{3b}$ vs. PdP$_{3c}$ and their orientifolds}
\label{sec:gauging}

In this section we discuss gauging of global symmetries in the toy models studied above, focusing on a detailed example. On one hand, this section is a bridge that allows us to fix the notation necessary for the more general analysis of the conformal dualities for multi-planarizable quivers studied below. On the other hand we will be able to give an explanation for the results obtained in \cite{Antinucci:2020yki}.
The models that we present here correspond indeed to stacks of D3-branes probing the complex cone over the pseudo del Pezzo surfaces PdP$_{3b}$ and PdP$_{3c}$ with appropriate orientifold projections, over specific toric phases that enjoy the $\mathbb{Z}_2$ needed for the projection.
The final models can be constructed by gauging the global symmetries of the toy models discussed above, by further requiring the cancellation of gauge anomalies and the vanishing of the beta functions.

As observed in \cite{Antinucci:2020yki} the quivers associated to the PdP$_{3b}$ and PdP$_{3c}$ coincide, and are 
given in Fig. \ref{fig:quiverPdP3bcHalf}.
\begin{figure}
\begin{subfigure}{0.4\textwidth}
\centering
{\begin{tikzpicture}[auto, scale= 0.3]
        \node [circle, draw=blue!50, fill=blue!20, inner sep=0pt, minimum size=5mm] (1) at (5,-2.5) {$2$};            
        \node [circle, draw=blue!50, fill=blue!20, inner sep=0pt, minimum size=5mm, nearly transparent] (4) at (-5,-2.5) {$4$};
        \node [circle, draw=blue!50, fill=blue!20, inner sep=0pt, minimum size=5mm, nearly transparent] (5) at (-5,2.5) {$5$}; 
        \node [circle, draw=blue!50, fill=blue!20, inner sep=0pt, minimum size=5mm] (2) at (5,2.5) {$1$}; 
        \node [circle, draw=blue!50, fill=blue!20, inner sep=0pt, minimum size=5mm] (0) at (0,6) {$0$};
        \node [circle, draw=blue!50, fill=blue!20, inner sep=0pt, minimum size=5mm] (3) at (0,-6) {$3$}; 
        \node (c2) at (0, 2.5) {};
        \node (c1) at (0, -2.5) {};
        \draw (0)  to node {} (1) [->, thick, ];
        \draw (1)  to node {} (2) [<-, thick, ];
        \draw (2)  to node {} (3) [->, thick, ];     
        \draw (3)  to node {} (4) [->, thick, nearly transparent];
        \draw (4)  to node {} (5) [->, thick, nearly transparent];
        \draw (5)  to node {} (0) [->, thick, nearly transparent];
        \draw (0)  to node {} (3) [-, thick, ];
        \draw (5)  to node {} (c2) [-, thick, nearly transparent];
        \draw (c1)  to node {} (1) [<-, thick];
        \draw (1)  to node {} (3) [->, thick, ];
        \draw (3)  to node {} (5) [->, thick, nearly transparent];
        \draw (2)  to node {} (c2) [<-, thick, ];
        \draw (c1) to node {} (4) [-, thick, nearly transparent];
        \draw (4)  to node {} (0) [->, thick, nearly transparent];
        \draw (0)  to node {} (2) [->, thick, ];

        \draw [thick, dashed, red] (0, -8) to node [pos=0.01] {$\Omega$} (0,8) ;
        \end{tikzpicture}}
        \subcap{The quiver of theories PdP$_{3b}$ and PdP$_{3c}$, with the orientifold represented with the dashed red line.} \label{fig:quiverPdP3bcHalf}
\end{subfigure}
\hfill
\begin{subfigure}{0.4\textwidth}
\centering
{\begin{tikzpicture}[auto]
         \node [diamond, draw=blue!50, fill=blue!20, inner sep=0pt, minimum size=6.5mm] (0) at (0,0) {$0$};
         \node [circle, draw=blue!50, fill=blue!20, inner sep=0pt, minimum size=5mm] (1) at (0,2) {$1$};
         \node [diamond, draw=blue!50, fill=blue!20, inner sep=0pt, minimum size=6.5mm] (3) at (2,0) {$3$};
         \node [circle, draw=blue!50, fill=blue!20, inner sep=0pt, minimum size=5mm] (2) at (2,2) {$2$};
         \draw (0) to node {} (1) [->, thick];
         \draw (0) to node {} (3) [-, thick];
         \draw (1) to [out=140, in=50, looseness=8] (1) [-, thick, green];
         \draw (2) to [out=140, in=50, looseness=8] (2) [-, thick, red];
         \draw (0) to node {} (2) [->, thick]; 
         \draw (1) to node {} (2) [->, thick]; 
         \draw (1) to node {} (3) [->, thick];
         \draw (2) to node {} (3) [->, thick];
         \node [above=0.7 cm of F] {$T_1$};
         \node [above=0.7 cm of G2] {$\widetilde{T}_2$};
    \end{tikzpicture}}\vspace{25pt}
        \subcap{The quiver of theories PdP$_{3b}$ and PdP$_{3c}$ after the orientifold projection.} \label{fig:quiverPdP3bcOrientifold}
\end{subfigure}
\end{figure}
However the two models have different superpotentials.
Explicitly:
\begin{eqnarray}
\label{WPdP3B}
W_b &=& 
X_{01} X_{13} X_{30}
-X_{13} X_{35} X_{51}
+X_{35} X_{50} X_{03}
-X_{50} X_{01} X_{12}X_{24} X_{45}  
\nonumber \\
&+&X_{12} X_{23} X_{34}X_{45} X_{51}  
+X_{24} X_{40} X_{02}
-X_{02} X_{23} X_{30}
-X_{40} X_{03} X_{34}
\end{eqnarray}
and
\begin{eqnarray}
\label{WPdP3C}
W_c &=& 
X_{01} X_{13} X_{30} 
-X_{13} X_{34} X_{45} X_{51}
+X_{35} X_{50} X_{03}
-X_{40} X_{01} X_{12}X_{24} 
\nonumber \\
&+&X_{12} X_{23} X_{35} X_{51}  
-X_{24} X_{45}X_{50} X_{02}
-X_{02} X_{23} X_{30}
+X_{40} X_{03} X_{34}\ .
\end{eqnarray}
This difference gives rise to two different dimers and thus two different toric diagrams. As a consequence the two models are \emph{not} related by any IR duality.

As discussed in the introduction in this case we refer to the quiver as multi-planarizable, because there exist two inequivalent periodic planar quivers (and consequently dimers), i.e.  there are two different consistent choices of toric data for the quiver
in Fig. \ref{fig:quiverPdP3bcHalf}.
In absence of orientifolds this degeneration does not have further implications for the gauge theories associated to the two singularities.

    \begin{figure}[ht!]
        \centering{
        \includegraphics[scale=0.55]{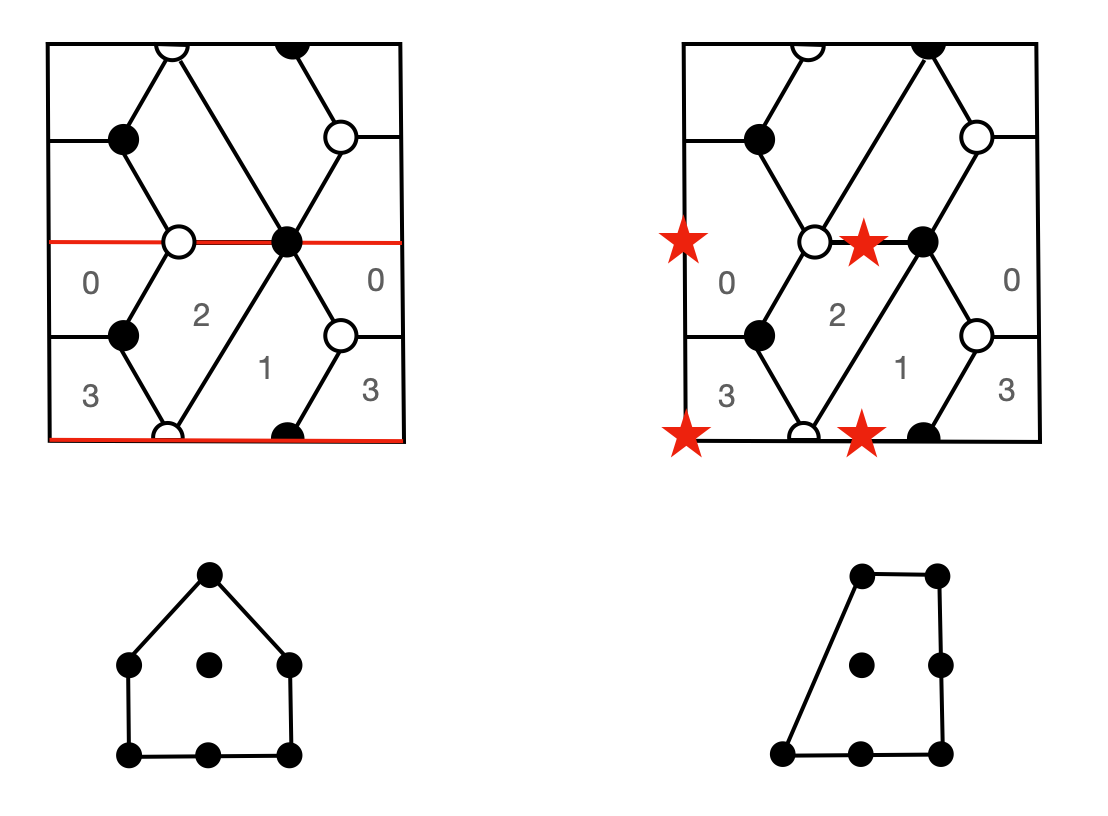}
        \caption{The dimers of Pseudo del Pezzo $3b$ and $3c$ with the fixed lines and fixed points projections, drawn in red.}
        \label{dimerpdp}}
    \end{figure}

The situation is more interesting if one studies orientifold projections.
When orientifolds are added to the previous brane setups symplectic and orthogonal gauge groups and two index symmetric and antisymmetric representations naturally arise after the projections and the quivers become unoriented.
This is a consequence of the fact that the orientifolds reverse the orientation of the strings, giving rise to a $\mathbb{Z}_2$ involution on the gauge theory 
\cite{Sagnotti:1987tw, Pradisi:1988xd, Bianchi:1990yu, Bianchi:1990tb, Polchinski:1995mt, Angelantonj:2002ct}.
On the brane tiling these projections can be represented by the general fixed points/fixed lines procedure spelled out in \cite{Franco:2007ii,Imamura:2008zz,Argurio:2020dko} (recently also projections that involve the Klein bottle have been considered \cite{Garcia-Valdecasas:2021znu}).
Orientifolds have many applications in various sectors, both at theoretical level \cite{Etxebarria:2021lmq,Garcia-Etxebarria:2016bpb,Garcia-Etxebarria:2012ypj,Garcia-Etxebarria:2013tba,Argurio:2019eqb, Argurio:2020dkg, Argurio:2020npm, Argurio:2022vfq}  and at phenomenological one  \cite{Ibanez:2001nd, Wijnholt:2007vn,Cicoli:2021dhg,Addazi:2014ila,Addazi:2015rwa,Addazi:2015hka, Addazi:2015yna}. 

Here we will not review the whole construction and refer the reader to \cite{Franco:2007ii,Argurio:2020dko} for more details and notations.
The idea is that one considers the action of the orientifold by looking at identifications on the dimer in terms of fixed points and fixed lines. The charges carried by the orientifolds reflect in consistent choices of charges associated to the fixed points and fixed lines.
As discussed in \cite{Antinucci:2020yki}, the non-trivial statement is that there exists a fixed-line projection on PdP$_{3b}$ that has the same quiver as a fixed-point projection of PdP$_{3c}$.
The action of such projections on the dimer is explicitly represented in Fig. \ref{dimerpdp}. The final quiver after the projection corresponds to the unshaded one in Fig. \ref{fig:quiverPdP3bcHalf}, which results in the quiver in Fig. \ref{fig:quiverPdP3bcOrientifold}.
The net effect of the projection is then to identify the gauge groups $SU(N_1)$ with $SU(N_5)$ and $SU(N_2)$ with $SU(N_4)$. As a consequence, bifundamental fields $X_{51}$ and $X_{24}$ are projected to two-index tensor representations, drawn in red and green. Finally, the remaining two gauge groups $SU(N_0)$ and $SU(N_3)$ are projected to $USp$ or $SO$ depending on the choice of charges associated to the fixed points. We use to draw the real gauge nodes as diamonds on the quiver. Observe that the quiver in Fig. \ref{fig:quiverPdP3bcOrientifold} is the simplest model that one can construct by gauging the flavour in the model discussed in Sec. \ref{sec:ToyModelReal} and Fig. \ref{fig:DualityFrameRealGroups}.

In the case of the fixed-point orientifold there are four charges, denoted with $\tau$ and labeled from $1$ to $4$ as in Fig. \ref{dimerpdp}. 
These charges are either $+1$ or $-1$ and then, from now on, we will associate to them the value $\tau_i = \pm 1$ with $i=1,\dots,4$. In the case at hand, $\tau_1$ projects the bifundamental $X_{51}$ to the conjugate tensor $\widetilde{T}_{11}$, $\tau_2$ projects the bifundamental $X_{42}$ to the tensor $T_{22}$, $\tau_3$ acts on the gauge factor $SU(N_3)$ and $\tau_4$ acts on the factor $SU(N_0)$. As discussed in \cite{Antinucci:2020yki} the product of these charges is constrained as $\prod_{i=1}^{4} \tau_i = (-1)^{n_W/2}$ where $n_W$ is the number of superpotential terms. We thus have $\prod_{i=1}^{4} \tau_i =1$.
The further requirement that PdP$_{3c}$ and PdP$_{3b}$ have the same quiver after the projections imposes that $\tau_1 = \tau_4$ and $\tau_2 = \tau_3$, as these are the charges of the two fixed lines.
The choices of $\tau_i$ consistent with the vanishing of the beta functions are constrained by $\tau_1+\tau_2=0$.
This requirement further fixes the gauge 
ranks properly.
There are two possible choices for $\tau_1=\pm 1$, but they are equivalent because of the $\mathbb{Z}_2$ reflection symmetry of the quiver. In general, for the more complicated cases that we will analyze below, this property does not hold anymore and the two choices for $\tau_1$ will give rise to different models.

Let us fix $\tau_1=1$ for concreteness. In this case we are left with an
\begin{equation}
    SO(N_0=n) \times SU(N_1=n-2) \times SU(N_2=n-2) \times USp(N_3=n-4) 
\end{equation}
gauge theory. Each pair of nodes is connected by a bifundamental field, where the representations are inherited from the parent theory consistently with the projection. Furthermore there are two tensors for the unitary gauge groups.
In the $SU(N_1)$ case there is a conjugate antisymmetric tensor, denoted $\widetilde{T}_{11}$ and colored in green in Fig. \ref{fig:quiverPdP3bcOrientifold}. In the $SU(N_2)$ case there is a symmetric tensor that we denote $T_{22}$ and color in red in Fig. \ref{fig:quiverPdP3bcOrientifold}. The same colors are used in the rest of the paper. In the general analysis in the following sections we will not specify the value of $\tau_1$. For this reason in this section we are using the same notation $T$ for symmetric and antisymmetric tensors.
This implies that in each case the correct representation will be specified by the choice of sign for $\tau_1$.\footnote{We are confident this will not generate any confusion in the attentive reader.} On the other hand we distinguish with a tilde the conjugate representation, because it is independent of the choice of sign for $\tau_1$.

The superpotentials \eqref{WPdP3B}  and \eqref{WPdP3C} are projected as
\begin{eqnarray}
\label{eq:OLb}
W_{3b}^{\Omega_\text{f.l.}} &=& 
X_{01}X_{13}X_{30}
-X_{02}X_{23}X_{30}
+\frac{1}{2}T_{22}X_{02}^2
\nonumber \\
&-&\frac{1}{2}T_{22}(X_{01}X_{12})^2
-\frac{1}{2} \widetilde{T}_{11}X_{13}^2
+\frac{1}{2}\widetilde{T}_{11}(X_{12}X_{23})^2
\end{eqnarray}
by fixed lines, and 
\begin{eqnarray}\label{eq:OPc}
W_{3c}^{\Omega_\text{f.p.}} &=& 
X_{01}X_{13}X_{30}
-X_{02}X_{23}X_{30} 
+ T_{22} X_{12} X_{01} X_{02} 
- \widetilde{T}_{11} X_{13} X_{12}X_{23}
\end{eqnarray}
by fixed points. Observe that, as in Sec. \ref{sec:toy}, in these formulas and in the rest of the paper all the gauge contractions will be understood.

Let us make an important remark regarding the factors of $1/2$ in formula \eqref{eq:OLb}. The presence of such factors has not been discussed in the original references on orientifolds and dimers \cite{Franco:2007ii}, because the authors were not interested in the structure of the exactly marginal deformations. Here we observe that these factors  appear each time a fixed line crosses some superpotential interaction in the dimer, which in turn arises from a disk amplitude in the five-brane picture and whose volume is halved by the fixed line. As we discussed above, they are crucial in our discussion about the conformal duality.
The analysis of the fixed point via $a$-maximization was performed in \cite{Antinucci:2020yki}. This analysis has shown that the two models differ only by exactly marginal deformations.  

Here we analyze the fixed points of the two models by turning on the superpotential terms one by one. We stress that a similar analysis can be performed in the other models considered in next sections. We start from $W=0$, where a fixed point exists with central charge $a \simeq 0.650$, and the unitarity bound is not violated. Let us first focus on the PdP$_{3b}$. At this point, at large $N$, the cubic operators without tensors in Eq. \eqref{eq:OLb} are marginal, the quintic operators are irrelevant and the cubic operators with tensors are relevant. We turn on the cubic operators and find a new fixed point with $a \simeq 0.609$, where the unitarity bound is satisfied. At this new fixed point, the quintic operators are marginal.

Studying the model PdP$_{3c}$, at the fixed point with $W=0$ the cubic operators in Eq. \eqref{eq:OPc} are marginal and the quartic operators, which involve tensors, are irrelevant. By turning on these operators we find a new fixed point with central charge $a \simeq 0.609$, the same value we just found for PdP$_{3b}$. The unitarity bound is satisfied and we see that the central charge has decreased. This suggests that the quartic operators are dangerously irrelevant. At this fixed point the operators appearing in the superpotential of PdP$_{3b}$ are classically marginal. One can check, from Tab. \ref{tab:PdP3bcglobalcharges}, that these operators are in fact exactly marginal, therefore the two theories live on the same conformal manifold.

\begin{center}
	\begin{tabular*}{0.8\textwidth}{@{\extracolsep{\fill}}c|cccccccc}
		\toprule
  & $X_{01}$ & $X_{02}$ & $X_{12}$ & $X_{13}$ & $X_{23}$ & $X_{30}$ & $\widetilde{T}_{11}$ & $T_{22}$ \\
		\midrule
 $U(1)_1$ & 0 & 1 & 1 & 1 & 0 & $-1$ & $-2$ & $-2$ \\
 $U(1)_2$ & 1 & 0 & $-1$ & 0 & 1 & $-1$ & 0 & 0 \\
		\bottomrule
	\end{tabular*}
	\captionof{table}{The global symmetry charges for the matter fields, at large $N$, both for PdP$_{3b}$ and for PdP$_{3c}$.}
	\label{tab:PdP3bcglobalcharges}
\end{center}

Here we will go one step further by applying a chain of tensor deconfinement `tricks' and Seiberg dualities to show that the two superpotentials are actually identical up to some sign factors. The interpretation of the conformal duality in this sense is more natural and fits with the general behaviors discussed in Sec. \ref{sec:toy}.
Observe that the simplest example of  such a conformal duality is the case of $\mathcal{N}=4$ SYM in presence of a $\beta$-deformation \cite{Leigh:1995ep}.
More generally the $\beta$-deformation is always an exactly  marginal deformation for any toric quiver gauge theory \cite{Benvenuti:2005wi,Imamura:2007dc}. Starting from the superpotential $W = h W_0$ of a toric quiver gauge theory the beta deformed superpotential is $W = h W_0 + \beta W_\beta$, where $W_\beta$ corresponds to the
toric superpotential but this time  taken with all plus signs.
In this sense a model with $W = h W_0$ is conformally dual to a model with $W = h W_\beta$. Here we will consider exactly marginal deformations of this type, but only by flipping the sign of a subset of superpotential interactions.

The next step consists in finding an auxiliary quiver without tensors. This corresponds to deconfining the symmetric tensor with an $SO(N_A=n+2)$ gauge node and the conjugate antisymmetric tensor with an $USp(N_B=n-6)$ gauge node. The new bifundamentals $X_{A1}$ and $X_{2B}$ will appear quadratically in the superpotentials, i.e. we substitute $T_{22} \rightarrow X_{2B}^2$ and $\widetilde T_{11} \rightarrow X_{A1}^2$.
(Indeed by confining these nodes the theory comes back to the original one with the tensor.) There are in addition non-perturbative contributions for the tensors but for generic $N$ they are irrelevant and we can ignore them.
In the deconfined model we can Seiberg dualize the unitary gauge nodes.
In this way we want to explicitly see the conformal duality applying the ideas and the  construction discussed in Sec. \ref{sec:toy}.

The models share two cubic superpotential terms, hence the chain of dualities we are going to discuss will affect them in the same way, for this reason we prefer to keep the discussion simple by not showing them in the following. 
Seiberg duality on  $SU(N_2=n-2)$ gives an $SU(\tilde N_2 = n)$ gauge node with dual superpotential 
\begin{eqnarray}
W_\text{dual} = 
M_{1B} Y_{B2} Y_{21}
+M_{13} Y_{32} Y_{21}
+M_{0B} Y_{B2} Y_{20}
+M_{03} Y_{32} Y_{20} \; ,
\end{eqnarray}
where $M$ are the mesons of this duality and $Y$ are the dual (bi-)fundamentals. 
The remaining superpotentials in the two cases are
\begin{eqnarray}
W_{3b} = 
M_{03} X_{30} - X_{02}X_{23}X_{30}
+\frac{1}{2} M_{0B}^2 -\frac{1}{2} M_{1B}^2 X_{01}^2
+\frac{1}{2} X_{A1}^2 M_{13}^2 -\frac{1}{2} X_{A1}^2 X_{13}^2
\end{eqnarray}
and 
\begin{eqnarray}
W_{3c} = 
M_{03} X_{30} - X_{02}X_{23}X_{30}
+M_{0B} M_{1B} X_{01} - X_{A1}^2 M_{13} X_{13}\; .
\end{eqnarray}
By integrating out the massive fields (only $M_{03}$ and  $X_{30}$ for the moment)
\begin{eqnarray}
M_{03} X_{30} -X_{02}X_{23}X_{30}+M_{03} Y_{32} Y_{20} \rightarrow  - X_{01}X_{13} Y_{32} Y_{20} \; ,
\end{eqnarray}
we have 
\begin{eqnarray}
W_{3b} &=& 
M_{1B} Y_{B2} Y_{21}
+M_{13} Y_{32} Y_{21}
+M_{0B} Y_{B2} Y_{20}
 \\
&-& X_{01}X_{13} Y_{32} Y_{20}
+\frac{1}{2} M_{0B}^2 -\frac{1}{2} M_{1B}^2 X_{01}^2
+\frac{1}{2} X_{A1}^2 M_{13}^2 -\frac{1}{2} X_{A1}^2 X_{13}^2 
\nonumber 
\end{eqnarray}
and
\begin{eqnarray}
W_{3c} &=& 
M_{1B} Y_{B2} Y_{21}
+M_{13} Y_{32} Y_{21}
+M_{0B} Y_{B2} Y_{20}
 \\
&-&
 X_{01}X_{13} Y_{32} Y_{20}
+M_{0B} M_{1B} X_{01} - X_{A1}^2 M_{13} X_{13}
\nonumber 
\end{eqnarray}
Next we can dualize node $2$.
Seiberg duality on  
$SU(N_1=n-2)$ gives an
$SU(\tilde N_1 =2n-4)$ gauge node. Using the letter $N$ for the mesons of this duality and $Z$ for the dual quarks, the dual superpotential of the gauge node reads
\begin{eqnarray}
W_\text{dual}
&=&
\sum_{i=1,2} 
N_{23}^{(i)} Z_{31}^{i} Z_{12} +
N_{A3}^{(i)} Z_{31}^{i} Z_{2A} +
N_{03}^{(i)} Z_{31}^{i} Z_{10}+
\nonumber \\
&+&N_{0B} Z_{B1} Z_{10}
+N_{2B} Z_{B1} Z_{10}
+N_{AB} Z_{B1} Z_{1A} \; , 
\end{eqnarray}
where the index $i$ refers to the mesons constructed from $X_{13}$ ($i=1$) or $M_{13}$ ($i=1$) .
The remaining superpotentials in the two cases are
\begin{eqnarray}
\label{final1}
W_{3b,\text{def}} &=& 
M_{0B} Y_{B2} Y_{20} + Y_{32} N_{23}^{(2)} + Y_{B2} N_{2B}^{(2)} - N_{03}^{(1)} Y_{32} Y_{20} + W_{\text{dual}}
\nonumber \\
 &+&\frac{1}{2} (M_{0B}^2-N_{0B}^2 + {N_{A3}^{(1)}}^2 -  {N_{A3}^{(2)}}^2 ) 
 \end{eqnarray}
 and
\begin{eqnarray}
\label{final2}
W_{3c,\text{def}} &=&  
M_{0B} Y_{B2} Y_{20} + Y_{32} N_{23}^{(2)} + Y_{B2} N_{2B}^{(2)} - N_{03}^{(1)} Y_{32} Y_{20} + W_{\text{dual}}
\nonumber \\
 &+& M_{0B} N_{0B} + {N_{A3}^{(1)}}  {N_{A3}^{(2)}}  \; .
 \end{eqnarray}
We can see that the difference between the two models corresponds then to the last line in (\ref{final1}) and 
(\ref{final2}) respectively. 

By rotating the fields $M_{0B}$ and $N_{0B}$ and the fields 
$N_{A3}^{(1)}$ and  $N_{A3}^{(2)}$ 
we arrive at a situation similar to the one discussed in Sec. \ref{sec:toy}. The main difference in this case is that the difference between the models resides in mass terms and not in cubic interactions. Nevertheless we can proceed by rotating these fields as discussed above and then integrate them out in both phases.\footnote{Observe that there are other massive terms that can be integrated out as well but they behave identically between the two phases.} The final difference then resides  only in the sign of some of the superpotential terms as in the discussion in Sec. \ref{sec:toy}.

Observe that we can rotate fields only if they carry the same charges under the global symmetry, otherwise we will generate non-marginal terms and break the global symmetry explicitly. As stressed in Sec.  \ref{sec:toy}, in addition to sharing the same quiver the models need to have the same global anomalies in order to be conformally dual. This is crucial, as it determines which theories that share the same quiver flow to the same conformal manifold. For example, the parent theories PdP$_{3b}$ and PdP$_{3c}$, without the orientifold projection, still share the same quiver but are not conformally dual because cancellation of the $\beta$-functions requires different solutions. In this sense, the action of the orientifold is pivotal to ensure anomaly matching.

\section{Embedding in string theory}
\label{sec:embedding}

In this section we give a general description of the 4d $\mathcal{N}=1$ toric quiver gauge theories that generalize the duality between PdP$_{3b}$ and PdP$_{3c}$  after the fixed-line/fixed-point orientifold projections.
The key fact that lends itself to a natural generalization is that both PdP$_{3b}$ and PdP$_{3c}$ can be represented by the same quiver. However the two models have different superpotentials that correspond to the two inequivalent planarizations of the quiver and they cannot be related by any low-energy duality. This reflects in the fact that the two toric diagrams 
are not related by any $SL(2,\mathbb{Z})$ transformation. In general we are looking for examples of such type: pairs (or sets) of models sharing the same quiver but with inequivalent planarizations. 
This is a necessary but not sufficient condition in our construction of conformally dual models.
We refer to such quivers as multi-planarizable. After obtaining classes of models with this property we discuss general aspects of their orientifold projections that give rise to conformal dualities.

\subsection{Multi-planarizable quivers}

In most  cases, given a planarizable quiver the latter admits a unique planarization in terms of plaquettes and therefore a unique dimer and tiling. However here we are interested in families of models with the same quiver but different planarizations. We identify such families in terms of their toric diagrams, that are inequivalent also up to $SL(2,\mathbb{Z}_2)$ transformations.
From the toric diagram we then identify the brane tiling using the inverse algorithm. This step is not unique, because toric dual phases can emerge by different choices of the intersections of the five-branes in the brane tilings that preserve the zig-zag paths. 
Nevertheless we observe that it is always possible to choose these brane tilings for pairs of inequivalent toric diagrams, in order to obtain an identical quiver.
This implies that the quivers obtained by this procedure are the multi-planarizable quivers that we are looking for. 
The simplest realization of models with such property corresponds to the case of PdP$_{3b}$ and PdP$_{3c}$.

\begin{minipage}{\linewidth}
      \centering
      \begin{minipage}{0.45\linewidth}
          \begin{figure}[H]
        \begin{tikzpicture}[auto, scale=1]
        \node [circle, fill=black!30!red, inner sep=0pt, minimum size=2mm] (l0) at (0,0) {};            
        \node [circle, fill=black!30!red, inner sep=0pt, minimum size=2mm] (l1) at (0,1) {};
        \node [circle, fill=black!30!red, inner sep=0pt, minimum size=2mm] (l2) at (0,2) {}; 

        \node [circle, fill=black!30!green, inner sep=0pt, minimum size=2mm] (c0) at (1,0) {}; 
        \node [circle, fill=black!30!green, inner sep=0pt, minimum size=2mm] (c1) at (1,1) {};
        \node [circle, fill=black!30!green, inner sep=0pt, minimum size=2mm] (c2) at (1,2) {};
        \node [circle, fill=black!30!green, inner sep=0pt, minimum size=2mm] (c3) at (1,3) {};
        \node [circle, fill=black!30!green, inner sep=0pt, minimum size=2mm] (c4) at (1,4) {};
        \node [circle, fill=black!30!green, inner sep=0pt, minimum size=2mm] (c5) at (1,5) {};
        \node [circle, fill=black!30!green, inner sep=0pt, minimum size=2mm] (c6) at (1,6) {};
        \node [circle, fill=black!30!green, inner sep=0pt, minimum size=2mm] (c7) at (1,7) {};
        \node [circle, fill=black!30!green, inner sep=0pt, minimum size=2mm] (c8) at (1,8) {};

        \node [circle, fill=black!30!yellow, inner sep=0pt, minimum size=2mm] (r0) at (2,0) {};
        \node [circle, fill=black!30!yellow, inner sep=0pt, minimum size=2mm] (r1) at (2,1) {};
        \node [circle, fill=black!30!yellow, inner sep=0pt, minimum size=2mm] (r2) at (2,2) {};
        \node [circle, fill=black!30!yellow, inner sep=0pt, minimum size=2mm] (r3) at (2,3) {};        

        \draw (l0) to node {} (l2) [-, thick];
        \draw (l2) to node {} (c8) [-, thick];
        \draw (c8) to node {} (r3) [-, thick];
        \draw (r3) to node {} (r0) [-, thick];
        \draw (r0) to node {} (l0) [-, thick];
                
        \node [left=0.2cm of l0] {$(0,\,0)$};
        \node [left=0.2cm of l2, black!30!red] {$(0,\,k_1)$};
        \node [left=0.2cm of c8, black!30!green] {$(1,\,k_3)$};
        \node [right=0.2cm of r3, black!50!yellow] {$(2,\,k_2)$};
\end{tikzpicture}\vspace{10pt}
    \caption{Generic toric diagram  labeled by $(k_1,k_2,k_3)$.}
    \label{FIG3}
          \end{figure}
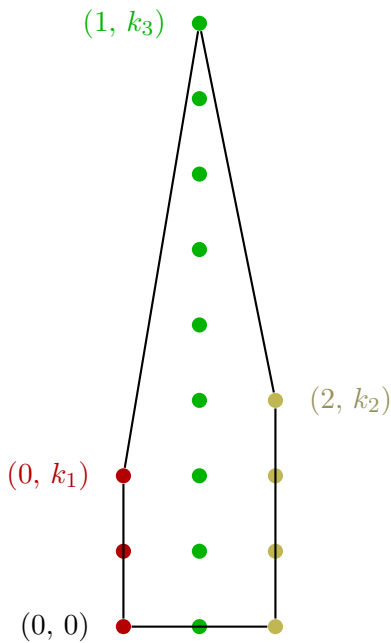
      \end{minipage}
      \hspace{0.05\linewidth}
      \begin{minipage}{0.45\linewidth}
          \begin{figure}[H]
           \centering
               \includegraphics[width=10cm,angle=90,origin=c,scale=.9]{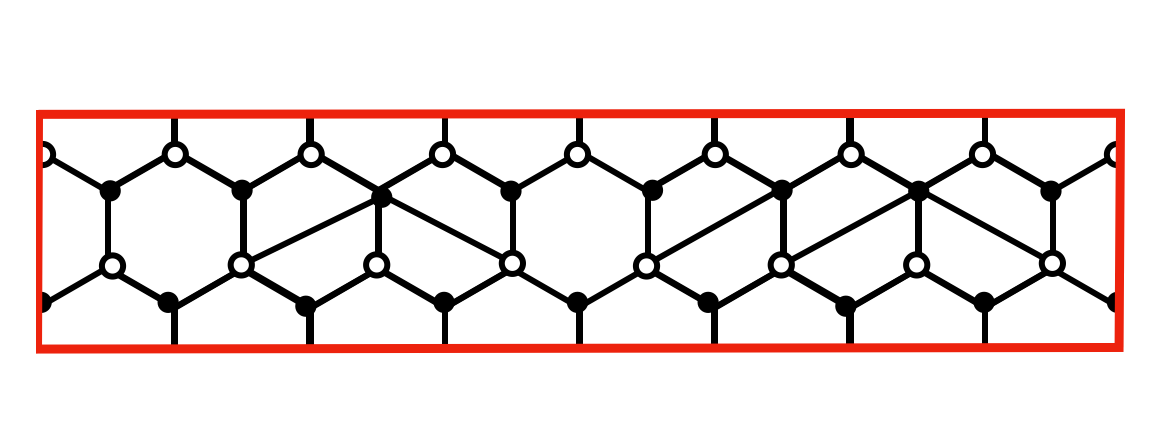}
\caption{Fundamental cell of the $(k_1,k_2,k_3)$ model.}
\label{FIG1}
          \end{figure}
      \end{minipage}
  \end{minipage}\vspace{15pt}
  
Let us start by identifying the toric diagrams: they are drawn in Fig. \ref{FIG3}. 
In general we label these toric diagrams by three integers 
$(k_1,k_2,k_3)$, where $k_1$ and $k_2$ represent the number of points in the two external lines and $k_3$ is the number of points in the central one, excluding the points at the base. Equivalently, $k_1$ and $k_2$ represent the length of the external lines, while $k_3$ the length of the central line.

It is possible to show that such toric diagrams always admit a dimer with a fundamental cell given in Fig. \ref{FIG1}.
This dimer has $2k_1$ squares obtained by cutting hexagons with the NW-SE orientation and $2k_2$ with the NE-SW orientation. Then, there are $k_3-k_2-k_1$ hexagons in the central part of the fundamental cell. The other column is on the other hand made of $k_3$ hexagons. We have to require $k_3 \geq k_2+k_1$, which is saturated when there are only squares in the central column. With this notation, the examples in the previous section correspond to $(1,1,2)$ for PdP$_{3b}$ and $(0,2,2)$ for PdP$_{3c}$.

\begin{figure}
\begin{center}
\includegraphics[width=8cm]{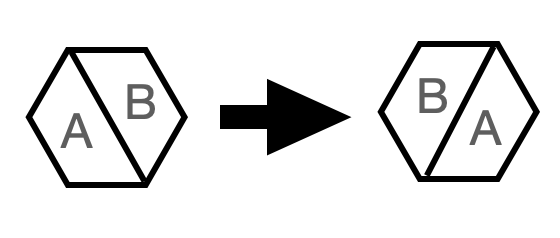}
\caption{In this picture we represent the flip on the dimer that preserves the quiver. We consider a pair of faces $A$ and $B$ obtained by cutting an hexagon with a NW-SE diagonal and we flip it into a NE-SW one. By relabeling the faces one can see that the quiver remains identical even if in general it is associated to a different toric diagram.}
\label{FIG2}
\end{center}
\end{figure}

Here is the key observation that allows us to construct multiplanarizable quivers. We consider the flip in figure \ref{FIG2}, where a hexagon cut by a diagonal with the NW-SE orientation is transformed into a NE-SW oriented one, while relabeling the faces as in Fig. \ref{FIG2}.
This flip does not give in general a new consistent brane tiling, and in many cases the flipped dimer 
shows one of the inconsistencies discussed in \cite{Hanany:2015tgh}.
However this is not the case  for the models labeled by $(k_1,k_2,k_3)$ considered here.
Indeed in such cases the flip does not modify the quiver even if it is associated in general to a consistent but inequivalent  model. In fact the latter has a different toric diagram.
Labeling the original toric diagram with the triple  $(k_1,k_2,k_3)$, the new toric diagram after such a flip is given by the new triple $(k_1+1,k_2-1,k_3)$.
In general we can iterate the flip described above and conclude that two models $(k_1,k_2,k_3)$ and $(k'_1,k'_2,k'_3)$ have the same quiver but different toric data if $k_3'=k_3$ and $k_1+k_2=k'_1+k'_2$ (observe that the choice $k_1 = k'_2$ is trivial in this sense).

\subsection{Orientifold projections}

In this sub-section we add the orientifolds to the previous brane setups 
and we focus on the relation between the orientifolds and the conformal symmetry. Typically when O-planes are considered it is not straightforward to preserve the conformal symmetry for stacks of $N$ D3- branes probing the singularity. 
On the field theory side suitable choices of gauge ranks have to be considered. 

As discussed in \cite{Antinucci:2020yki}
there are three possible scenarios if one considers the fate of conformal invariance 
after the projection.
The first scenario corresponds to the case in which  there is a fixed point after the projection  and there are $\mathcal{O}(1/N)$ 
corrections on the  physical observables
due to the presence of the orientifold. 
The second scenario corresponds to the
case in which the theory obtained after the projection does not have any conformal fixed point. 
The third possible scenario corresponds to 
the case in which there is a new fixed point in the unoriented model, but the corrections are not of order $\mathcal{O}(1/N)$ anymore.

In the example of \cite{Antinucci:2020yki} it was shown that 
there is an interplay between the first and the third scenario in the orientifold projections of the two inequivalent models. 
Here we will extend this relation in the models parameterized by $(k_1,k_2,k_3)$.
Namely we will find a general assignment of gauge ranks such that some unoriented conformal theories
will belong to the first scenario and some other to the third scenario. We checked that the central charges of these orientifold models match and that only the cases with $k_1=k_2$ belong to the first scenario.
Furthermore we claim that by applying the orientifold and assigning such ranks  to pairs of inequivalent models sharing the same quiver we obtain unoriented theories differing only by exactly marginal deformations as in  the case studied in \cite{Antinucci:2020yki} that we have reviewed above.

Concretely, by analyzing a large amount of models, we have indeed observed that the conformal duality obtained in \cite{Antinucci:2020yki} can be generalized by requiring that the final models have two real gauge groups (and equivalently two two-index tensors for two of the unitary groups). This restricts the possibilities, implying that  $k_1+k_2=2k$  with $k$ integer, and $k_3 \geq 2k$. We have explicitly seen that relaxing this condition allows for models which are not conformally dual, but it is not clear, at least for the moment, what is the reason behind it, both from the field theory perspective and from the string theory side.
In this way one of the possible models is always $(k,k,k_3 \geq 2k)$, that admits a fixed-line projection. This fixes also the possible signs of the fixed-point projections in the other models in the same family of conformally dual theories.
In order to construct a family of fixed-point conformally dual orientifolds we proceed as follows. First we identity the 
model with $(0,2k,k_3)$ as the seed of the family. This model and the $(k,k,k_3)$ model must be chosen in a Seiberg dual phase that allows the fixed-point or fixed-line projections to give rise to the same final quiver. Such a phase is not unique, there can also be other phases that give rise to the same final quiver, and there can be also other (inequivalent) choices. As we will discuss, those inequivalent choices are no longer related by chains of Seiberg dualities after the orientifold projection. 
Once this is fixed we can increase $k_1$ using the flip in figure \ref{FIG2} by two units preserving the sum $k_1+k_2$. A generic model in this case is $(2p+1,2k-2p-1,k_3)$.
Depending on the parity of $k$ the sequence of flips terminates with $(k,k,k_3)$ or $(k-1,k+1,k_3)$. We will give evidence that all of the models constructed by fixed-point orientifolds in this way are conformally dual to the fixed-line\footnote{For odd $k$, the $(k,k,k_3)$ case admits also a fixed-point projection. We will see in the examples below that this choice has to be considered as well because it can give rise to new types of conformally dual models.
Observe that  
there are also other  choices of signs if $(k,k,k_3)$ has fixed points but we will not discuss such possibility below.} orientifold of $(k,k,k_3)$ (i.e. that there is at least one Seiberg dual phase that gives rise to such a conformally dual theory, once the gauge ranks are suitably chosen).
Observe that the minimal $k=1$ corresponds to the case studied in \cite{Antinucci:2020yki}, where indeed there are only two possibilities $(1,1,k_3=2)$ and $(0,2,k_3=2)$.
Furthermore, the case $k=2$ cannot give rise to any intermediate case either, since only $(2,2,k_3=4)$ and $(0,4,k_3=4)$ are allowed, while $(1,3,k_3=4)$ does not admit any toric dual phases that can be projected consistently with the other two cases.
Hence, the first non-trivial case with more than two models has necessarily $k=3$.

\begin{figure}[ht!]
\begin{center}
\includegraphics[scale=.7]{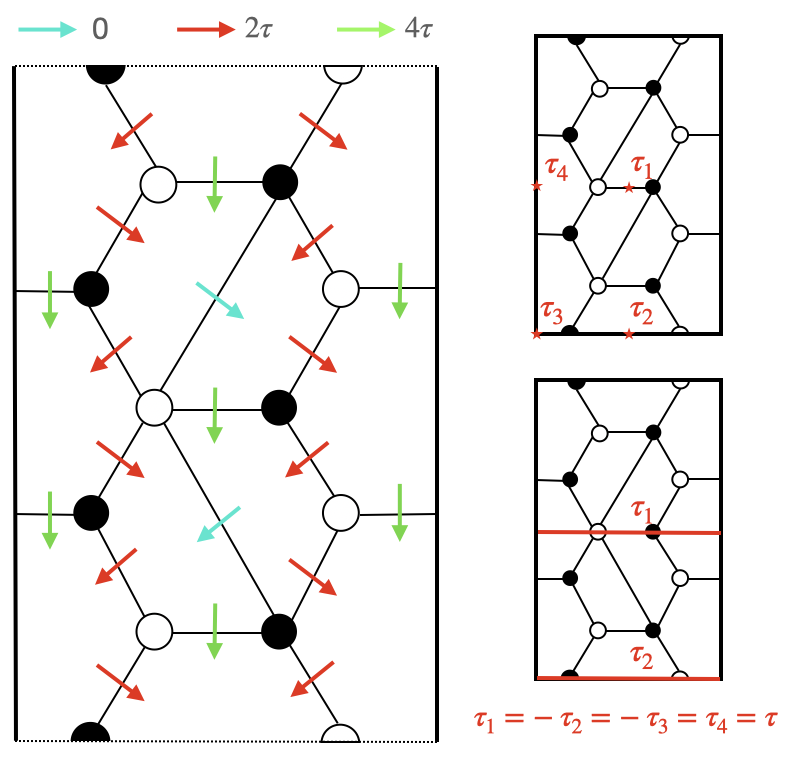}
\caption{On the LHS  we summarize the general rank assignation on the dimer that we have considered in the whole paper. We start by considering the same rank for each node and then we shift the ranks by $j \tau$ where $j={\color{cyan}0},{\color{red}2},{\color{green}4}$ following the coloring of the arrows in the figure. On the RHS we consider the choices of the charges $\tau$ fo a generic fixed line and fixed point projection.}
\label{rankass}
\end{center}
\end{figure}

By inspection we have also observed that there is a unique assignation of ranks that gives rise to the generalization of the conformal duality found in the original PdP$_{3b/c}$ case. This choice is explained as follows. Consider the limiting case with $k_1 = k_2$ and its fixed-line orientifold projection, so that there are only two charges whose signs are $\tau_1$ and $\tau_2$. On the reduced fundamental cell delimited by the fixed lines, $\tau_1$ is the charge associated to the line at the top, while $\tau_2$ at the bottom, as in Fig. \ref{rankass}. The vanishing of the beta functions always requires that $\tau_1 = - \tau_2 = \tau$, consistently with the prototypical case of PdP models. Consider the case in which the gauge factors projected by the top line and the bottom lines lie on the side of the dimer, and label them with 0 and $3k$, respectively. The hexagons on the sides of the reduced cell are labelled with $3\ell $, $\ell= 0,\ldots,\,k$. As a consequence, gauge factors from the central line in the reduced cell of the dimer are labelled with $3\ell+1$ and $3\ell+2$ for a pair of squares, $3\ell+1$ (or $3\ell+2$ equivalently) for hexagons. Having set up the notation for the labels, we assign rank $N_0=n$ to the first gauge factor\footnote{Or better, the zeroth one.} and the choice of the ranks is summarized as 
\begin{align}
    &N_{3\ell} = n - 4 \ell  \, \tau_1 \; , \label{eq:shiftonsideofdimer} \\[5pt]
    &N_{3\ell+1} = N_{3\ell+2} = n - 2 - 4 \ell \, \tau_1 \; , \label{eq:shiftoncenterofdimer}
\end{align}
with $\ell=0, \ldots, \, k$, where the shifts are understood as in Fig. \ref{rankass}. 

If the projected gauge factors lie on the central part of the dimer instead, we only need to shift the labels by 3, with $\ell \to \ell+1$, and the rank assignement remains as in Eq.~\eqref{eq:shiftonsideofdimer} for the gauge factors on the side of the dimer, and Eq.~\eqref{eq:shiftoncenterofdimer} for the gauge factors on the central part of the dimer. Colored fields in red and green represent tensors and conjugated tensors, while diamond nodes are real gauge groups. Finally, the general quiver is drawn as in Fig. \ref{fig:GenericQuiver} and wherever there are hexagons in the central line of the dimer, their contribution to the quiver is drawn in an example in Fig. \ref{fig:GenericQuiverHexagons}, where we may think of it as part of the diagram has been folded.

\begin{figure}[ht!]
    \centering{ 
        \begin{tikzpicture}[auto]
        \node [diamond, draw=blue!50, fill=blue!20, inner sep=0pt, minimum size=5mm] (0) at (0,0) {};
		\node [circle, draw=blue!50, fill=blue!20, inner sep=0pt, minimum size=4mm] (2) at (1.25,-1) {};
		\node [circle, draw=blue!50, fill=blue!20, inner sep=0pt, minimum size=4mm] (1) at (1.25,1) {};
		\node [circle, draw=blue!50, fill=blue!20, inner sep=0pt, minimum size=4mm] (3) at (2.5,0) {};
  		\node [circle, draw=blue!50, fill=blue!20, inner sep=0pt, minimum size=4mm] (5) at (3.75,1) {};
            \node [circle, draw=blue!50, fill=blue!20, inner sep=0pt, minimum size=4mm] (4) at (3.75,-1) {};
    
            \node [circle, draw=blue!50, fill=blue!20, inner sep=0pt, minimum size=4mm] (3i) at (5,0) {};
		\node [circle, draw=blue!50, fill=blue!20, inner sep=0pt, minimum size=4mm] (3i2) at (6.25,-1) {};
		\node [circle, draw=blue!50, fill=blue!20, inner sep=0pt, minimum size=4mm] (3i1) at (6.25,1) {};
		\node [circle, draw=blue!50, fill=blue!20, inner sep=0pt, minimum size=4mm] (3i3) at (7.5,0) {};
		\node [circle, draw=blue!50, fill=blue!20, inner sep=0pt, minimum size=4mm] (22) at (8.75,-1) {};
		\node [circle, draw=blue!50, fill=blue!20, inner sep=0pt, minimum size=4mm] (11) at (8.75,1) {};
		\node [diamond, draw=blue!50, fill=blue!20, inner sep=0pt, minimum size=5mm] (33) at (10,0) {};
		\node [left=0.2cm of 0] {\scriptsize{$0$}};
  		\node [above=0.2cm of 1] {\scriptsize{$1$}};
    	\node [left=0.2cm of 2] {\scriptsize{$2$}};
   		\node [below=0.2cm of 3] {\scriptsize{$3$}};
            \node [below=0.2cm of 4] {\scriptsize{$4$}};
    	\node [above=0.2cm of 5] {\scriptsize{$5$}};
            \node [left=0.3cm of 3i] {\normalsize{$\ldots \quad \ldots$}};
            \node [left=0.7cm of 3i2] {\normalsize{$\ldots$}};
		\node [below=0.2cm of 3i] {\scriptsize{$3\ell$}};
		\node [below=0.2cm of 3i2] {\scriptsize{$3\ell+2$}};
		\node [above=0.2cm of 3i1] {\scriptsize{$3\ell+1$}};
		\node [below=0.2cm of 3i3] {\scriptsize{$3\ell+3$}};
            \node [right=0.2cm of 3i3] {\normalsize{$\ldots \quad \ldots$}};
            \node [right=0.7cm of 3i1] {\normalsize{$\ldots$}};
		\node [right=0.2cm of 22] {\scriptsize{$3k-1$}};
		\node [above=0.2cm of 11] {\scriptsize{$3k-2$}};
		\node [right=0.2cm of 33] {\scriptsize{$3k$}};
        \draw (2) to [out=320, in=220, looseness=8] (2) [-, thick, red];
        \draw (0) to (1) [->, thick];
        \draw (1) to (3) [->, thick];
        \draw (0) to (2) [->, thick];
        \draw (2) to (3) [->, thick];
        \draw (1) to (2) [->, thick];
        \draw (3) to (0) [->, thick];
        \draw (3) to (4) [->, thick];
        \draw (3) to (5) [->, thick];
        \draw (4) to (3i) [->, thick];
        \draw (5) to (3i) [->, thick];
        \draw (4) to (5) [->, thick];
        \draw (5) to (1) [->, thick];
        \draw (3i) to (3i1) [->, thick];
        \draw (3i) to (3i2) [->, thick];
        \draw (3i1) to (3i3) [->, thick];
        \draw (3i2) to (3i3) [->, thick];
        \draw (3i1) to (3i2) [->, thick];
        \draw (3i3) to (3i) [->, thick];
        \draw (3i3) to (11) [->, thick];
        \draw (3i3) to (22) [->, thick];
        \draw (11) to (33) [->, thick];
        \draw (22) to (33) [->, thick];
        \draw (11) to (22) [->, thick];
        \draw (22) to [out=320, in=220, looseness=8] (22) [-, thick, green];
    \end{tikzpicture}}
    \caption{The generic quiver associated to a toric theory with $(k_1,\, k_2,\, k_3)$ and $k_3 = k_1 + k_2 = 2k$ and $\ell=0, \ldots, k$.}
    \label{fig:GenericQuiver}
\end{figure}

\begin{figure}[ht!]
    \centering{ 
        \begin{tikzpicture}[auto]
            \node [diamond, draw=blue!50, fill=blue!20, inner sep=0pt, minimum size=5mm] (m1) at (-1.25,1) {};
            \node [circle, draw=blue!50, fill=blue!20, inner sep=0pt, minimum size=4mm] (0) at (0,0) {};
		\node [circle, draw=blue!50, fill=blue!20, inner sep=0pt, minimum size=4mm] (1) at (1.25,-1) {};
		\node [circle, draw=blue!50, fill=blue!20, inner sep=0pt, minimum size=4mm] (2) at (1.25,1) {};
		\node [circle, draw=blue!50, fill=blue!20, inner sep=0pt, minimum size=4mm] (3) at (2.5,0) {};
  		\node [circle, draw=blue!50, fill=blue!20, inner sep=0pt, minimum size=4mm] (4) at (3.75,1) {};
            \node [circle, draw=blue!50, fill=blue!20, inner sep=0pt, minimum size=4mm] (5) at (3.75,-1) {};
    
            \node [circle, draw=blue!50, fill=blue!20, inner sep=0pt, minimum size=5mm] (3i) at (5,0) {};
		\node [circle, draw=blue!50, fill=blue!20, inner sep=0pt, minimum size=4mm] (3i1) at (6.25,1) {};
		\node [circle, draw=blue!50, fill=blue!20, inner sep=0pt, minimum size=4mm] (3i3) at (7.5,0) {};
		\node [circle, draw=blue!50, fill=blue!20, inner sep=0pt, minimum size=4mm] (22) at (8.75,-1) {};
		\node [circle, draw=blue!50, fill=blue!20, inner sep=0pt, minimum size=4mm] (11) at (8.75,1) {};
		\node [diamond, draw=blue!50, fill=blue!20, inner sep=0pt, minimum size=5mm] (33) at (10,0) {};
            \node [left=0.2cm of m1] {\scriptsize{$1$}};
		\node [left=0.2cm of 0] {\scriptsize{$3$}};
  		\node [below=0.2cm of 1] {\scriptsize{$4$}};
    	\node [above=0.2cm of 2] {\scriptsize{$5$}};
   		\node [below=0.2cm of 3] {\scriptsize{$6$}};
            \node [above=0.2cm of 4] {\scriptsize{$7$}};
    	\node [below=0.2cm of 5] {\scriptsize{$8$}};
            \node [left=0.3cm of 3i] {\normalsize{$\ldots \quad \ldots$}};
            \node [left=0.7cm of 3i1] {\normalsize{$\ldots$}};
		\node [below=0.2cm of 3i] {\scriptsize{$3\ell+3$}};
		\node [above=0.2cm of 3i1] {\scriptsize{$3\ell+4$}};
		\node [below=0.2cm of 3i3] {\scriptsize{$3\ell+6$}};
            \node [right=0.2cm of 3i3] {\normalsize{$\ldots \quad \ldots$}};
            \node [right=0.7cm of 3i1] {\normalsize{$\ldots$}};
		\node [right=0.2cm of 22] {\scriptsize{$3k+2$}};
		\node [above=0.2cm of 11] {\scriptsize{$3k+1$}};
		\node [right=0.2cm of 33] {\scriptsize{$3k+3$}};
        \draw (0) to [out=190, in=280, looseness=8] (0) [-, thick, red];
        \draw (m1) to (0) [->>, thick];
        \draw (2) to (m1) [->, thick];
        \draw (0) to (1) [->, thick];
        \draw (1) to (3) [->, thick];
        \draw (0) to (2) [->, thick];
        \draw (2) to (3) [->, thick];
        \draw (1) to (2) [->, thick];
        \draw (3) to (0) [->, thick];
        \draw (3) to (4) [->, thick];
        \draw (3) to (5) [->, thick];
        \draw (4) to (3i) [->, thick];
        \draw (5) to (3i) [->, thick];
        \draw (4) to (5) [->, thick];
        \draw (5) to (1) [->, thick];
        \draw (3i) to (3i1) [->>, thick];
        \draw (3i1) to (3i3) [->>, thick];
        \draw (3i3) to (3i) [->, thick];
        \draw (3i3) to (11) [->, thick];
        \draw (3i3) to (22) [->, thick];
        \draw (11) to (33) [->, thick];
        \draw (22) to (33) [->, thick];
        \draw (11) to (22) [->, thick];
        \draw (22) to [out=320, in=220, looseness=8] (22) [-, thick, green];
    \end{tikzpicture}}
    \caption{The generic quiver associated to a toric theory with $(k_1,\, k_2,\, k_3)$ and $k_3 > k_1 + k_2 = 2k$ and $\ell=1, \ldots, (k+1)$.}
    \label{fig:GenericQuiverHexagons}
\end{figure}
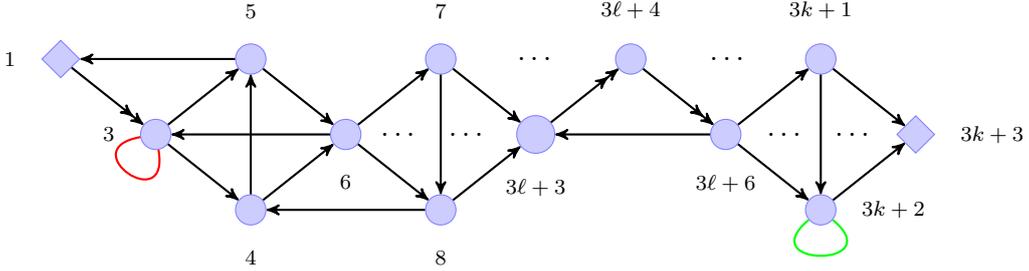

\section{Case study}
\label{sec:study}
\label{sec:example}

In this section we apply the construction discussed above for obtaining conformally dual models by suitable orientifold projections of multi-planarizable toric quiver gauge theories. We will discuss explicitly two cases, both with $k=2$, where all the properties and possibilities discussed in Sec. \ref{sec:toy} show up. A larger set of examples is included in App. \ref{sec:apx}. In the following analysis we will keep the same conventions for the quivers and for the projections discussed in the previous sections. We checked, by following the same analysis discussed for the toy models and the PdPs, that each model has a fixed point that satisfies the unitarity bound. We will then show in each case that, by applying chains of tensor matter deconfinements, dualities and rotations, all the conformally dual models can be transformed into models with the same superpotential up to some signs in the interactions, making the conformal duality completely explicit.

\subsection{$(2,2,5)$ vs. $(0,4,5)$}
\label{sec:(225)(045)}

\begin{figure}[ht!]
\begin{center}
\includegraphics[scale=0.6]{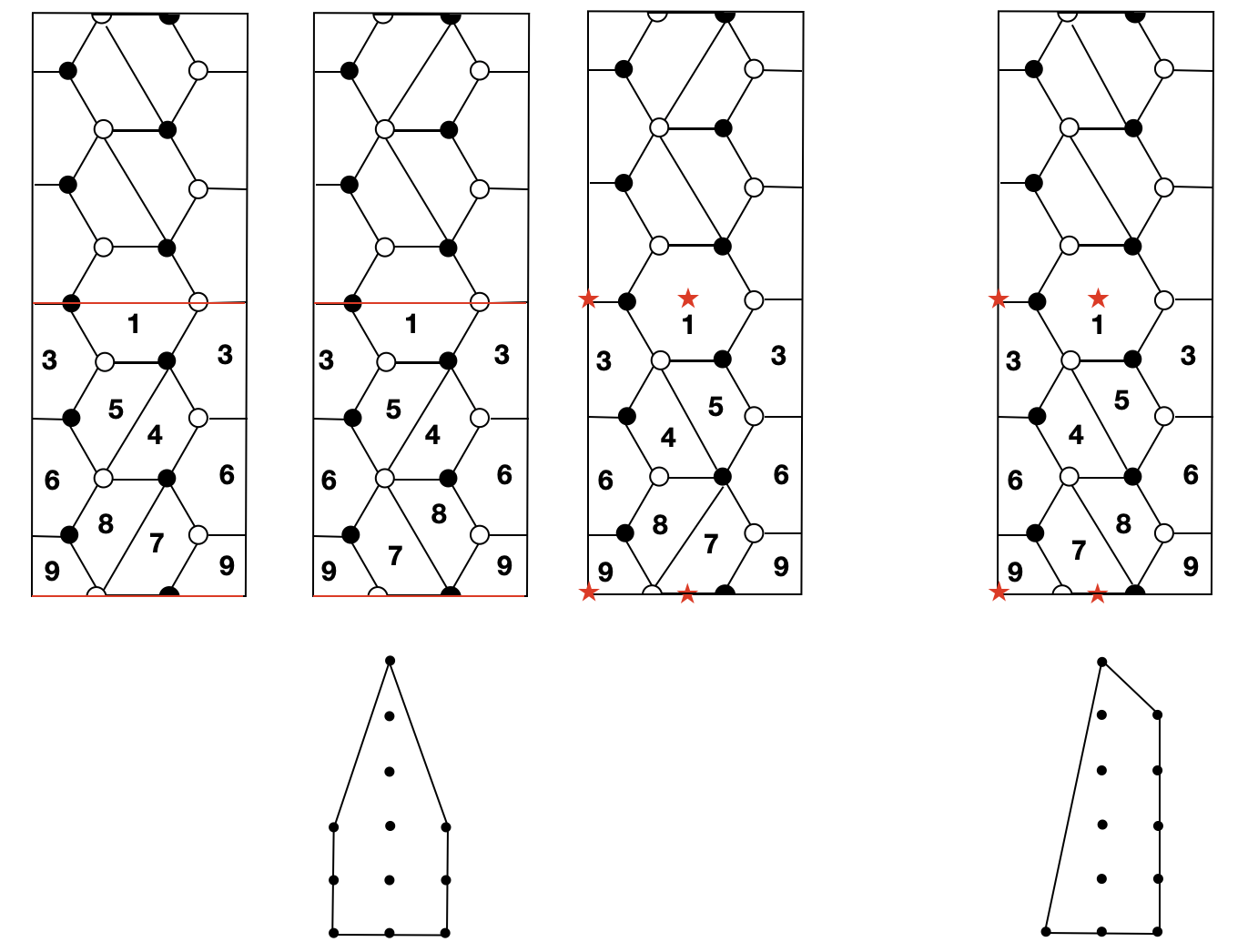}
\caption{The dimers representing the orientifold projection of toric diagrams $(2,2,5)$, on the left, and $(0,4,5)$ on the right. For $(2,2,5)$, there are both fixed lines and fixed points projections.}
\label{FIG336156}
\end{center}
\end{figure}

Let us start the analysis with the dimers and the toric diagrams depicted in Fig. \ref{FIG336156}. There are three dimers associated to the $(2,2,5)$ model and one dimer
associated to the $(0,4,5)$ model. Even if the three dimers associated to $(2,2,5)$ are Seiberg dual before the orientifold, we will analyze the various possibilities because the conformal duality model emerges in different ways.

While in the $(0,4,5)$ case we have only a single fixed-point orientifold projection that adapts to our goals, in the $(2,2,5)$ case there are three orientifold projections that we can consider. Two of them are fixed-line projections while the last is one by fixed points.
 
\begin{figure}
    \centering{
    \begin{tikzpicture}[auto]
        \node [circle, draw=blue!50, fill=blue!20, inner sep=0pt, minimum size=5mm] (3) at (0,0) {$3$};
		\node [circle, draw=blue!50, fill=blue!20, inner sep=0pt, minimum size=5mm] (4) at (1.25,-1) {$4$};
		\node [circle, draw=blue!50, fill=blue!20, inner sep=0pt, minimum size=5mm] (5) at (1.25,1) {$5$};
		\node [circle, draw=blue!50, fill=blue!20, inner sep=0pt, minimum size=5mm] (6) at (2.5,0) {$6$};
		\node [diamond, draw=blue!50, fill=blue!20, inner sep=0pt, minimum size=6mm] (1) at (-1.25,1) {$1$};
		\node [circle, draw=blue!50, fill=blue!20, inner sep=0pt, minimum size=5mm] (8) at (3.75,-1) {$8$};
            \node [circle, draw=blue!50, fill=blue!20, inner sep=0pt, minimum size=5mm] (7) at (3.75,1) {$7$};
            \node [diamond, draw=blue!50, fill=blue!20, inner sep=0pt, minimum size=6mm] (9) at (5,0) {$9$};
        \draw (1) to (3) [->>, thick];
        \draw (3) to (5) [->, thick];
        \draw (5) to (1) [->, thick];
        \draw (3) to (4) [->, thick];
        \draw (4) to (5) [->, thick];
        \draw (4) to (6) [->, thick];
        \draw (5) to (6) [->, thick];
        \draw (6) to (3) [->, thick];
        \draw (9) to (6) [->, thick];
        \draw (6) to (7) [->, thick];
        \draw (6) to (8) [->, thick];
        \draw (8) to (9) [->, thick];
        \draw (7) to (9) [->, thick];
        \draw (7) to (8) [->, thick];
        \draw (8) to (4) [->, thick];
         \draw (3) to [out=190, in=280, looseness=8] (3) [-, thick, red];
         \draw (7) to [out=140, in=40, looseness=8] (7) [-, thick, green];
    \end{tikzpicture}}
    \caption{The quiver for the theories $(2,2,5)$ and $(0,4,5)$, after the orientifold projection.}
    \label{fig:Quiver225}
\end{figure}
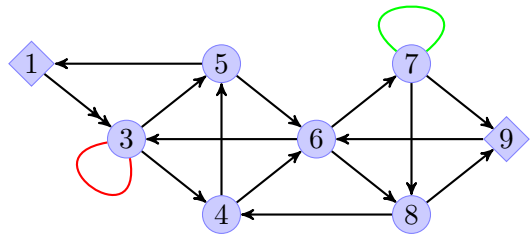

The projections give rise to different superpotentials. For the first fixed-line projection of the $(2,2,5)$ model we have
\begin{align}\label{eq:W225FixedLines1}
W_{(2,2,5)}^{\Omega_\text{f.l.}^1} &=
\frac{1}{2}T_{33} (X_{13}^2 - Y_{13}^2) + Y_{13}X_{35}Y_{51} - X_{13}X_{34}X_{45}X_{51} \nonumber \\[5pt] 
&+ X_{34}X_{46}X_{63} - X_{35}X_{56}X_{63} + X_{56}X_{68}X_{84}X_{45} - X_{46}X_{67}X_{78}X_{84} \nonumber \\[5pt]
&+ X_{67}X_{79}X_{96} - X_{68}X_{89}X_{96} +\frac{1}{2} \widetilde{T}_{77} (X_{78}^2 X_{89}^2 - X_{79}^2) \; ,
\end{align}
while the second fixed-line projection of the $(2,2,5)$ model we have
\begin{align}\label{eq:W225FixedLines2}
W_{(2,2,5)}^{\Omega_\text{f.l.}^2} &=
\frac{1}{2}T_{33} (X_{13}^2 - Y_{13}^2) + Y_{13}X_{35}Y_{51} - X_{13}X_{34}X_{45}X_{51} \nonumber \\[5pt] 
&+ X_{34}X_{46}X_{63} - X_{35}X_{56}X_{63} + X_{56}X_{67}X_{78}X_{84}X_{45} - X_{46}X_{68}X_{84} \nonumber \\[5pt]
&+ X_{68}X_{89}X_{96} - X_{67}X_{79}X_{96} +\frac{1}{2} \widetilde{T}_{77} (X_{79}^2X_{78}^2 - X_{89}^2) \; .
\end{align}
On the other hand the fixed-point projection of the $(2,2,5)$ model gives 
\begin{align}\label{eq:eq:W225FixedPoints1}
W_{(2,2,5)}^{\Omega_\text{f.p.}} &=
\frac{1}{2}T_{33}X_{13}Y_{13} + Y_{13}X_{34}X_{45}X_{51} - X_{13}X_{35}Y_{51} \nonumber \\[5pt] 
&+ X_{35}X_{56}X_{63} - X_{34}X_{46}X_{63} + X_{46}X_{68}X_{84} - X_{56}X_{67}X_{78}X_{84}X_{45} \nonumber \\[5pt]
&+ X_{67}X_{79}X_{96} - X_{68}X_{89}X_{96} + \widetilde{T}_{77} X_{79}X_{78}X_{89} \; .
\end{align}
Lastly, the fixed-point projection for the 
$(0,4,5)$ model gives
\begin{align}\label{eq:eq:W045FixedPoints1}
W_{(0,4,5)}^{\Omega_\text{f.p.}} &=
T_{33} X_{13}Y_{13} + Y_{13}X_{34}X_{45}X_{51} - X_{13}X_{35}Y_{51} \nonumber \\[5pt] 
&+ X_{35}X_{56}X_{63} - X_{34}X_{46}X_{63} + X_{46}X_{67}X_{78}X_{84} - X_{56}X_{68}X_{84}X_{45} \nonumber \\[5pt]
&+ X_{68}X_{89}X_{96} - X_{67}X_{79}X_{96} +\widetilde{T}_{77}X_{78}X_{89}X_{79} \; .
\end{align}
The ranks of the gauge groups consistent with the anomaly cancellations and the existence of a conformal fixed point are
\begin{eqnarray}
\label{ranks336}
&&
N_1 = n, \, N_3 = n - 2 \tau , \, N_4 = N_5 = n - 4 \tau , \nonumber \\
&&
N_6 = n - 6 \tau , \, N_7 = N_8 = n - 8 \tau , \, N_8 = n - 10 \tau \; .
\end{eqnarray}
Let us remind the reader that the two possible choices of sign $\tau=\pm 1$ give rise to two different quivers, that in this case are inequivalent and need to be studied separately. 

The two different projections indeed yield either $SO(N_1)$ and $USp(N_9)$ or $USp(N_1)$ and $SO(N_9)$. Furthermore the tensors $T_{33}$ and $\widetilde T_{77}$ are either  symmetric and conjugate anti-symmetric or conjugate symmetric and anti-symmetric, respectively.

We have checked that both choices of $\tau$ give consistent SCFT by $a$-maximizing with the ranks in formula 
\eqref{ranks336}. Furthermore we have computed the various 't Hooft anomalies and have seen that they coincide between the models. The relation between these models is then very similar to one between the projections of PdP$_{3b}$ and PdP$_{3c}$.

In the following we will see that indeed, by applying the general analysis of Sec. \ref{sec:toy}, we can distinguish (in some Seiberg dual phase) the models only by some superpotential signs.

Let us begin by comparing the superpotential of the two fixed-line projections of the model $(2,2,5)$, Eqs. \eqref{eq:W225FixedLines1}-\eqref{eq:W225FixedLines2}. They differ only in the last terms in the second line and the first two terms in the last line. While the latter two can be reabsorbed by a change of sign in $X_{96}$, the former two can not, and we proceed as follow. Observe that before the orientifold projection the two theories represent two different Seiberg dual phases associated with the same toric geometry. After the projection they are conformally dual, and one might think they are actually Seiberg dual. We shall see that this is \emph{not} the case. Performing Seiberg duality on the gauge node labeled by 5, the superpotentials become\footnote{Also, some fields become massive in both models and are integrated out, but their $F$-terms do not affect the terms we are interested in. }
\begin{align}
    &W_{(2,2,5)}^{\Omega_\text{f.l.}^1} = (\ldots) + M_{46}X_{68}X_{84} - X_{46}X_{67}X_{78}X_{84}
    \; , \nonumber \\[5pt]
    &W_{(2,2,5)}^{\Omega_\text{f.l.}^2} = (\ldots) + M_{46}X_{67}X_{78}X_{84} - X_{46}X_{68}X_{84} 
    \; , 
\end{align}
where we have highlighted only the unequal terms  
and where $M_{46}$ is one of the mesons in the duality.
If we consider the combinations $N_{46}^{(\pm)}= X_{46} \pm M_{46}$ we can see that the two superpotentials differ only in a sign, as discussed in Sec. \ref{sec:toy}. Hence, the theories after the orientifold projection are conformally dual but not Seiberg dual (as one could expect, since usually the orientifold does not ``commute'' with Seiberg duality.)\footnote{This happens when one performs Seiberg duality on a gauge factor ``close'' to the fixed locus. On the brane tiling, Seiberg duality corresponds to moving NS5-branes across each other. This operation does not commute with the orientifold projection when one needs to cross the orientifold plane. On the other hand, before and after the projection the duality is preserved when the NS5-branes do not cross the orientifold plane and the same operation is performed on the NS5's on the other side, preserving the $\mathbb{Z}_2$ symmetry.} However, in larger dimer structures there may be chains of Seiberg dualities preserved by the orientifold projection, as we show in App. \ref{sec:apx}. Observe that one can equivalently deconfine the tensor $\widetilde{T}_{77}$ and dualize node 7, obtaining the same result. 

If we compare the orientifold projection with fixed points on the model $(2,2,5)$ to the fixed-line ones, we see from Eqs. \eqref{eq:eq:W225FixedPoints1}-\eqref{eq:W225FixedLines1} that we need to act upon the terms with the tensors and the same pair of terms of the previous case, hence the discussion holds in the same way. Furthermore, Eqs. \eqref{eq:eq:W225FixedPoints1}-\eqref{eq:W225FixedLines2} differ only in the terms where tensors appear, and we are back to the case studied in Sec. \ref{sec:ToyModelReal}.

These three theories obtained by projecting the $(2,2,5)$ models can also be related to the fixed-point projection of the $(0,4,5)$ case
corresponding to the superpotential 
\eqref{eq:eq:W045FixedPoints1}. The proof is straightforward because it uses the same dualities and field redefinitions discussed already in the $(2,2,5)$ case.

The models discussed in this subsection summarize the core of the mechanism of conformal duality as it is introduced in Sec. \ref{sec:toy}, i.e. no field redefinition reabsorbs the relative sign between parts of a pair of superpotential and transform one into the other. It is the flip move, introduced in Sec. \ref{sec:embedding} from one dimer to the other that generates the sign flips. Some of them can be fixed by the redefinition of an horizontal edge ($X_{96}$ in \eqref{eq:W225FixedLines1}-\eqref{eq:W225FixedLines2}), but this is not enough to fix the whole superpotential. The reason is that the orientifold theory inherits the structure of the interaction from the parent theory and the pattern of the signs is strongly constrained by the toric condition.

\subsection{$(2,2,6)$ vs. $(0,4,6)$}
\label{sec:(226)vs(046)}

\begin{figure}[ht!]
\begin{center}
\includegraphics[scale=0.7]{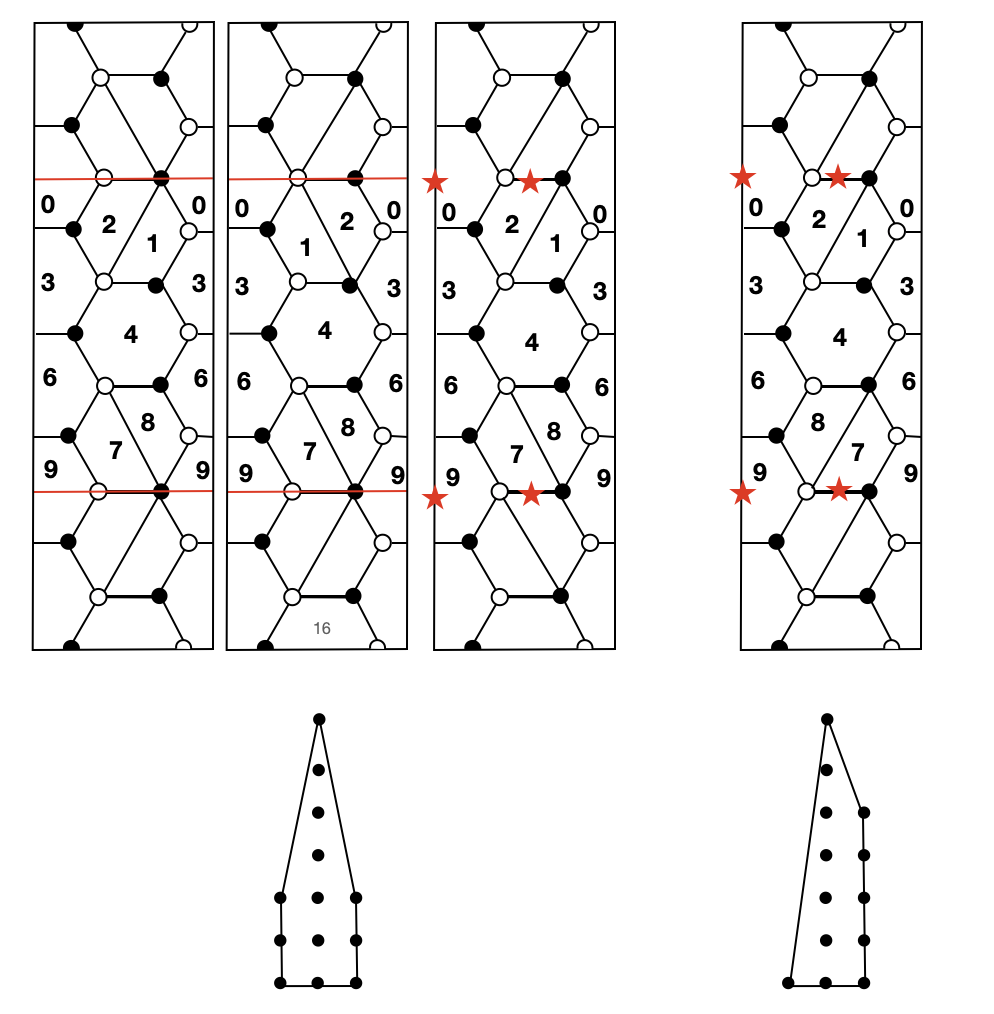}
\caption{The dimers representing the orientifold projection of toric diagrams $(2,2,6)$, on the left, and $(0,4,6)$ on the right. For $(2,2,6)$, there are both fixed-line and fixed-point projections.}
\label{FIG337157}
\end{center}
\end{figure}

We conclude this section with a second example. The two models before the projections correspond to the toric diagrams identified by $(k_1,k_2,k_3) =(2,2,6)$ and $(k_1,k_2,k_3) =(0,4,6)$.
We refer to the projections considered here as ``first case'', because there is a second possibility discussed in the App. \ref{sec:apx}. The models and the projections are 
summarized in Fig. \ref{FIG337157}.

\begin{figure}
    \centering{
    \begin{tikzpicture}[auto]
        \node [diamond, draw=blue!50, fill=blue!20, inner sep=0pt, minimum size=6mm] (0) at (0,0) {$0$};
	\node [circle, draw=blue!50, fill=blue!20, inner sep=0pt, minimum size=5mm] (1) at (1.25,-1) {$1$};
	\node [circle, draw=blue!50, fill=blue!20, inner sep=0pt, minimum size=5mm] (2) at (1.25,1) {$2$};
	\node [circle, draw=blue!50, fill=blue!20, inner sep=0pt, minimum size=5mm] (3) at (2.5,0) {$3$};
	\node [circle, draw=blue!50, fill=blue!20, inner sep=0pt, minimum size=6mm] (4) at (3.75,-1) {$4$};
        \node [circle, draw=blue!50, fill=blue!20, inner sep=0pt, minimum size=6mm] (6) at (5,0) {$6$};
        \node [circle, draw=blue!50, fill=blue!20, inner sep=0pt, minimum size=5mm] (7) at (6.25,1) {$7$};
        \node [circle, draw=blue!50, fill=blue!20, inner sep=0pt, minimum size=5mm] (8) at (6.25,-1) {$8$};
        \node [diamond, draw=blue!50, fill=blue!20, inner sep=0pt, minimum size=6mm] (9) at (7.5,0) {$9$};
        \draw (0) to (1) [->, thick];
        \draw (0) to (2) [->, thick];
        \draw (1) to (3) [->, thick];
        \draw (2) to (3) [->, thick];
        \draw (1) to (2) [->, thick];
        \draw (3) to (0) [->, thick];
        \draw (3) to (4) [->>, thick];
        \draw (4) to (6) [->>, thick];
        \draw (6) to (3) [->, thick];
        \draw (6) to (7) [->, thick];
        \draw (6) to (8) [->, thick];
        \draw (7) to (9) [->, thick];
        \draw (8) to (9) [->, thick];
        \draw (7) to (8) [->, thick];
        \draw (9) to (6) [->, thick];
        \draw (8) to (4) [->, thick];
        \draw (4) to (1) [->, thick];
        \draw (2) to [out=140, in=40, looseness=8] (2) [-, thick, red];
        \draw (7) to [out=140, in=40, looseness=8] (7) [-, thick, green];
    \end{tikzpicture}}
    \caption{The quiver for the theories $(2,2,6)$ and $(0,4,6)$, after the orientifold projection.}
    \label{fig:Quiver226}
\end{figure}
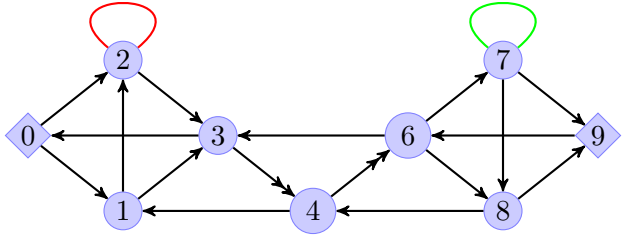

The projections give rise to different superpotentials. For the first fixed-line projection of the $(2,2,6)$ model we have
\begin{align}
\label{W226L1}
W_{(2,2,6)}^{\Omega_\text{f.l.}^1} &= \frac{1}{2} T_{22} \left( X_{02}^2 - X_{01}^2X_{12}^2 \right) + X_{01}X_{13}X_{30} -X_{02}X_{23}X_{30} \nonumber \\[5pt] 
&+ X_{23}Y_{34}X_{41}X_{12} - X_{13}X_{34}X_{41} + X_{34}X_{46}X_{63} - Y_{34}Y_{46}X_{63} \nonumber \\[5pt]
&+ Y_{46}X_{67}X_{78}X_{84} - X_{46}X_{68}X_{84} + X_{68}X_{89}X_{96} - X_{67}X_{79}X_{96} \nonumber \\[5pt]
&+ \frac{1}{2} \widetilde{T}_{77} \left( X_{79}^2 - X_{78}^2X_{89}^2 \right) \; ,
\end{align}
while the second fixed-line projection of the $(2,2,6)$ model we have
\begin{align}
\label{W226L2}
W_{(2,2,6)}^{\Omega_\text{f.l.}^2} &= \frac{1}{2} T_{22} \left( X_{01}^2X_{12}^2 - X_{02}^2 \right) + X_{02}X_{23}X_{30} -X_{01}X_{13}X_{30} \nonumber \\[5pt] 
&+ X_{13}Y_{34}X_{41} - X_{23}X_{34}X_{41}X_{12} + X_{34}X_{46}X_{63} - Y_{34}Y_{46}X_{63} \nonumber \\[5pt]
&+ Y_{46}X_{67}X_{78}X_{84} - X_{46}X_{68}X_{84} + X_{68}X_{89}X_{96} - X_{67}X_{79}X_{96} \nonumber \\[5pt]
&+ \frac{1}{2} \widetilde{T}_{77} \left( X_{79}^2 - X_{78}^2X_{89}^2 \right) \; .
\end{align}
On the other hand the fixed-point projection of the $(2,2,6)$ model gives
\begin{align}
\label{W226P}
W_{(2,2,6)}^{\Omega_\text{f.p.}} &= T_{22}X_{02}X_{01}X_{12} + X_{01}X_{13}X_{30} -X_{02}X_{23}X_{30} \nonumber \\[5pt] 
&+ X_{23}Y_{34}X_{41}X_{12} - X_{13}X_{34}X_{41} + X_{34}X_{46}X_{63} - Y_{34}Y_{46}X_{63} \nonumber \\[5pt]
&+ Y_{46}X_{67}X_{78}X_{84} - X_{46}X_{68}X_{84} + X_{68}X_{89}X_{96} - X_{67}X_{79}X_{96} \nonumber \\[5pt]
&+ \widetilde{T}_{77}X_{79}X_{78}X_{89} \; .
\end{align}
Lastly, the fixed-point orientifold projection for the $(0,4,6)$ model yields
\begin{align}
\label{W046P}
W_{(0,4,6)}^{\mathcal{O}_{P}} &= T_{22}X_{01}X_{12}X_{02} + X_{01}X_{13}X_{30} -X_{02}X_{23}X_{30} \nonumber \\[5pt] 
&+ X_{23}Y_{34}X_{41}X_{12} - X_{13}X_{34}X_{41} + X_{34}X_{46}X_{63} - Y_{34}Y_{46}X_{63} \nonumber \\[5pt]
&+ Y_{46}X_{68}X_{84} - X_{46}X_{67}X_{78}X_{84} + X_{67}X_{79}X_{96} - X_{68}X_{89}X_{96} \nonumber \\[5pt]
&+ \widetilde{T}_{77}X_{79}X_{78}X_{89} \; .
\end{align}
The ranks of the gauge groups consistent with the anomaly cancellations and the existence
of a conformal fixed point are
\begin{eqnarray}
\label{eq:ranks226046}
&&
N_0 = n,
\,
N_1 = N_2 = n - 2 \tau,
\,
N_3 =  n - 4 \tau \nonumber \\
&&
N_4 = n - 6 \tau, 
\,
N_6 = n - 8 \tau,
\,
N_7 = N_8 = n - 10 \tau,
\,
N_9 =  n - 12 \tau \; . 
\end{eqnarray}
Again the two possible choices of sign $\tau=\pm 1$ give two different quivers, that in this case are inequivalent and need to be studied separately. The two different projections indeed give either $SO(N_0)$ and $USp(N_9)$ or $USp(N_0)$ and $SO(N_9)$. Furthermore the tensors $T_{22}$ and $\widetilde T_{77}$ are either symmetric and conjugate anti-symmetric or conjugate symmetric and anti-symmetric, respectively. We have checked that both choices of $\tau$ give a consistent SCFT by maximizing the $a$ central charge with the ranks in formula \eqref{eq:ranks226046}. Furthermore we have computed the various 't Hooft anomalies and we have seen that they coincide among the models.

Similarly to the previous example we can study the conformal duality between these four models by applying the discussion in Sec. \ref{sec:toy}. 
Again the first three cases are toric dual before the projections and even if such a duality is broken by the projections the
models remain conformally dual.

Let us discuss the map between each pair of superpotentials explicitly. The superpotentials of the two fixed-line projections, given in formulae \eqref{W226L1} and \eqref{W226L2}, differ by a sign\footnote{Also by a sign in the first line, but this can be reabsorbed redefining $X_{30}$ and $T_{22}$, similarly as in the previous section.} by considering the combinations $N_{34}^{(\pm)} = X_{34} \pm Y_{34}$. The superpotentials of the two fixed-point projections, given in formula \eqref{W226P} and \eqref{W046P}, differ by a sign by considering the combinations $N_{46}^{(\pm)} = X_{46}\pm Y_{46}$. The map among one of the superpotentials obtained by a fixed-line projection and one obtained from a fixed-point projection requires also to apply one Seiberg duality on node $SU(N_1)$ and one on node $SU(N_8)$.

The transformations of the two superpotentials obtained by a fixed-line projection after this duality corresponds to
\begin{eqnarray}
\frac{1}{2} T_{22} (X_{02}^2 - X_{01}^2X_{12}^2)
    &\rightarrow&
  \frac{1}{2} T_{22} (X_{02}^2 - M_{02}^2) \; ,
\nonumber \\ 
\frac{1}{2} \widetilde{T}_{55} (X_{79}^2 - X_{78}^2X_{89}^2)
     &\rightarrow& 
   \frac{1}{2} \widetilde{T}_{55} (X_{79}^2 - M_{79}^2) \; ,
\end{eqnarray}
while those of the two superpotentials obtained by a fixed-point projection after this duality corresponds to
\begin{eqnarray}
T_{22} X_{01} X_{12} X_{02}
    &\rightarrow&
T_{22} M_{02} X_{02} \; ,
\nonumber \\ 
\widetilde{T}_{77} X_{79} X_{78} X_{89}
      &\rightarrow&
\widetilde{T}_{77} X_{79} M_{79} \;.
\end{eqnarray}

After applying these dualities we consider the combinations of fields $X_{02} \pm M_{02}$ and $X_{79} \pm M_{79}$. We can see explicitly that combining these operations and the field redefinitions $X_{46}\pm Y_{46}$ and $X_{34}\pm Y_{34}$ when necessary, each superpotential obtained by a fixed-line projection differs only in sign choices from the superpotentials obtained by a fixed-point projection.

This concludes the analysis for this case, showing indeed that the conformal duality can be reformulated along the lines of the discussion in Sec. \ref{sec:toy}.

We have studied many other examples, and they all behave as the two examples studied here. For completeness we briefly discuss these models in App. \ref{sec:apx}.

\section{Conclusions}
\label{sec:conc}

In this paper we have generalized the result obtained in \cite{Antinucci:2020yki}, where it was shown  that
two  quiver gauge theories describing stacks of D3-branes probing different toric CY$_3$  singularities
reside on the same conformal manifold  once suitable orientifold projections  are considered.
The two singularities studied in  \cite{Antinucci:2020yki} are the PdP$_{3b}$ and PdP$_{3c}$ surfaces
and the orientifolds are implemented by a fixed-line and a fixed-point projection on the dimer, respectively.

The key point allowing for a natural generalization of this conformal duality is that the two models share the same quiver even
before considering the orientifold.
These quivers have different superpotentials however, and this gives rise to different dimers and toric diagrams.
Nonetheless there exists a ``flip'' on the  dimer (see Fig. \ref{FIG2}) that can be used to transform one toric diagram into the other, by reversing the orientation of one zig-zag path in a consistent way.
We have found infinite families of dimers associated to different toric diagrams but sharing the same quiver 
by the application of this flip. It is enough to require that the flip does not spoil the quiver structure and that the zig-zag paths intersect consistently. 
We have referred to quivers with this property as multi-planarizable, i.e. the same quiver admits different 
planar periodic quivers associated, in general, to different CY singularities.
We have then studied the orientifold projections in terms of fixed-line and fixed-point identifications on the dimer.
By analyzing a large set of models we have found infinite classes of models that behave as the ancestral
case of PdP$_{3b}$ and PdP$_{3c}$.
These models are specified by the reflection symmetry property of the toric diagram, in the phase projected by fixed lines.

We have further studied the conformal dualities among the models obtained after the projection by iterative applications of ordinary Seiberg dualities. Our strategy consists in considering two conformally dual phases and applying the same dualities on both. This procedure preserves the fact that the two models have the same quivers
and that all the fields have the same global charges.
We have found that a judicious application of Seiberg dualities transforms the superpotential interactions of the two phases in a remarkable way. There always exists a phase where two conformally dual models  have the same quiver and the same superpotential interaction. The only difference regards the sign of the coupling: some of these may indeed have a different sign. This sign difference cannot be reabsorbed in a phase for the fields and thus represents a genuine exactly marginal deformation connecting the two models.

We left open many questions and possible directions of future investigation.
Here we have found infinite families of quivers that admit different periodic planar descriptions.
It is natural to wonder if these families exhaust (up to Seiberg or toric dualities) the possible quivers with this peculiar property.
If other such models exist then one should check if there are orientifold projections giving rise
to conformally dual models for suitable and consistent choices of gauge ranks. One may also add flavor branes to the configuration along the lines of \cite{Bianchi:2013gka}, which will provide an even richer structure on the gauge theory and more freedom for the choice of the gauge ranks, but at the same time one would lose the description of the orientifold projection from the dimer construction.

One can also wonder whether there are quivers that coincide only after the orientifold projection. If true, it should be possible to find other conformal dualities for these cases as well.

In general it is worth mentioning that all the conformally dual models found here and in \cite{Antinucci:2020yki,Antinucci:2021edv,Amariti:2021lhk,Amariti:2022dyi} 
share a similar property when looking at their toric diagrams, i.e. they have the same number of internal and external points.
This signals that the gauge theories have the same number of gauge groups and of non-anomalous global symmetries.
The role of the orientifold is to further break the global symmetry, and to identify the 't Hooft 
anomalies of the surviving ones.
This idea can be used to generalize the construction to more general cases.
One may in principle find examples of conformal dualities with different lagrangians and quivers by studying
orientifolds of models with the same number of gauge groups and global symmetries.
The final goal of this program consists in finding a predictive stringy recipe for obtaining conformal dualities in presence of orientifolds.
One may think of such a top-down approach as complementary to the bottom-up one used in \cite{Razamat:2019vfd} to find conformal dualities.
A key role in their setup seems related to the presence of orientifolds, since most of the examples feature real gauge groups and tensor matter fields.

One last natural question regards the holographic interpretation of the dualities found here.\footnote{See also the recent reference \cite{Manzoni:2022htx} for related comments.} In particular, it is necessary to understand the role played by exactly marginal deformations in the gravity dual. 
As we stressed in the discussion the marginal deformation connecting the models obtained here is related to the $\beta$-deformation of SYM, because 
it can always be reformulated
as a sign flip of some superpotential terms. 
 Observe that, despite the similarities among  models related by such a sign flip, there are non trivial differences.
 For example they differ in the spectrum of chiral primaries operators \cite{Berenstein:2000hy,Berenstein:2000ux}. 
See also \cite{Rossi:2005mr,Rossi:2006mu} for an analysis of the chiral matter superfield propagator in the $\beta$-deformed case.
The holographic dual mechanism for the case of the $\beta$-deformed $\mathcal{N}=4$ SYM was originally interpreted in \cite{Lunin:2005jy} as a  TsT transformation on the AdS side. Here we are considering more complicated cases, because the sign change involves only a subset of couplings
and because we are in presence of orientifolds. A general discussion on the gravitational origin of marginal deformations for toric quiver gauge theories appeared in \cite{Imamura:2007dc}. In absence of orientifolds, the intimate relations between the marginal deformations and the zig-zag paths recently obtained in \cite{Ashmore:2021mao,Tasker:2021lek} may provide a useful guideline to address the problem from the gravity side. Generalizing such constructions in presence of orientifolds represents a first step to investigate in this direction.

Finally it would be desirable  to find the origin of the conformal dualities in terms of string dualities, starting from two inequivalent CY$_3$ singularities and considering orientifolds of those which are known to give rise to conformally dual gauge theories.

\section*{Acknowledgments}

We would like to thank Riccardo Argurio for interesting discussions and comments. MF would like to thank SISSA, Trieste for hospitality during the initial stages of this work. MF and SM would like to thank the University of Milano for the warm hospitality while working on this manuscript. MF gratefully acknowledges support from the Simons Center for Geometry and Physics (workshops ``5d $\mathcal{N}=1$ SCFTs and Gauge Theories on Brane Webs'' and ``Supersymmetric Black Holes, Holography and Microstate Counting''), Stony Brook University at which some of the research for this paper was performed. The work of AA and SR is supported in part by MIUR-PRIN contract 2017CC72MK-003. The work of MB, SM and FR is partially supported by the MIUR PRIN Grant 2020KR4KN2 ``String Theory as a bridge between Gauge Theories and Quantum Gravity''. The work of MF is supported in part by the Knut and Alice Wallenberg Foundation under grant KAW 2021.0170, the VR grant 2018-04438, the Olle Engkvists Stiftelse grant No. 2180108, and in part by the European Union's Horizon 2020 research and innovation programme under the Marie Skłodowska-Curie grant agreement No. 754496 - FELLINI. 

\appendix
\section{Further examples}\label{sec:apx}

In table \ref{tab:cases} is the list of cases that we have analyzed in detail. 
\begin{table}[ht!]
\centering
\begin{tabular}{lcc}
$\!\!n_G \rightarrow n_G^{\Omega}$&$(k_1,k_2,k_3)$&Comments  \\\hline
 $6\rightarrow 4 $ &$(1,1,2)$ - $(0,2,2)$ & Original PdP$_{3b/c}$ case, see \cite{Antinucci:2020yki}\\
 $8\rightarrow 5 $& $(1,1,3)$ - $(0,2,3)$ & See App. \ref{sec224}
 \\
 $10\rightarrow 6 $& $(1,1,4)$ - $(0,2,4)$ & Two cases, see App. \ref{sec225}\\
 $12\rightarrow  7$& $(1,1,5)$ - $(0,2,5)$ & See App. \ref{(2,2,6)(1,3,6)}\\
\hline 
 $12\rightarrow  7$& $(2,2,4)$ - $(0,4,4)$ & Also $\Omega_\text{f.p.}$ for (2,2,4), see App. \ref{sec155}\\
 $14\rightarrow 8 $& $(2,2,5)$ - $(0,4,5)$ & See Sec. \ref{sec:(225)(045)}\\
  $16\rightarrow 9  $& $(2,2,6)$ - $(0,4,6)$ & Two cases, Sec. \ref{sec:(226)vs(046)} and App. \ref{(337)(157)bis} \\
\hline
$ 18\rightarrow 10 $& $(3,3,6)$ - $(2,4,6)$ - $(0,6,6)$ &  Intermediate case, see App. \ref{177}\\
\end{tabular}
\caption{List of cases (ordered by increasing number $n_G^\Omega$ of gauge groups after the projection) analyzed in detail in the paper.}
\label{tab:cases}
\end{table}
Two examples have been studied in the core of the paper. Here for completeness we briefly survey the other cases. We 
will give the brane tilings, the orientifold  projections, specifying the choices of fixed lines or fixed points. 

For each case we will also provide the choice of ranks (consistent with the general one discussed in Fig. \ref{rankass}) that allows to find the conformally dual models together with the projected  superpotentials. 

\subsection{$(1,1,3)$ - $(0,2,3)$}
\label{sec224}

\begin{figure}
\begin{center}
\includegraphics[scale=0.6]{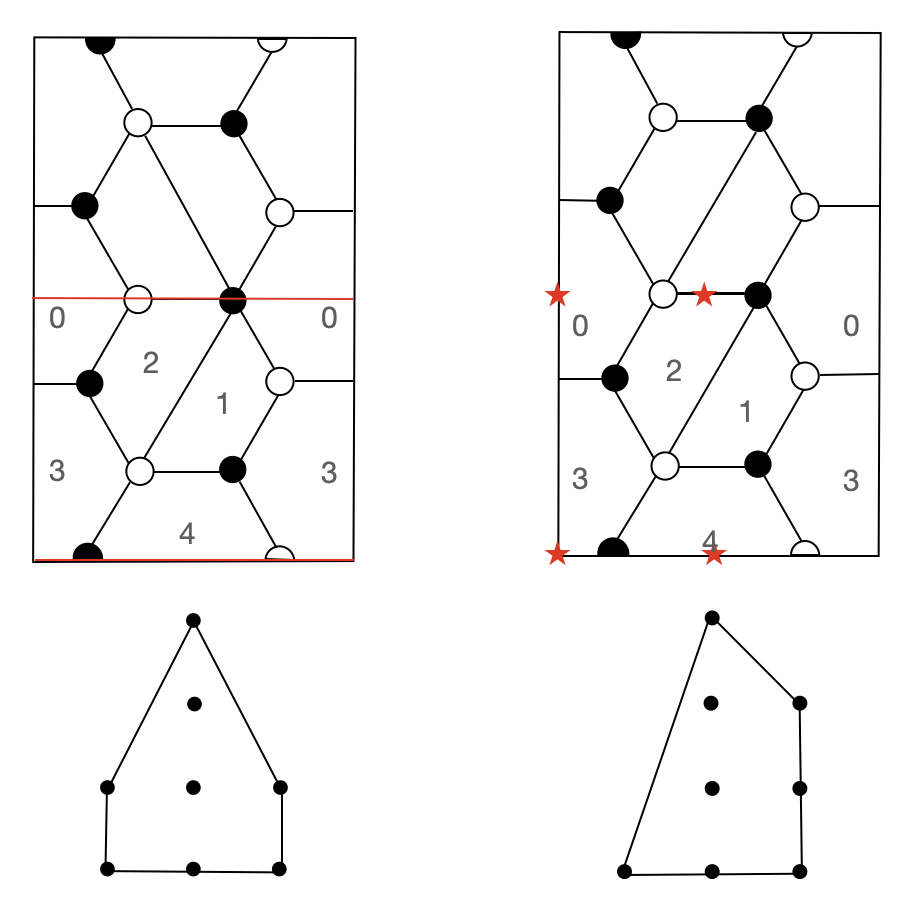}
\caption{Fixed-line and fixed-point projections for the $(1,1,3)$ and the $(0,2,3)$ models.}
\label{FIG224133}
\end{center}
\end{figure}

The dimers and the toric diagrams are given in Fig. \ref{FIG224133}.
The $(k_1, k_2, k_3)= (1,1,3)$ model is projected with fixed lines while the $(k_1, k_2, k_3)= (0,2,3)$ model is projected with fixed points.
We assign the charges  to the fixed points and lines as discussed in the body of the paper.
Again we do not distinguish between $\tau_1=1$ and $\tau_1=-1$ and consider the two possibilities in a uniform language. 
The superpotential for the $(1,1,3)$ model after the fixed lines projection is
\begin{eqnarray}
\label{W113}
W_{(1,1,3)}^{\Omega_\text{f.l.}} &=&
\frac{1}{2}  T_{22} (X_{02}^2 - X_{01}^2 X_{12}^2)
 +  X_{30} X_{01} X_{13} - X_{30} X_{02}  X_{23}
\nonumber \\
&+&
 X_{41}X_{12}X_{23}Y_{34} - X_{41}X_{13}X_{34}
+ \frac{1}{2} \widetilde T_{33} \left(X_{34}^2 - Y_{34}^2 \right) \; .
\end{eqnarray}
The superpotential for the $(0,2,3)$ model after the fixed points projection is
\begin{eqnarray}
\label{W023}
W_{(0,2,3)}^{\Omega_\text{f.p.}} &=&
  T_{22} X_{01} X_{12} X_{02} 
  +  X_{30} X_{01} X_{13} - X_{30} X_{02}  X_{23}
\nonumber \\
&+&
 X_{41}X_{12}X_{23}Y_{34} - X_{41}X_{13}X_{34}
+  \widetilde T_{33} X_{34}Y_{34} \; ,
\end{eqnarray}
where, as in the body of the paper, the gauge contractions are taken opportunely. The choice of gauge ranks that gives rise to the conformal duality is
\begin{equation}
N_0=n, \, 
N_1 = N_2 = n - 2 \tau, \, 
N_3 = n - 4 \tau, \,
N_4 = n - 6 \tau \; .
\end{equation}

The conformal duality can be associated to a pair of superpotentials with different signs by first applying Seiberg duality on node $SU(N_1)$. Then we can consider the combinations of the fields $X_{34}$ with $Y_{34}$ and $X_{02}$ with $M_{02}$, where $M_{02}$ in the dual phase is the meson $X_{01} X_{12}$, that is considered as elementary field after the duality on $SU(N_1)$. After these operations, the two superpotentials \eqref{W113} and \eqref{W023} become identical up to some signs. The conformal duality between the two models becomes then manifest and explicit, fitting with the discussion of Sec. \ref{sec:toy}.

\subsection{$(1,1,4)$ - $(0,2,4)$}
\label{sec225}
The next case corresponds to the fixed line projection of $(k_1, k_2, k_3)= (1,1,4)$
and to the fixed point projection of 
$(k_1, k_2, k_3)= (0,2,4)$.
In this case there are two possible projections for each models. These projections give different quivers and they are studied separately.

\subsubsection*{First case}
\begin{figure}[ht!]
\begin{center}
\includegraphics[scale=0.6]{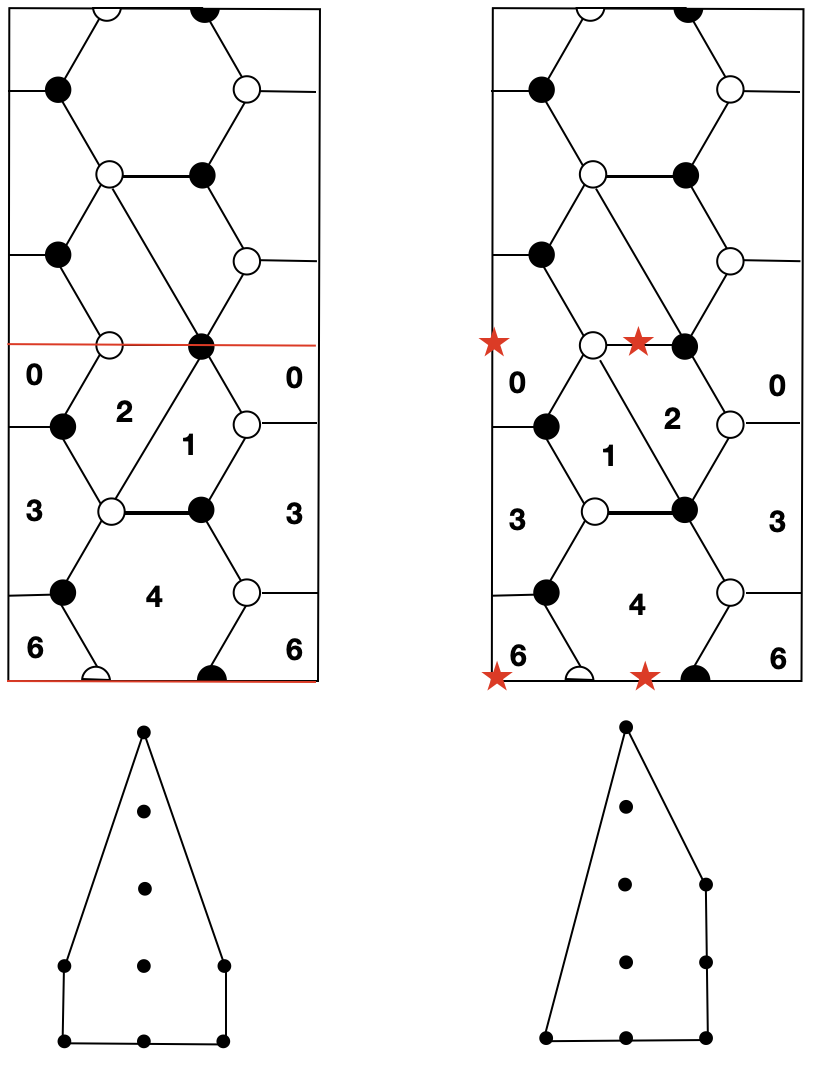}
\caption{Fixed lines and fixed points projections for $(1,1,4)$ and  $(0,2,4)$. This is the first conformal duality obtained projecting these models.}
\label{FIG225135[1]}
\end{center}
\end{figure}

The first possibility is represented in Fig. \ref{FIG225135[1]}.
The superpotential for the $(1,1,4)$ model after the fixed-line projection is
\begin{eqnarray}
\label{W114}
W_{(1,1,4)}^{\Omega_\text{f.l.}} &=&
\frac{1}{2}
T_{22} \left(X_{02}^2-X_{01}^2 X_{12}^2\right)
  +  X_{30} X_{01} X_{13} - X_{30} X_{02}  X_{23}
\nonumber \\[5pt]
&+&
 X_{41}X_{12}X_{23}Y_{34} - X_{41}X_{13}X_{34}
+ X_{63}X_{34}X_{46} - X_{63} Y_{34} Y_{46} 
\nonumber \\[5pt]
&+&
\frac{1}{2}
\widetilde T_{44} \left(X_{46}^2-Y_{46}^2 \right) \; .
\end{eqnarray}
The superpotential for the $(0,2,4)$ model after the fixed points projection is
\begin{eqnarray}
\label{W024}
W_{(0,2,4)}^{\Omega_\text{f.p.}} &=&
\frac{1}{2}
T_{22} X_{01}X_{12}X_{02}
  +  X_{30} X_{02} X_{23} - X_{30} X_{01}  X_{13}
\nonumber \\[5pt]
&+&
 X_{41}X_{13}Y_{34} - X_{41}X_{12}X_{23}X_{34}
+ X_{63}X_{34}X_{46} - X_{63} Y_{34} Y_{46} 
\nonumber \\[5pt]
&+&
\frac{1}{2}
\widetilde T_{44} X_{46}Y_{46} \; .
\end{eqnarray}
The choice of gauge ranks that gives rise to the conformal duality is
\begin{eqnarray}
N_0 = n, \,
N_1 = N_2 = n - 2 \tau, \,
N_3 = n - 4 \tau, \,
N_4 = n - 6 \tau, \,
N_6 = n - 8 \tau \; .
\end{eqnarray}
We can use the usual tools, i.e. Seiberg duality on  $SU(N_1)$ and suitable field redefinitions, such that \eqref{W114} and \eqref{W024} become identical up to some signs. The conformal duality between the two models becomes again explicit as in the general discussion of Sec. \ref{sec:toy}.

\subsubsection*{Second case}
\begin{figure}[ht!]
\begin{center}
\includegraphics[scale=0.6]{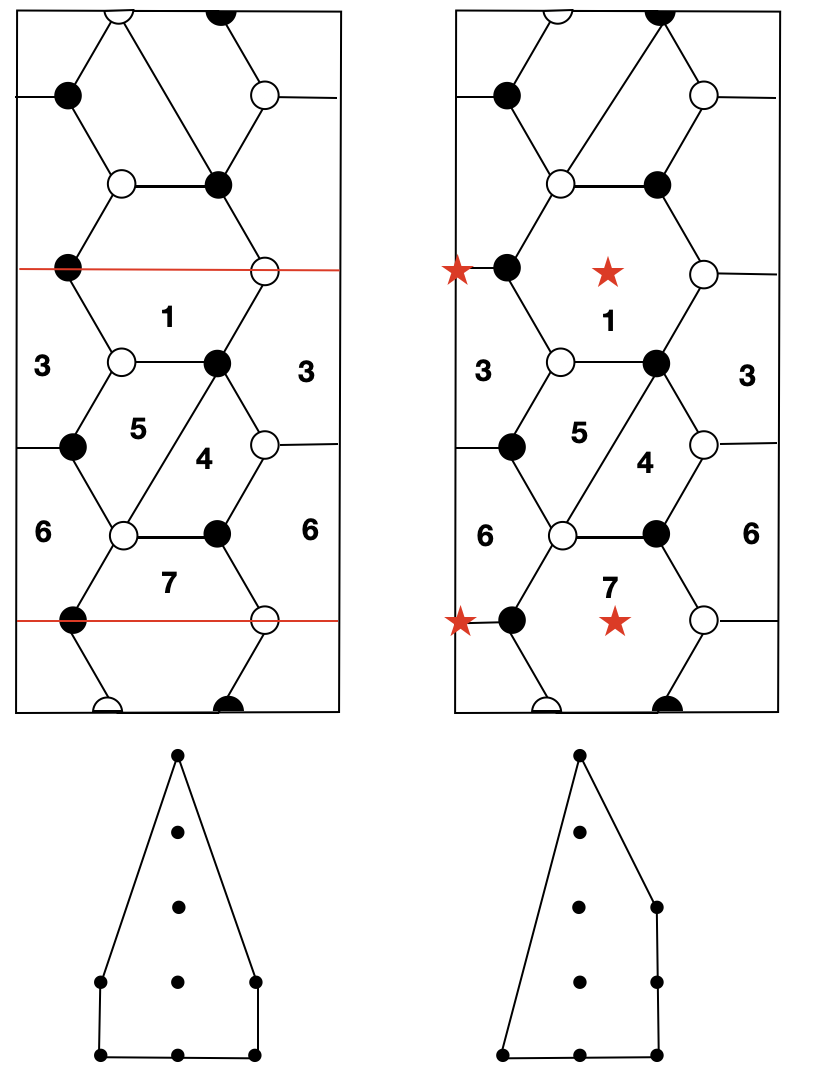}
\caption{Fixed-line and fixed-point projections for $(1,1,4)$ and  $(0,2,4)$. This is the second conformal duality obtained projecting these models.}
\label{FIG225135[2]}
\end{center}
\end{figure}

The first possibility is represented in Fig. \ref{FIG225135[2]}.

The superpotential for the $(1,1,4)$ model after the fixed lines projection is
\begin{align}\label{W114II}
     W_{(1,1,4)}^{\Omega_\text{f.l.}} &= \frac{1}{2}T_{33} \left( X_{13}^2 - Y_{13}^2 \right) + Y_{13}X_{35}X_{51} - X_{13}X_{34}X_{45}X_{51} \nonumber \\[5pt]
     &+ X_{34}X_{46}X_{63} - X_{35}X_{56}X_{63} + X_{56}Y_{67}X_{74}X_{45} - X_{46}X_{67}X_{74} \nonumber \\[5pt]
     &+ \frac{1}{2}T_{33} \left( X_{67}^2 - Y_{67}^2 \right) \; .
\end{align}
The superpotential for the $(0,2,4)$ model after the fixed points projection is
\begin{align}\label{W024II}
     W_{(1,1,4)}^{\Omega_\text{f.l.}} &= T_{33}X_{13}Y_{13} + Y_{13}X_{35}X_{51} - X_{13}X_{34}X_{45}X_{51} \nonumber \\[5pt]
     &+ X_{34}X_{46}X_{63} - X_{35}X_{56}X_{63} + X_{56}Y_{67}X_{74}X_{45} - X_{46}X_{67}X_{74} \nonumber \\[5pt]
     &+ T_{33}X_{67}Y_{67} \; .
\end{align}
The choice of gauge ranks that gives rise to the conformal duality is
\begin{eqnarray}
N_1 = n, \,
N_3 = n - 2 \tau, \,
N_4 = N_5 = n - 4 \tau, \,
N_6 = n - 6 \tau, \,
N_7 = n - 8 \tau \; .
\end{eqnarray}
In this case, it can be shown that the two superpotentials in Eqs. \eqref{W114II}-\eqref{W024II} to be identical without any further duality. Indeed by substituting the combinations $X_{13}\pm Y_{13}$ and $X_{67}\pm Y_{67}$, the superpotentials \eqref{W114II} and \eqref{W024II} become identical up to some signs, making the conformal duality explicit.

\subsection{$(1,1,5)$ - $(0,2,5)$}
\label{(2,2,6)(1,3,6)}

\begin{figure}[ht!]
\begin{center}
\includegraphics[scale=0.6]{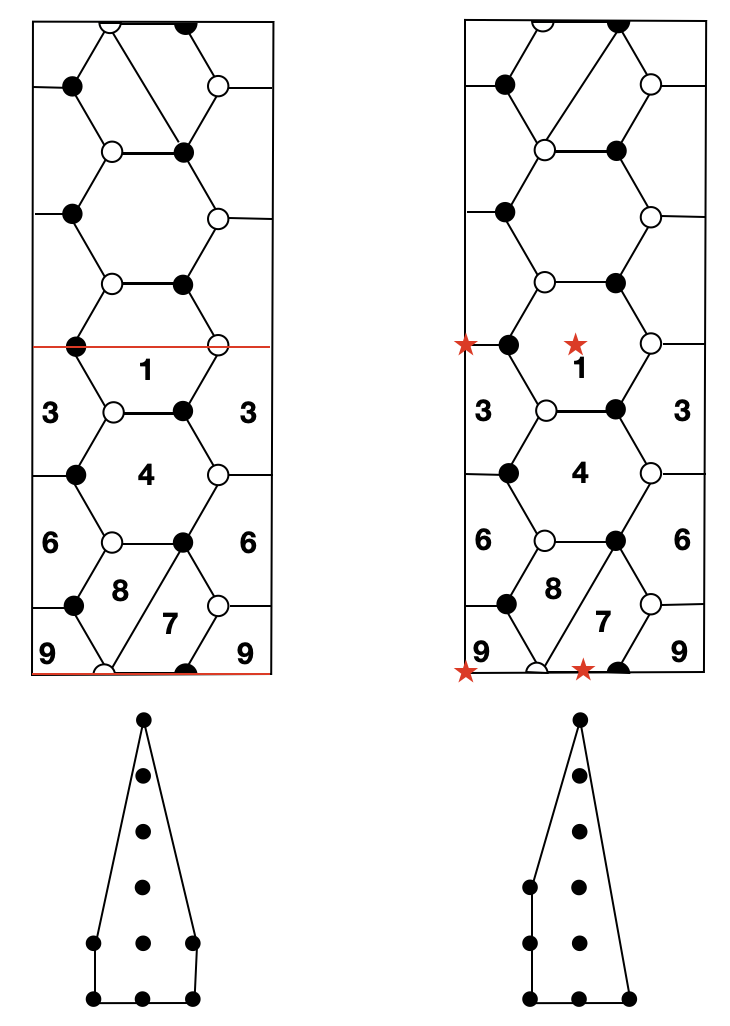}
\caption{Fixed lines and fixed points projections for $(1,1,5)$ and  $(0,2,5)$.}
\label{FIG226136}
\end{center}
\end{figure}

The next case corresponds to the fixed line projection of $(k_1, k_2,  k_3)= (1,1,5)$ and to the fixed-point projection of $(k_1, k_2, k_3)= (0,2,5)$. The dimers and the toric diagrams are given in Fig. \ref{FIG226136}. The are also other models where the diagonal appears in the first hexagon on the tiling, but conformal duality works similarly and we show only these two dimers for simplicity.
The superpotential for the $(1,1,5)$ model after the fixed-line projection is
\begin{align}\label{eq:W115}
    W_{(1,1,5)}^{\Omega_\text{f.l.}} &= \frac{1}{2} T_{33} \left( X_{13}^2 - Y_{13}^2 \right) + Y_{13}Y_{34}X_{41} - X_{13}X_{34}X_{41} \nonumber \\[5pt]
    &+ X_{34}X_{46}X_{63} - Y_{34}Y_{46}X_{63} + Y_{46}X_{68}X_{84} - X_{46}X_{67}X_{78}X_{84} \nonumber \\[5pt]
    &+ X_{67}X_{79}X_{96} - X_{68}X_{89}X_{96} + \frac{1}{2} \widetilde{T}_{77} \left( X_{78}^2X_{89}^2 - X_{79}^2 \right) \; .
\end{align}
The superpotential for the $(0,2,5)$ model after the fixed points projection is
\begin{align}\label{eq:W025}
    W_{(0,2,5)}^{\Omega_\text{f.l.}} &= T_{33} X_{13}Y_{13} + Y_{13}Y_{34}X_{41} - X_{13}X_{34}X_{41} \nonumber \\[5pt]
    &+ X_{34}X_{46}X_{63} - Y_{34}Y_{46}X_{63} + Y_{46}X_{68}X_{84} - X_{46}X_{67}X_{78}X_{84} \nonumber \\[5pt]
    &+ X_{67}X_{79}X_{96} - X_{68}X_{89}X_{96} + \widetilde{T}_{77}  X_{78}^2X_{89}X_{79} \; .
\end{align}
The choice of gauge ranks that gives rise to the conformal duality is
\begin{eqnarray}
&&
N_1 = n, \, N_3 = n - 2\tau, \, N_4 = n - 4 \tau, \,
\nonumber \\
&&
N_6 = n - 6 \tau, \, N_7 = N_8 = n - 8 \tau, \, N_9 = n - 10 \tau \; .
\end{eqnarray}
The conformal duality between the two models becomes  explicit as in Sec. \ref{sec:toy}, by applying a Seiberg duality on $SU(N_8)$, and considering the combinations $X_{13}\pm Y_{13}$ and $X_{79} \pm M_{79}$, where $M_{79} \equiv X_{78}X_{89}$.

\subsection{$(2,2,4)$ - $(0,4,4)$}
\label{sec155}

\begin{figure}[ht!]
\begin{center}
\includegraphics[scale=0.9]{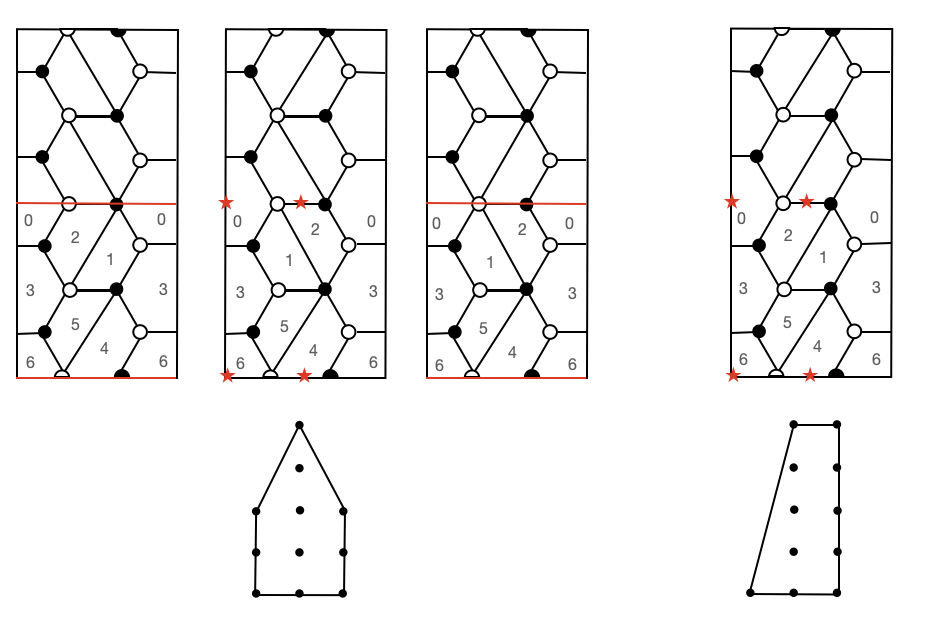}
\caption{Fixed lines and fixed-point projections for $(2,2,4)$ and  (0,4,4).}
\label{FIG335155}
\end{center}
\end{figure}

Here we study two fixed line projections and one fixed point projection for $(k_1, k_2, k_3)= (2,2,4)$ and a fixed point projection of $(k_1,  k_2, k_3)= (0,4,4)$. The projections are summarized in Fig. \ref{FIG335155}.
The superpotential for the $(2,2,4)$ model after the first fixed lines projection is
\begin{align}
    W_{(2,2,4)}^{\Omega_\text{f.l.}^1} &= \frac{1}{2} T_{00}\left(X_{02}^2-X_{01}^2X_{12}^2 \right) + X_{01}X_{13}X_{30} - X_{02} X_{23}X_{30} \nonumber \\[5pt] 
    & + X_{23}X_{35}X_{51}X_{12} - X_{13}X_{34}X_{45}X_{51} + X_{34}X_{46}X_{63} - X_{35}X_{56}X_{63} \nonumber \\[5pt]
    &+ \frac{1}{2} \widetilde{T}_{44}\left(X_{45}^2X_{56}^2-X_{46}^2 \right) \; .
\end{align}
The superpotential for the $(2,2,4)$ model after the fixed-point projection is
\begin{align}
    W_{(2,2,4)}^{\Omega_\text{f.p.}} &= T_{00}X_{02}X_{01}X_{12} + X_{02}X_{23}X_{30} - X_{01} X_{13}X_{30} \nonumber \\[5pt] 
    & + X_{13}X_{35}X_{51} - X_{23}X_{34}X_{45}X_{51}X_{12} + X_{34}X_{46}X_{63} - X_{35}X_{56}X_{63} \nonumber \\[5pt]
    &+ \widetilde{T}_{44}X_{45}X_{56}X_{46} \; .
\end{align}
The superpotential for the $(2,2,4)$ model after the second fixed-line projection is
\begin{align}
    W_{(2,2,4)}^{\Omega_\text{f.l.}^2} &= \frac{1}{2} T_{00}\left(X_{01}^2X_{12}^2 - X_{02}^2\right) + X_{02}X_{23}X_{30} - X_{01} X_{13}X_{30} \nonumber \\[5pt] 
    & + X_{13}X_{35}X_{51} - X_{23}X_{34}X_{45}X_{51}X_{12} + X_{34}X_{46}X_{63} - X_{35}X_{56}X_{63} \nonumber \\[5pt]
    &+ \frac{1}{2} \widetilde{T}_{44}\left(X_{45}^2X_{56}^2-X_{46}^2 \right) \; .
\end{align}
The superpotential for the $(0,4,4)$ model after the fixed-point projection is
\begin{align}
    W_{(0,4,4)}^{\Omega_\text{f.p.}} &= T_{00}X_{02}X_{01}X_{12} + X_{01}X_{13}X_{30} - X_{02} X_{23}X_{30} \nonumber \\[5pt] 
    & + X_{23}X_{35}X_{51}X_{12} - X_{13}X_{34}X_{45}X_{51} + X_{34}X_{46}X_{63} - X_{35}X_{56}X_{63} \nonumber \\[5pt]
    &+ \widetilde{T}_{44}X_{45}X_{56}X_{46} \; .
\end{align}
The choice of gauge ranks that gives rise to the conformal duality is
\begin{eqnarray}
N_0 = n,\, N_1=N_2 = n - 2 \tau ,\, N_3 = n - 4 \tau ,\, N_4 = N_5 = n - 6 \tau ,\, N_6 = n - 8 \tau
\end{eqnarray}
The situation here is similar to the one in sub-section \ref{sec:(225)(045)}. Indeed there are three cases that are Seiberg dual before the projections, and that become conformally dual after the projection. We can make the conformal dualities among the various model more explicit along the lines of the discussion in Sec. \ref{sec:toy}. By combining the dualities on node $SU(N_{1,2,3})$ and the suitable field redefinitions we can indeed prove that each pair of superpotentials become identical up to signs.

\subsection{$(2,2,6)$ - $(0,4,6)$: second case}
\label{(337)(157)bis}

\begin{figure}
\begin{center}
\includegraphics[scale=0.7]{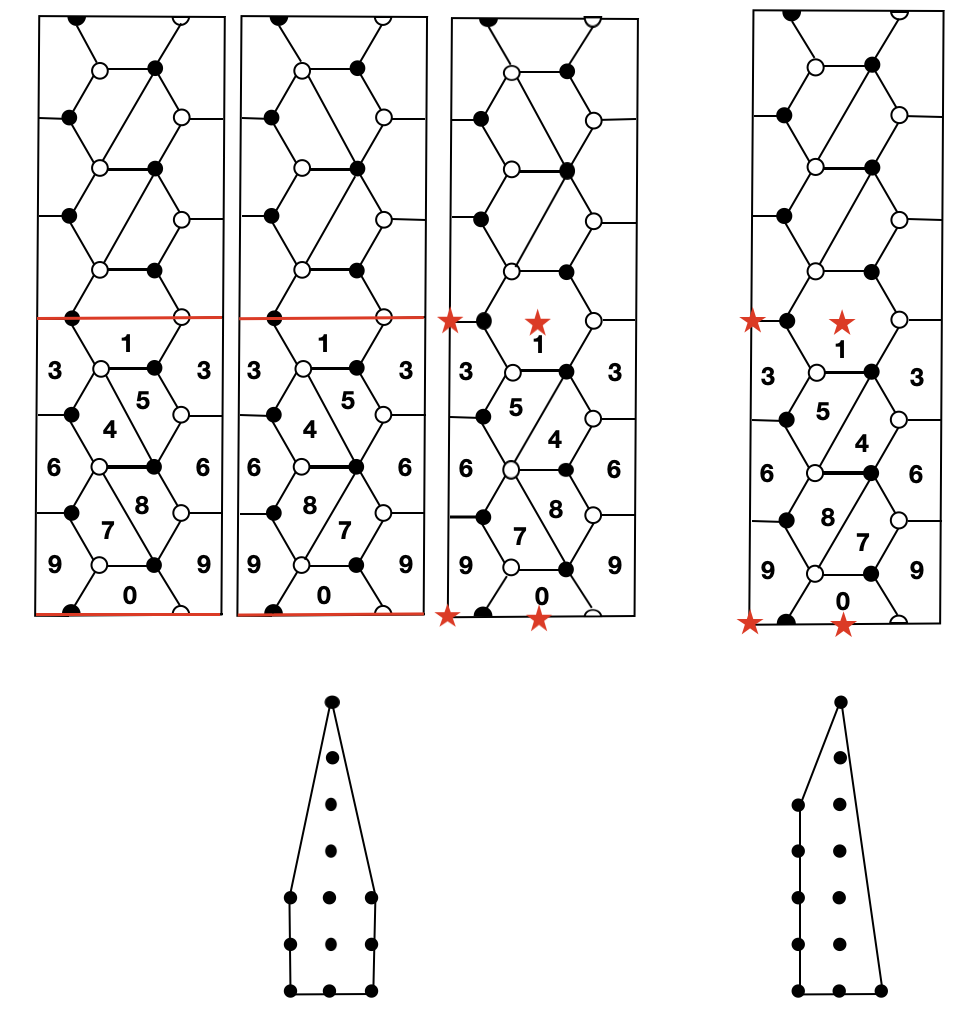}
\caption{Fixed-line and fixed-point projections for $(2,2,6)$ and $(0,4,6)$. This is the second conformal duality obtained projecting these models. The first one has been discussed in the body of the paper in Sec. \ref{sec:(226)vs(046)}.}
\label{FIG337157bis}
\end{center}
\end{figure}

Here we study another case involving the $(2,2,6)$ and $(0,4,6)$ models.  Here we have two fixed line projections and one fixed point projection for $(k_1, k_2, k_3)= (2, 2, 4)$ and a fixed point projection of $(k_1, k_2, k_3)= (0,4,4)$. The projections are summarized in Fig. \ref{FIG337157bis}.
The superpotential for the $(2,2,6)$ model after the first fixed-line projection is
\begin{align}
    W_{(2,2,6)}^{\Omega_\text{f.l.}^1} &= \frac{1}{2} T_{33} \left(X_{13}^2 - Y_{13}^2\right) + Y_{13}X_{34}X_{45}X_{51} - X_{13}X_{35}X_{51} \nonumber \\[5pt]
    & + X_{35}X_{56}X_{63} - X_{34}X_{46}X_{63} + X_{46}X_{67}X_{78}X_{84} - X_{56}X_{68}X_{84}X_{45} \nonumber \\[5pt]
    & + X_{68}X_{89}X_{96} - X_{67}X_{79}X_{96} + X_{79}Y_{90}X_{07} - X_{89}X_{90}X_{07}X_{78} \nonumber \\[5pt]
    &+ \frac{1}{2}
\widetilde T_{99}(X_{90}^2 - Y_{90}^2) \; .
\end{align}
The superpotential for the $(2,2,6)$ model after the second fixed lines projection is
\begin{align}
    W_{(2,2,6)}^{\Omega_\text{f.l.}^2} &= \frac{1}{2} T_{33} \left(X_{13}^2 - Y_{13}^2\right) + Y_{13}X_{34}X_{45}X_{51} - X_{13}X_{35}X_{51} \nonumber \\[5pt]
    & + X_{35}X_{56}X_{63} - X_{34}X_{46}X_{63} + X_{46}X_{68}X_{84} - X_{56}X_{67}X_{78}X_{84}X_{45} \nonumber \\[5pt]
    & + X_{67}X_{79}X_{96} - X_{68}X_{89}X_{96} + X_{89}Y_{90}X_{07}X_{78} - X_{79}X_{90}X_{07} \nonumber \\[5pt]
    &+ \frac{1}{2}
\widetilde T_{99}\left( X_{90}^2 - Y_{90}^2 \right) \; .
\end{align}

The superpotential for the $(2,2,6)$ model after the fixed points projection is
\begin{align}
    W_{(2,2,6)}^{\Omega_\text{f.p.}} &= T_{33}Y_{13}X_{13} + Y_{13}X_{35}X_{51} - X_{13}X_{34}X_{45}X_{51}  \nonumber \\[5pt]
    & + X_{34}X_{46}X_{63} - X_{35}X_{56}X_{63} + X_{56}X_{67}X_{78}X_{84}X_{45} - X_{46}X_{68}X_{84} \nonumber \\[5pt]
    & + X_{68}X_{89}X_{96} - X_{67}X_{79}X_{96} + X_{79}Y_{90}X_{07} - X_{89}X_{90}X_{07}X_{78} \nonumber \\[5pt]
    &+ \widetilde T_{99}Y_{90}X_{90} \; .
\end{align}
The superpotential for the $(0,4,6)$ model after the fixed points projection is
\begin{align}
    W_{(0,4,6)}^{\Omega_\text{f.p.}} &= T_{33}Y_{13}X_{13} + Y_{13}X_{35}X_{51} - X_{13}X_{34}X_{45}X_{51} \nonumber \\[5pt]
    & + X_{34}X_{46}X_{63} - X_{35}X_{56}X_{63} + X_{56}X_{68}X_{84}X_{45} - X_{46}X_{67}X_{78}X_{84} \nonumber \\[5pt]
    & + X_{67}X_{79}X_{96} - X_{68}X_{89}X_{96} + X_{89}Y_{90}X_{07}X_{78} - X_{79}X_{90}X_{07} \nonumber \\[5pt]
    &+ \widetilde T_{99}Y_{90}X_{90} \; .
\end{align}
The choice of gauge ranks that gives rise to the conformal duality is
\begin{eqnarray}
&&
N_1 = n,\, N_3 = n - 2 \tau,\, N_4 = N_5 = n - 4 \tau, \, \\
&&
N_6 = n - 6 \tau , \, N_7 = N_8 = n - 8 \tau ,\, N_9 = n - 10 \tau , N_0 = n - 12 \tau \;.
\nonumber 
\end{eqnarray}
The situation here is similar to the one in Sec. \ref{sec:(226)vs(046)}. The two cases with fixed lines are related by a field redefinition. Analogously the two cases with fixed points are related by a field redefinition. The cases with fixed lines are connected to the cases with fixed points by first considering Seiberg dualities on $SU(N_2)$ ad $SU(N_5)$ and then by suitable field redefinitions.

\subsection{$(3,3,6)$ - $(2,4,6)$ - $(0,6,6)$}
\label{177}

The last case that we present in this appendix correspond to the fixed lines projection of $(3,3,6)$ and to the fixed point projections of both $(2,4,6)$ and $(0,6,6)$.

\begin{figure}
\begin{center}
\includegraphics[scale=0.7]{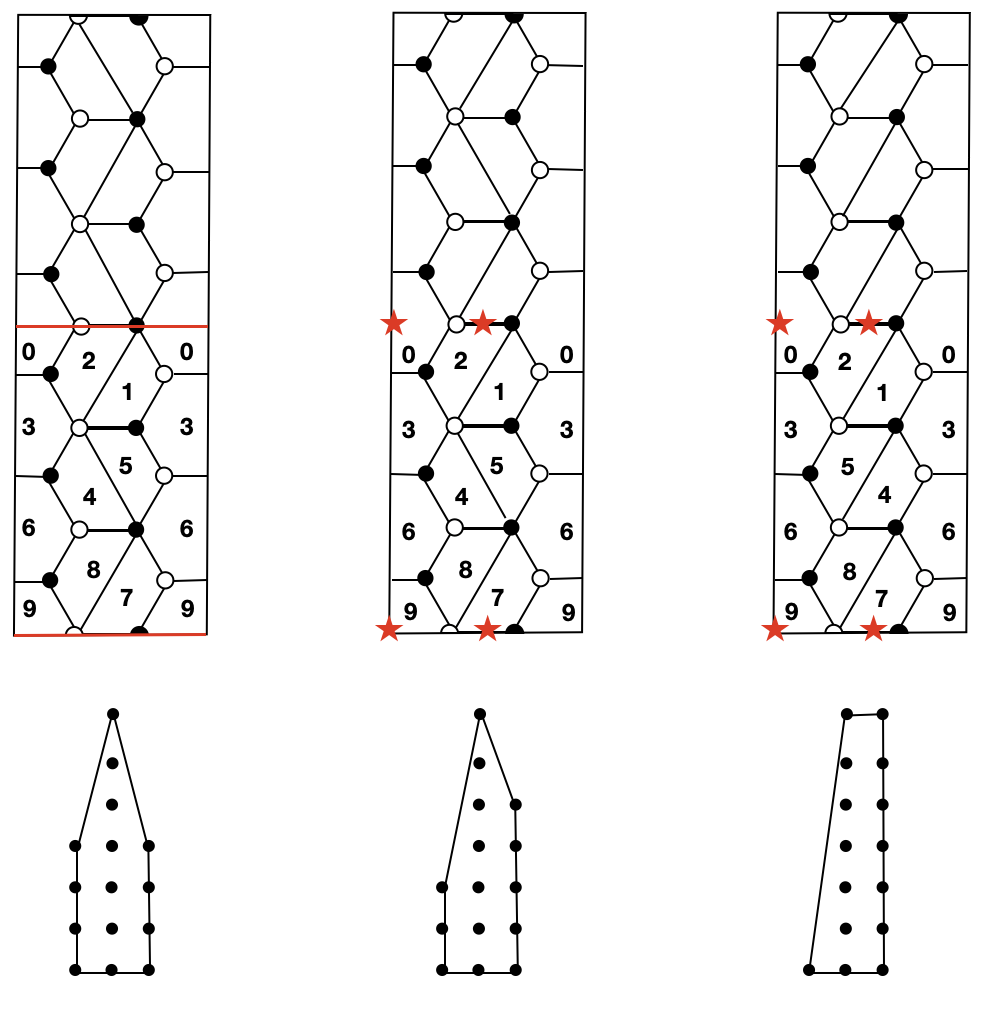}
\caption{Fixed-line projection of $(3,3,6)$ and fixed-point projections of $(2,4,6)$ and $(0,6,6)$.}
\label{FIG447357177}
\end{center}
\end{figure}

This case is interesting because it is the first one where there are three different toric diagrams that give origin to conformal dualities after the projection. There are degenerate choices of $(3,3,6)$ and $(2,4,6)$ but here for simplicity we study only a single possibility. The other Seiberg dual case become conformally dial after the projection as in the cases discussed above. The toric diagrams, the dimers and the projections are represented in Fig. \ref{FIG447357177}.
The fixed-line projection of the $(3,3,6)$ model under investigation has superpotential 
\begin{align}
    W_{(3,3,6)}^{\Omega_\text{f.l.}} &= \frac{1}{2}T_{22} (X_{02}^2-X_{01}^2X_{12}^2) + X_{01}X_{13}X_{30} - X_{02}X_{23}X_{30} \nonumber \\[5pt]
    &+ X_{23}X_{34}X_{45}X_{51}X_{12} - X_{13}X_{35}X_{51} + X_{35}X_{56}X_{63} - X_{34}X_{46}X_{63} \nonumber \\[5pt]
    &+ X_{46}X_{68}X_{84} - X_{56}X_{67}X_{78}X_{84}X_{45} + X_{67}X_{79}X_{96} - X_{68}X_{89}X_{96} \nonumber \\[5pt]
    &+ \frac{1}{2} \widetilde{T}_{77}(X_{78}^2 X_{89}^2- X_{79}^2) \; .
\end{align}
The fixed point projection of the $(2,4,6)$ model under investigation has superpotential 
\begin{align}
    W_{(2,4,6)}^{\Omega_\text{f.p.}} &= T_{22}X_{02}X_{01}X_{12} + X_{01}X_{13}X_{30} - X_{02}X_{23}X_{30} \nonumber \\[5pt]
    &+ X_{23}X_{34}X_{45}X_{51}X_{12} - X_{13}X_{35}X_{51} + X_{35}X_{56}X_{63} - X_{34}X_{46}X_{63} \nonumber \\[5pt]
    &+ X_{46}X_{68}X_{84} - X_{56}X_{67}X_{78}X_{84}X_{45} + X_{67}X_{79}X_{96} - X_{68}X_{89}X_{96} \nonumber \\[5pt]
    &+ \widetilde{T}_{77}X_{78}X_{89}X_{79} \; .
\end{align}
The fixed-point projection of the $(0,6,6)$ model has superpotential 
\begin{align}
    W_{(0,6,6)}^{\Omega_\text{f.p.}} &= T_{22}X_{02}X_{01}X_{12} + X_{01}X_{13}X_{30} - X_{02}X_{23}X_{30} \nonumber \\[5pt]
    &+ X_{23}X_{35}X_{51}X_{12} - X_{13}X_{34}X_{45}X_{51} + X_{34}X_{46}X_{63} - X_{35}X_{56}X_{63} \nonumber \\[5pt]
    &+ X_{56}X_{68}X_{84}X_{45} - X_{46}X_{67}X_{78}X_{84} + X_{67}X_{79}X_{96} - X_{68}X_{89}X_{96} \nonumber \\[5pt]
    &+ \widetilde{T}_{77}X_{78}X_{89}X_{79} \; .
\end{align}
Finally, the ranks read
\begin{eqnarray}
&&
N_0 = n, \quad
N_1 = N_2 = n - 2 \tau, \, \quad
N_3 = n - 4 \tau, \, \quad
N_4 = N_5 = n - 6 \tau,  \\
&&
N_6 = n - 8\tau, \, \quad
N_7 = N_8 = n - 10 \tau, \, \quad
N_9 = n - 12 \tau \; .
\end{eqnarray}
The first two cases are related by Seiberg dualities on $SU(N_1)$ and $SU(N_8)$ and by suitable field redefinitions. The second and the third case are related by Seiberg duality on $SU(N_4)$ and field redefinitions. The first and the third cases are then related by Seiberg duality on $SU(N_{1,4,8})$ and field redefinitions.

This concludes the analysis showing that the superpotentials of the three cases become identical up to sign factors, fitting with the analysis of section \ref{sec:toy}.

\section{Counting phases}\label{sec:apxcounting}

Given a general model $(k_1,k_2,k_3)$ there could be various configurations with the same numbers, i.e. various choices to combine the squares and the hexagons in the reduced cell of the dimer after the orientifold projection. For simplicity, let us call the reduced cell $C_{\Omega}$. A dual family of models can be denoted with two numbers $(k, k_3)$, where $k_1 + k_2 = k$; also, denote $k_3 = 2k + p$, so that we can distinguish between two cases: $p=0$ and $p\neq0$.

Let us focus first with $p=0$, so the dual family is identified by $(k,2k)$. There are only $k$ pairs of squares in the central column of $C_{\Omega}$, a certain number $n_{WE}$ with a long edge oriented along the $WE$ axis and another $n_{EW}$ along $EW$, related to $k_1$ and $k_2$ as explained above. We ask the question: how many configurations are there with a certain number of $n_{WE}$ and $n_{EW}$? In other words, how many conformally dual models in a family $(k,k_3)$?. We need to count the combinations of these elements that fill $k$ position. Since there are 2 choices per position, $n_{EW} = k - n_{WE}$ and in total there are $2^{k}$ configurations. These can be counted as
\begin{align}
    2^k = \sum_{n_{WE}=0}^{k} \binom{k}{n_{WE}} \; ,
\end{align}
where each summand $\binom{k}{n_{WE}}$ is the number of configurations of a model with $k$ squares of which $n_{WE}$ are $WE$ oriented, the remaining $n_{EW} = k - n_{WE}$ are $EW$. However, some of them are equivalent, i.e. they give the same quiver and superpotential with some indices recast. First, we can exchange $n_{WE} \leftrightarrow n_{EW}$, which means the cell $C_{\Omega}$ is reflected along a vertical axis and the property of the binomial coefficient
\begin{align}
    \binom{k}{n_{WE}} = \binom{k}{k - n_{WE}} = \binom{k}{n_{EW}} 
\end{align}
account for this reflection. We need to count only the configurations until $n_{WE} = \lfloor \frac{k}{2} \rfloor$. Second, each set of configurations $\binom{k}{n_{WE}}$ has an equivalence class, for we can reflect along a horizontal axis and a configuration is mapped to another one. If $k$ is odd, there is always one mapped to itself. The number of inequivalent configurations of a cell $N_{C_{\Omega}}$ with $k$ pairs of squares and no hexagons is then 
\begin{align}\label{eq:countingNoHex}
    &N_{C_{\Omega}} = \binom{k}{0} + \sum_{n_{WE}=1}^{\lfloor \frac{k}{2} \rfloor} \frac{1}{2} \binom{k}{n_{WE}} \; , \quad k \quad \mathrm{even} \; , \nonumber \\[5pt]
    &N_{C_{\Omega}} = \binom{k}{0} + \sum_{n_{WE}=1}^{\lfloor \frac{k}{2} \rfloor} \frac{1}{2} \left[ \binom{k}{n_{WE}} + 1 \right]\; , \quad k \quad \mathrm{odd} \; .
\end{align}
It is crucial to note that the orientifold projection is not yet chosen, i.e. the $\mathbb{Z}_2$ involution is not specified. Each inequivalent arrangement of squares in $\binom{k}{n_{WE}}$ gives a cell $C_{\Omega}$, and we choose a fixed point projection if the $\mathbb{Z}_2$ image cell $\overline{C}_{\Omega} = C_{\Omega}$. The copy has the same number of $WE$ pairs of squares $\overline{n}_{WE} = n_{WE}$, so that $k_1 = 2 n_{WE}$ and $k_2 = 2k - k_1 = 2 n_{EW}$. Thus, a projection with fixed points must have $k_1$ and $k_2$ even. We can express Eq. \ref{eq:countingNoHex} in terms of $k_1$ and count the number of orientifold projections with fixed points as
\begin{align}\label{eq:countingNoHexFP}
    &N_{C_{\Omega}}^\text{f.p.} = \binom{k}{0} + \sum_{k_1=2}^{ 2 \lfloor \frac{k}{2} \rfloor} \frac{1}{2} \binom{k}{\frac{k_1}{2}} \; , \quad k \quad \mathrm{even} \; , \nonumber \\[5pt]
    &N_{C_{\Omega}}^\text{f.p.} = \binom{k}{0} + \sum_{k_1=2}^{2 \lfloor \frac{k}{2} \rfloor} \frac{1}{2} \left[ \binom{k}{\frac{k_1}{2}} + 1 \right]\; , \quad k \quad \mathrm{odd} \; .
\end{align}

Each summand in Eq. \ref{eq:countingNoHexFP} counts the number of inequivalent fixed point projections with the same $(k_1, 2k - k_1, 2k)$. We can depict Eq. \ref{eq:countingNoHexFP} as the sum of the possible fixed points projections of a toric diagram with $(k_1, 2k - k_1, 2k)$. 

On the other hand, the fixed lines projection is given by $\overline{C}_{\Omega} = - C_{\Omega}$. This $\mathbb{Z}_2$ involution changes the direction of the squares, so it sends $n_{WE} \to \overline{n}_{EW}$ and $n_{EW} = k - n_{WE} \to \overline{n}_{WE} = k - \overline{n}_{EW}$. As a consequence, $k_1 = n_{WE} + \overline{n}_{WE} = k_2 = k$, with no restriction on $k_1$ being even or odd. The number of models projected with fixed lines is 

\begin{align}\label{eq:countingNoHexFL}
    &N_{C_{\Omega}}^\text{f.l.} = \binom{k}{0} + \sum_{n_{WE}=1}^{\lfloor \frac{k}{2} \rfloor} \frac{1}{2} \binom{k}{n_{WE}} \; , \quad k \quad \mathrm{even} \; , \nonumber \\[5pt]
    &N_{C_{\Omega}}^\text{f.l.} = \binom{k}{0} + \sum_{n_{WE}=1}^{\lfloor \frac{k}{2} \rfloor} \frac{1}{2} \left[ \binom{k}{n_{WE}} + 1 \right]\; , \quad k \quad \mathrm{odd} \; ,
\end{align}
all of them with $(k_1, k_2, k_3)=(k, k, 2k)$.

Let us show two examples. Consider $k=1$, which gives the PdP family of models. The unique $\binom{k}{0}$ configuration gives the reduced cell $C_{\Omega}$. Then we can perform the projection with fixed points, so that the $\mathbb{Z}_2$ copy $\overline{C}_{\Omega} = C_{\Omega}$ and we have PdP$_{3c}^{\Omega}$, $(k_1,k_2,k_3)=(0,2,2)$. Or, we can perform the orientifold projection with fixed lines, so that $\overline{C}_{\Omega} = - C_{\Omega}$ and we have PdP$_{3b}^{\Omega}$, $(k_1,k_2,k_3)=(1,1,2)$.

Consider now $k=2$, studied in Sec. \ref{sec155}. The first configuration is $\binom{2}{0}$, and we can decide for fixed-point projection $\overline{C}_{\Omega}=C_{\Omega}$ or fixed-line projection $\overline{C}_{\Omega}=-C_{\Omega}$, respectively the rightmost $(k_1,k_2,k_3)=(0,4,4)$ and the leftmost $(k_1,k_2,k_3)=(2,2,4)$ dimers in Fig. \ref{FIG335155}. Then, we have $\binom{2}{1}$, whose reduced cell $C_{\Omega}$ is drawn in the central dimers of Fig. \ref{FIG335155}, whose central left is given by a fixed-point orientifold $\overline{C}_{\Omega}=C_{\Omega}$ and central right by fixed-line orientifold $\overline{C}_{\Omega}=-C_{\Omega}$, both $(k_1,k_2,k_3)=(2,2,4)$. 

When $p \neq 0$, we need to distinguish between two orientifold projections, $\Omega_A$ and $\Omega_B$. The former consist in the projection where projected gauge factors come from the side of the cell $C_{\Omega}$, whereas the latter in the projection where at least one projected gauge factor comes from the central line in $C_{\Omega}$. When we add hexagons to the central lines of the cell $C_{\Omega}$, i.e. $p\neq0$, both types of the projection are allowed. If $p$ is odd the number of complete hexagons $n_h = (p-1)/2$, as half hexagon is added at one of the boundary of the reduced cell $C_{\Omega}$, in this case the projection is always $\Omega_B$. If $p$ is even we have two choices, projection $\Omega_A$ and $n_h = p/2$ complete hexagons in $C_{\Omega}$, or projection $\Omega_B$ and $n_h = p/2 - 1$ complete hexagons in $C_{\Omega}$ and one half hexagons on each boundary of the reduced cell. Once we have defined the number $n_h$, we have $k + n_h = n_{WE} + n_{EW} + n_h$ slots to be filled with squares of the two types and hexagons. The three numbers $(n_{WE},n_{EW}, n_h)$ defines a model, since they are related to $k_1$, $k_2$ and $k_3$. The number of possible configurations is the number of ways we can fill the slots, i.e. the number of permutations of 3 elements
\begin{align}\label{eq:countingHex}
    &P_{(k+n_h)}^{n_{WE},(k-n_{WE}),n_h} = \frac{(k+n_h)!}{n_{WE}!(k-n_{EW})!n_h!} \; , \nonumber \\[5pt]
    &N_{C_{\Omega}} = P_{(k+n_h)}^{0,k,n_h} + \sum_{n_{WE}=1}^{\lfloor \frac{k}{2} \rfloor} \frac{1}{2} P_{(k+n_h)}^{n_{WE},(k-n_{WE}),n_h} \; , \quad (k+n_h) \quad \mathrm{even} \; , \nonumber \\[5pt]
    &N_{C_{\Omega}} = P_{(k+n_h)}^{0,k,n_h} + \sum_{n_{WE}=1}^{\lfloor \frac{k}{2} \rfloor} \frac{1}{2} \left[ P_{(k+n_h)}^{n_{WE},(k-n_{WE}),n_h} + 1 \right] \; , \quad (k+n_h) \quad \mathrm{odd} \; , 
\end{align}
where we have already accounted for the exchange $n_{WE} \leftrightarrow n_{EW}$ and the reflection of $C_{\Omega}$ around an horizontal axis. Note that for $p=0$, i.e. $n_h = 0$, Eq. \ref{eq:countingNoHex} reduces to Eq. \ref{eq:countingHex}.

Now, we need to choose the orientifold projection, either with fixed points $\overline{C}_{\Omega} = C_{\Omega}$ or with fixed lines $\overline{C}_{\Omega} = - C_{\Omega}$. As before, for fixed points orientifolds $k_1 = 2 n_{WE}$, $k_2 = 2k - k_1 = n_{EW}$ and by definition $p = k_3 - 2k$. Again, we can express Eq. \ref{eq:countingHex} as a sum over $k_1$ even
\begin{align}\label{eq:countingHexFP}
    &N_{C_{\Omega}}^\text{f.p.} = P_{(k+n_h)}^{0,k,n_h} + \sum_{k_1=2}^{2 \lfloor \frac{k}{2} \rfloor} \frac{1}{2} P_{(k+n_h)}^{(k_1/2),(k-k_1/2),n_h} \; , \quad (k+n_h) \quad \mathrm{even} \; , \nonumber \\[5pt]
    &N_{C_{\Omega}}^\text{f.p.} = P_{(k+n_h)}^{0,k,n_h} + \sum_{k_1=2}^{2 \lfloor \frac{k}{2} \rfloor} \frac{1}{2} \left[ P_{(k+n_h)}^{k_1/2,(k-k_1/2),n_h} + 1 \right] \; , \quad (k+n_h) \quad \mathrm{odd} \; , 
\end{align}
where each summand with the same value of $k_1$ gives the number of conformally dual models with $(k_1, 2k-k_1, 2k + p)$. 
For fixed lines orientifolds, $k_1 = k_2 = k$ and 
\begin{align}\label{eq:countingHexFL}
    &N_{C_{\Omega}}^\text{f.l.} = P_{(k+n_h)}^{0,k,n_h} + \sum_{n_{WE}=1}^{\lfloor \frac{k}{2} \rfloor} \frac{1}{2} P_{(k+n_h)}^{n_{WE},(k-n_{WE}),n_h} \; , \quad (k+n_h) \quad \mathrm{even} \; , \nonumber \\[5pt]
    &N_{C_{\Omega}}^\text{f.l.} = P_{(k+n_h)}^{0,k,n_h} + \sum_{n_{WE}=1}^{\lfloor \frac{k}{2} \rfloor} \frac{1}{2} \left[ P_{(k+n_h)}^{n_{WE},(k-n_{WE}),n_h} + 1 \right] \; , \quad (k+n_h) \quad \mathrm{odd} \; , 
\end{align}
and all are conformally dual models with $(k,k,2k+p)$.

\bibliographystyle{JHEP}
\bibliography{biblio}

\end{document}